\documentclass[aps,prx,amsmath,twocolumn,groupedaddress,letterpaper,floatfix]{revtex4}
\usepackage{graphicx,color}
\usepackage{verbatim}
\usepackage{amssymb}   
\usepackage{amsmath}
\usepackage{amsfonts}
\usepackage{mathdots}
\usepackage{hyperref}
\usepackage{epsfig}
\usepackage{braket}
\usepackage{bm}
\begin{document}
	\title{Valley-Polarized Quantum Anomalous Hall State in Moir\'e MoTe$_2$/WSe$_2$ Heterobilayers}
	
	\author{Ying-Ming Xie$^1$}
	\author{Cheng-Ping Zhang$^1$}
	\author{Jin-Xin Hu$^1$}
	\author{Kin Fai Mak$^{2,3,4}$}
	\author{K. T. Law$^1$} \thanks{Corresponding author.\\phlaw@ust.hk}
	
	\affiliation{$^1$Department of Physics, Hong Kong University of Science and Technology, Clear Water Bay, Hong Kong, China} 	
	\affiliation{$^2$School of Applied and Engineering Physics, Cornell University, Ithaca, NY, USA}
	\affiliation{$^3$Kavli Institute at Cornell for Nanoscale Science, Ithaca, NY, USA}
	\affiliation{$^4$Laboratory of Atomic and Solid State Physics, Cornell University, Ithaca, NY, USA}

	\date{\today}
\begin{abstract}
	Moir\'e heterobilayer transition metal dichalcogenides (TMDs) emerge as an ideal system for simulating the single-band Hubbard model and interesting correlated phases have been observed in these systems. Nevertheless, the moir\'e bands in heterobilayer TMDs were believed to be topologically trivial. Recently, it was reported that both a quantum valley Hall insulating state at filling $\nu=2$ (two holes per moir\'e unit cell)  and a valley-polarized quantum anomalous Hall state at filling $\nu=1$ were observed in AB stacked moir\'e MoTe$_2$/WSe$_2$ heterobilayers. However, how the topologically nontrivial states emerge is not known. In this work, we propose that the pseudo-magnetic fields induced  by lattice relaxation in moir\'e MoTe$_2$/WSe$_2$ heterobilayers could naturally give rise to moir\'e bands with finite Chern numbers. We show that a time-reversal invariant quantum valley Hall insulator is formed at full-filing $\nu=2$, when two moir\'e  bands with opposite Chern numbers are filled. At half-filling $\nu=1$, Coulomb interaction lifts the valley degeneracy  and results in a valley-polarized quantum anomalous Hall state, as observed in the experiment. Our theory identifies a new way to achieve topologically non-trivial states in heterobilayer TMD materials.
\end{abstract}


\maketitle
{\emph {Introduction.}}--- Recently, there is an intense study on the moir\'e superlattices such as in twisted bilayer graphene \cite{Cao2018,Cao2018sc,Yankowitz2019,David2019, Kerelsky2019,Xie2019,Young2020,Bistritzer2011, Po2018,Liang2018} and twisted bilayer  transition metal dichalcogenides (TMDs) \cite{Wang2020,Zhang2020,Jain2018,fengcheng2019,Bi2021,Zhange_hetero_2017,Yaowang2017,fengcheng2018,Fengwang2019,Xiaoqin2019,Xiaodong2019,Shimazaki2020,Fai_hubbard2020,Fengwang2020,Fai_fractional2020,Huang_fractional2021,Yangzhang2020,Fai_strip2021,Fai2021}. The narrow moir\'e bands together with strong electron-electron interactions give rise to various interesting  quantum states of matter.  

The  moir\'e superlattices formed by TMD heterobilayers are particularly interesting \cite{Zhange_hetero_2017,Yaowang2017,fengcheng2018,Fengwang2019,Xiaoqin2019,Xiaodong2019,Shimazaki2020,Fai_hubbard2020,Fengwang2020,Fai_fractional2020,Huang_fractional2021,Yangzhang2020,Fai_strip2021,Fai2021,Macdonald2021}.   A moir\'e TMD heterobilayer is formed by stacking two different 2H-structure monolayer TMDs  MX$_2$ and $\tilde{\text{M}}\tilde{\text{X}}_2$. Due to the large energy offset of the valence bands of the two TMDs, the electrons near the Fermi energy are mostly originated from the TMD layer with higher valence band energy. Therefore, a moir\'e TMD heterobilayer can be approximately treated as a monolayer TMD with an additional moir\'e potential which is created through interlayer couplings. Moreover, due to the large Ising spin-orbit coupling \cite{XiaoDi2012,Xi2016,Lu2015}, spin degeneracy is lifted while the valley degeneracy plays the role of pseudo-spin. In the presence of Coulomb interactions, a moir\'e TMD heterobilayer can be treated as a single-band Hubbard model simulator \cite{fengcheng2018,Fai_hubbard2020} with parameters highly tunable through the twist angle and the displacement field. Several interesting correlated phenomena such as Mott insulating states\cite{Fai_hubbard2020},  Wigner crystal states\cite{Fengwang2020}, Pomeranchuk effect and continuous Mott transition \cite{Fai2021} have been observed in AA stacked moir\'e TMD heterobilayers. However, so far the moir\'e bands in heterobilayers are expected to be topologically trivial and topology does not play a role in these correlated phases.

Surprisingly, in a recent experiment with AB stacked moir\'e MoTe$_2$/WSe$_2$ heterobilayers, a quantum valley Hall insulator state at full-filling $\nu=2$, i.e., two holes per moir\'e unit cell, and a quantum anomalous Hall state at half-filling $\nu=1$ were observed \cite{Fai_ex2021}. As the quantized Hall resistance strongly correlates with valley polarization through  magnetic circular dichroism measurements, it is strongly suggestive that the quantum anomalous Hall state is a valley-polarized anomalous Hall state\cite{Fai_ex2021}. Although it was predicted that the quantum valley Hall state can emerge in homobilayer TMDs \cite{fengcheng2019} which can be described by a Kane-Mele model \cite{Kane2005}, the low energy description of heterobilayers is dramatically different due to the large offset of the energy of the bands which is estimated to be around 300 meV \cite{Fai2021,Fai_ex2021} and large differences in interlayer tunnelling strength. Therefore, the origin of the topologically non-trivial bands in heterobilayers is not known. 

In this work, we point out that a periodically modulated pseudo-magnetic field, which could emerge spontaneously under lattice relaxation, can give rise to topologically nontrivial moir\'e bands. Specifically, (i) the pseudo-magnetic field can create Chern bands with opposite Chern numbers at opposite valleys. As a result, a quantum valley Hall insulating phase would form at $\nu=2$, when the topmost moir\'e bands at two valleys are filled; (ii) Importantly,  at half-filling $\nu=1$, based on a self-consistent Hartree-Fock calculation, we found the Coulomb interactions could lift the degeneracies  of the two valleys. It results in an interactions-driven valley-polarized quantum anomalous Hall phase as observed in the recent experiments. Our theory identifies a new way to achieve topologically non-trivial states in heterobilayer TMD materials which were considered topologically trivial. 



{\emph {Model.}}--- As pointed out in Ref.~\cite{fengcheng2018}, due to the large Ising spin-orbit coupling which breaks the spin-degeneracy and the layer asymmetry, a TMD heterobilayer can be described by a single-band Hubbard model with the valley degrees of freedom playing the role of pseudo-spins. However, the resulting moir\'e bands are topologically trivial. One important element which was not considered in the original model of TMD heterobilayer \cite{fengcheng2018} is lattice relaxation. Indeed, it has been shown that local strain can result in lattice relaxation and even lattice reconstructions which are important in twisted bilayer graphene \cite{Koshino2017,Shihaohao2020} and twisted TMDs \cite{Falko2020,Jain2021,Falko2021,Weston2020, Linnian2021,Li_Crommie2021}.  Importantly, the lattice relaxation can generate periodically modulated pseudo-magnetic fields which play an important role in the moir\'e band structure \cite{Koshino2017}.

%


To capture the effects of periodic pseudo-magnetic fields $B(\bm{r})$, we include an additional gauge field $\bm{A}$ with $B(\bm{r})=\nabla \times \bm{A}(\bm{r})$ into the previously proposed model Hamiltonian for the moir\'e TMD heterobilayer  \cite{fengcheng2018,Yangzhang2020} as   $H=\int d\bm{r}\psi^{\dagger}_{\tau}(\bm{r})\mathcal{H}_{\tau} (\bm{r})\psi_{\tau}(\bm{r})$. Here,
\begin{equation}
	\mathcal{H}_{\tau}(\bm{r})=-\frac{(\hat{\bm{p}}+\tau e\bm{A})^2}{2m^*}+V(\bm{r}),\label{moire}
\end{equation}
where the momentum operator $\hat{\bm{p}}=-i\hbar\nabla$,   $m^*$ is the valence band effective mass,  $\tau=\pm$ for $\pm K$ valley.  The moir\'e potential is  $V(\bm{r})=2V_0\sum_{j=1,3,5}\cos(\bm{G}_j\cdot\bm{r}+\phi)$ with moir\'e wave vectors $\bm{G}_{j}=\frac{4\pi}{\sqrt{3}L_{M}}(\cos(\frac{(j-1)\pi}{3}),\sin(\frac{(j-1)\pi}{3}))$, $L_M\approx a/\sqrt{\delta^2+\theta^2}$ is the moir\'e lattice constant  with a lattice constant mismatch $\delta=(a-a')/a$ and a twist angle $\theta$. To be specific, we adopted the parameters: $m^*=0.6m_0$ with $m_0$ the electron mass, $a=3.565$\AA, $a'=3.317$\AA\ \cite{Mounet2018,Meckbach2020,li2018type} for the band structure calculations of TMD MoTe$_2$/WSe$_2$  heterobilayers, where the top valence  moir\'e bands originate from MoTe$_2$ layer \cite{Fai2021, Fai_ex2021}.  The model Hamiltonian $H$ respects $C_3$ symmetry and time-reversal symmetry $T=\tau_x\hat{K}$ with $\hat{K}$ as complex conjugate operation, and the moir\'e Hamiltonians of the two valleys are related by time-reversal symmetry: $T\mathcal{H}_{\tau}(\bm{r})T^{-1}=\mathcal{H}_{-\tau}(\bm{r})$. 
%

We  first consider the case  without the pseudo-magnetic fields $B(\bm{r})$, i.e., $\bm{A}=0$, the moir\'e Hamiltonian exhibits a spinless time-reversal symmetry: $T'\mathcal{H}_{\tau}(\bm{r})T'^{-1}=\mathcal{H}_{\tau}(\bm{r})$ with $T'=\hat{K}$. This spinless time-reversal symmetry enforces  the Berry curvature to be an odd function: $\Omega(\bm{k})=-\Omega(-\bm{k})$ (Supplementary Material (SM) Sec.~I\cite{Supp}). As a result, the Chern number of each moir\'e band is zero. To break this spinless time-reversal symmetry,   an additional periodically modulated pseudo-magnetic field $\bm{B}$ is introduced in the moir\'e Hamiltonian (\ref{moire}). Evidently, in this case $T'\mathcal{H}_{\tau}(\bm{r})T'^{-1}\neq\mathcal{H}_{\tau}(\bm{r})$. Hence, moir\'e bands with finite Chern numbers are allowed. 
\begin{figure}
	\centering
	\includegraphics[width=1\linewidth]{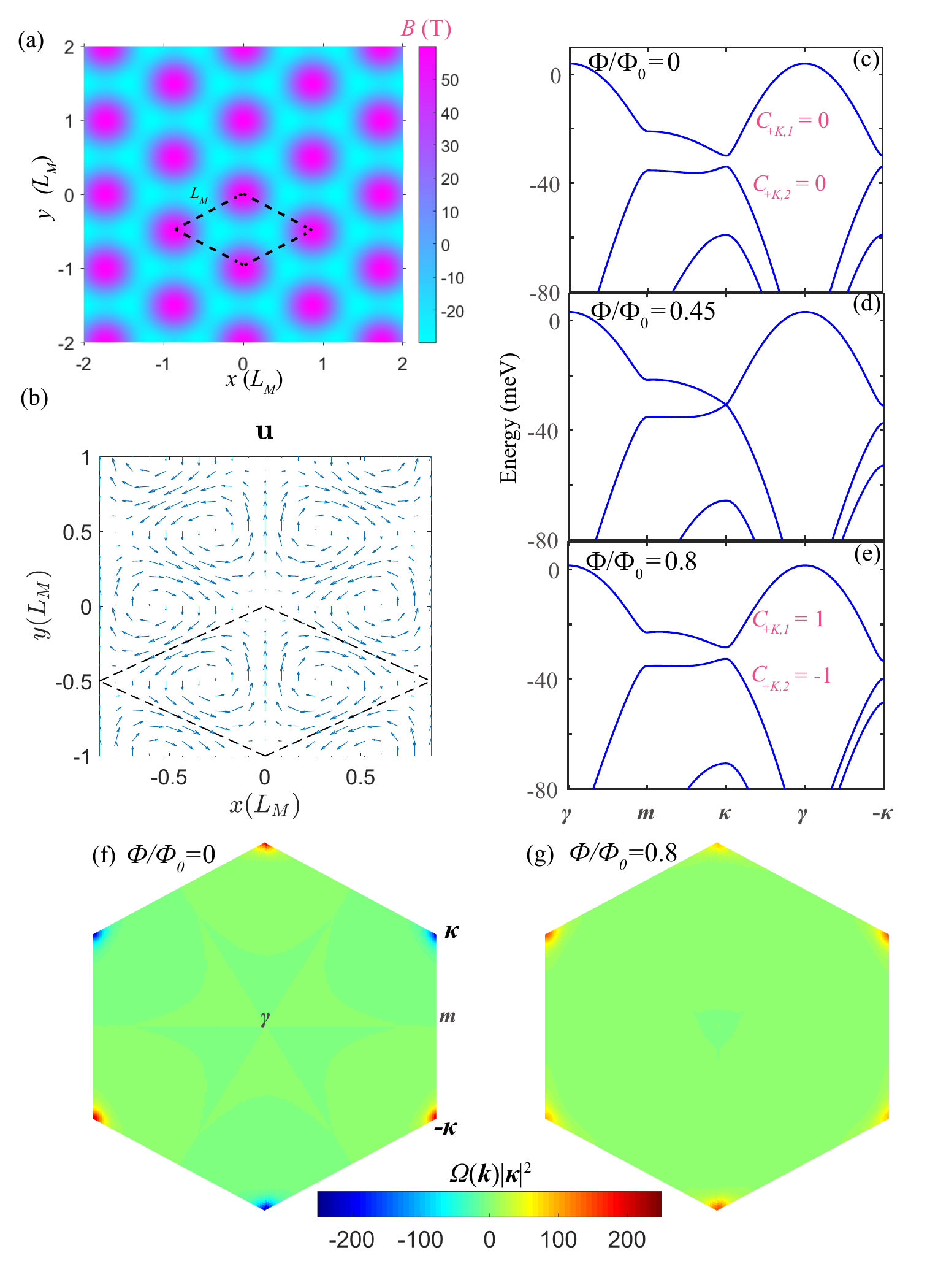}
	\caption{(a) and (b) show the topography of a $C_3$ symmetric periodic pseudo-magnetic field ($B_0=20$T) and the corresponding  strain displacement field $\bm{u}$, respectively. (c), (d) and (e) show the moir\'e band structures at valley $K$ calculated at $\Phi/\Phi_0=0, 0.45, 0.8$ respectively, where the moir\'e potential parameters are taken as $V_0=10$ meV, $\phi=0.3\pi$,  $\theta=0.53^{\circ}$. The top two moir\'e bands at $\Phi/\Phi_0$ with Chern number $C_{K,1}$ and $C_{K,2}$ are highlighted. The corresponding  distributions of the Berry curvature within moir\'e Brillouin zone at $\Phi/\Phi_0=0$ and $\Phi/\Phi_0=0.8$ are shown in (f) and (g), respectively.}
	\label{fig:fig1}
\end{figure}

To be specific, we consider a $C_3$-invariant periodic pseudo-magnetic field: $B(\bm{r})=B_0\sum_{j=1,3,5}\cos(\bm{G}_j\cdot \bm{r})$, which is expected to emerge  in an AB stacked moir\'e TMD bilayer under lattice relaxation as shown in Ref.~\cite{Falko2020} or can be generated by some out of plane corrugation effects \cite{Mao2020, Fu_zhang2021}.   The topography of this pseudo-magnetic field is shown in Fig.~\ref{fig:fig1}(a). It  displays the same period as the moir\'e superlattice, and it is important to note that the net flux in each moir\'e unit cell is zero, as we will see later the topology of this system can be understood in terms of the Haldane model \cite{Haldane1988}.  The corresponding gauge field of $B(\bm{r})$ is derived as $\bm{A}(\bm{r})=A_0[\bm{a_2}\sin (\bm{G_1\cdot r})-\bm{a_1}\sin(\bm{G_3}\cdot\bm{r})-\bm{a_3}\sin(\bm{G_5}\cdot\bm{r})]$, where $\bm{a_1}=(\frac{\sqrt{3}}{2},\frac{1}{2})L_M$, $\bm{a_2}=(0,1)L_M$,  $\bm{a_3}=\bm{a_2}-\bm{a_1}$, and  $A_0=\sqrt{3}B_0/4\pi$. This gauge field can be generated by a two-dimensional strain field  $u_{ij}(\bm{r})=(\partial_i u_j(\bm{r})+\partial_j u_i(\bm{r}))/2$ \cite{Levy2010,Guinea2010,Mao2020} with $\bm{A}=\alpha (2u_{xy}, u_{xx}-u_{yy})$. 
The strain displacement field  that gives rise to the periodic $B(\bm{r})$ is plotted in Fig.~\ref{fig:fig1}(b) (see the detail in Supplementary Material(SM)\cite{Supp}). The periodic strain displacement has been observed in moir\'e TMD bilayers \cite{Weston2020, Linnian2021,Li_Crommie2021}. Physically, there are different types of local stacking configurations which result in different lattice relaxation within a moir\'e unit cell  \cite{Falko2020, Supp}. As presented in SM \cite{Supp} such lattice relaxation could generate the periodic pseudomagnetic fields we introduced.  

Inserting $\bm{A}(\bm{r})$ into Eq.~(\ref{moire}), we obtained $\mathcal{H}_{\tau}(\bm{r})=-\frac{(\hat{\bm{p}}+\tau \frac{\Phi}{\Phi_0}\bm{\tilde{A}})^2}{2m^*}+V(\bm{r})$, 
where  $\Phi_0=\frac{h}{e}$ is a flux quantum and $\Phi=\frac{\sqrt{3}}{2}B _0L_M^2$ has the dimension of magnetic flux, $\tilde{\bm{A}}=\frac{4\pi\bm{A}(\bm{r})}{\sqrt{3}B_0L_M^2}$. We can then diagonalize  $\mathcal{H}_{\tau}(\bm{r})$ with plane wave basis to obtain the moir\'e bands (SM Sec.~IVA \cite{Supp}). 


As an illustration,  in Fig.~\ref{fig:fig1}(c) to (e), we show  the energy spectrum of $+K$ valley ($\tau=+1$) at a commensurate angle $\theta=0.53^{\circ}$ \cite{Yangzhang2020} but with different strength of pseudo-magnetic field: $\Phi/\Phi_0=0,0.45,0.8$. The corresponding  Berry curvature distribution of top moir\'e bands at $\Phi/\Phi_0=0$ and $\Phi/\Phi_0=0.8$ are shown in Fig.~\ref{fig:fig1}(f) and Fig.~\ref{fig:fig1}(g).   Without the pseudo-magnetic field, the  moir\'e band carries zero Chern number (labeled in Fig.~\ref{fig:fig1}(c)), although there appears to be finite Berry curvature at  moir\'e Brillouin zone corners $\pm \kappa$ (Fig.~\ref{fig:fig1}(f)). With an increase in the pseudo-magnetic field, it can be seen that the gap at $\kappa$ is closed at $\Phi/\Phi_0=0.45$  and reopens when $\Phi/\Phi_0>0.45$, which results in finite Chern numbers for the top two moir\'e bands (labeled in Fig.~\ref{fig:fig1}(e)). It is clear from Fig.~\ref{fig:fig1}(g) that, for the band with a finite Chern number, the Berry curvature has the same sign in the whole Brillouin zone. 

{\emph {Topological phase diagram.}}---To understand the topological phase transition, we derived an effective Hamiltonian near $\pm \kappa$ by performing perturbation theory at three moir\'e Brillouin corners  connected by the  reciprocal lattice vectors for the moir\'e pattern (SM Sec.~IIA \cite{Supp}).  The resulting effective Hamiltonian is
\begin{widetext}
	\begin{equation}
		H_{\pm\kappa}^{eff}(\bm{k})=\begin{pmatrix}
			\epsilon_0+2V_0\cos \phi&\pm \frac{1}{2}v(k_x+ik_y)&\pm \frac{1}{2}v(k_x-ik_y)\\
			\pm\frac{1}{2}v(k_x-ik_y)&\epsilon_0+2V_0\cos (\frac{2\pi}{3}+\phi)\mp \epsilon_B&\pm\frac{i}{2}v(k_x+ik_y)\\
			\pm\frac{1}{2}v(k_x+ik_y)&\mp \frac{i}{2}v(k_x-ik_y)&\epsilon_0+2V_0\cos (\frac{4\pi}{3}+\phi)\pm \epsilon_B
		\end{pmatrix},
	\end{equation}
\end{widetext}
where $\bm{k}$ is expanded near $\pm \kappa$,  $\epsilon_0=-\kappa^2/2m^*$, $v=\kappa/m^*$,	$\epsilon_B=\frac{\sqrt{3}{v}}{4L_M}\frac{\Phi}{\Phi_0}=\frac{\hbar eB_0}{4m^*}$ is a magnetic energy due to the presence of pseudo-magnetic fields. Interestingly, this three-band Hamiltonian exhibits as  a Dirac Hamiltonian at every two-band subspace. Moreover, the $\epsilon_B$ shifts the Dirac mass in opposite way  at $+ \kappa$ and $-\kappa$. This feature maps the effective Hamiltonian back to the Haldane model \cite{Haldane1988} and  moir\'e bands with finite Chern number would thus  be created when the  Dirac mass term changes sign at  $+ \kappa$ or $-\kappa$. The topological phase transition boundary lines are  obtained as: \begin{eqnarray}
	L_1: \frac{\epsilon_B}{V_0}&&=\pm 2[\cos(\frac{2\pi}{3}+\phi)-\cos\phi];\label{line1}\\
	L_2: \frac{\epsilon_B}{V_0}&&=\pm 2[\cos\phi-\cos(\frac{4\pi}{3}+\phi)];\label{line2}\\
	L_3: \frac{\epsilon_B}{V_0}&&=\pm [\cos(\frac{2\pi}{3}+\phi)-\cos(\frac{4\pi}{3}+\phi)]\label{line3}.
\end{eqnarray} 
Surprisingly, the  topological phase transition boundary lines $L_j$ only rely on the ratio $\frac{\epsilon_B}{V_0}$ and the phase $\phi$ of the moir\'e potential in the perturbative regime, where the moir\'e bandwidth $\epsilon_W$ is much larger than the magnetic energy $\epsilon_B$.  

\begin{figure}
	\centering
	\includegraphics[width=1\linewidth]{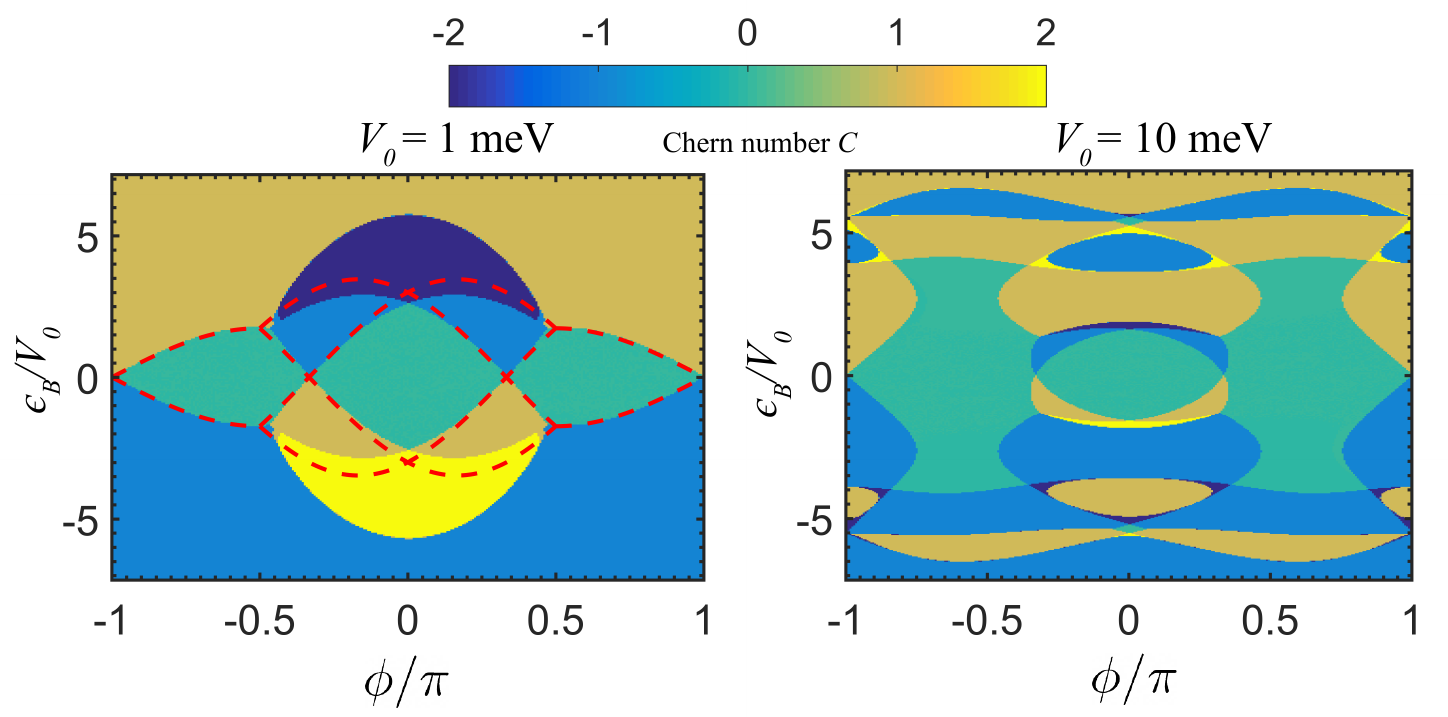}
	\caption{ (a) and (b) respectively, display the Chern number $C$ as a function of phase $\phi$ and the strength of the pseudo-magnetic fields characterized by the ratio of $\epsilon_B/V_0$ with $V_0=1$ meV and $V_0=10$ meV. The red dashed lines represent the phase boundaries given by the effective Hamiltonian.  }
	\label{fig:fig2}
\end{figure}

To determine the possible nontrivial topological regions,  the Chern number $C$ of the top moir\'e band with various $\phi$ and ratio $\epsilon_B/V_0$   is calculated numerically (SM Sec.~IVB \cite{Supp}). The typical topological phase diagram within ($V_0=1$ meV) and beyond ($V_0=10$ meV) the perturbation region are displayed in Fig.~\ref{fig:fig2}(a) and Fig.~\ref{fig:fig2}(b), respectively, where  $C$ is found to be able to take the value of $0$, $\pm 1$ and $\pm  2$. The  phase boundaries given  by the effective Hamiltonian (2) are  also depicted in Fig.~\ref{fig:fig2} as red dashed lines. In Fig.~\ref{fig:fig2}(a), impressively, most of the phase  boundaries in numerical results match the results from the effective Hamiltonian. As shown in Fig.~\ref{fig:fig2}(b), the phase boundaries become more complicated beyond the perturbative regime. Nevertheless,  there is still a large proportion of parameter space that exhibits Chern number $C=\pm 1$. 

We now discuss the  accessibility of  parameter space with finite Chern numbers. As shown in Fig.~\ref{fig:fig2},  the optimal $\phi$ is near $\pi/3$ and $\pi$, and a large magnetic energy $\epsilon_B$ at the order of $V_0$ is desired. Note that the magnitude of $\epsilon_B=\frac{\hbar eB_0}{4m^*}\sim 0.05 B_0$ meV/T is determined by the strength of pseudo-magnetic fields $B_0$. Considering a $B_0$  of  tens of T, the magnetic energy $\epsilon_B$ is estimated to be several meV which is achievable in heterobilayer TMDs and other moir\'e materials \cite{Falko2020}. As shown in Fig.~~\ref{fig:fig2}, this $\epsilon_B$ is certainly sufficient to drive the system to be topological for  $\phi$ near $\pi/3$ and  $\pi$, while   for $\phi$ far from these regions, it would depend on the magnitude of $V_0$.  A large $V_0$ would tend to make the system  trivial  since it would enhance the trivial energy gap between moir\'e bands. 
In the case of a large $V_0$ (tens of meVs), the topological region may still be  achievable  through a displacement field \cite{Fai2021}, because the displacement field effectively tunes the interlayer tunneling so that $V_0$ could be effectively changed.

%
%

\begin{figure}
	\centering
	\includegraphics[width=1\linewidth]{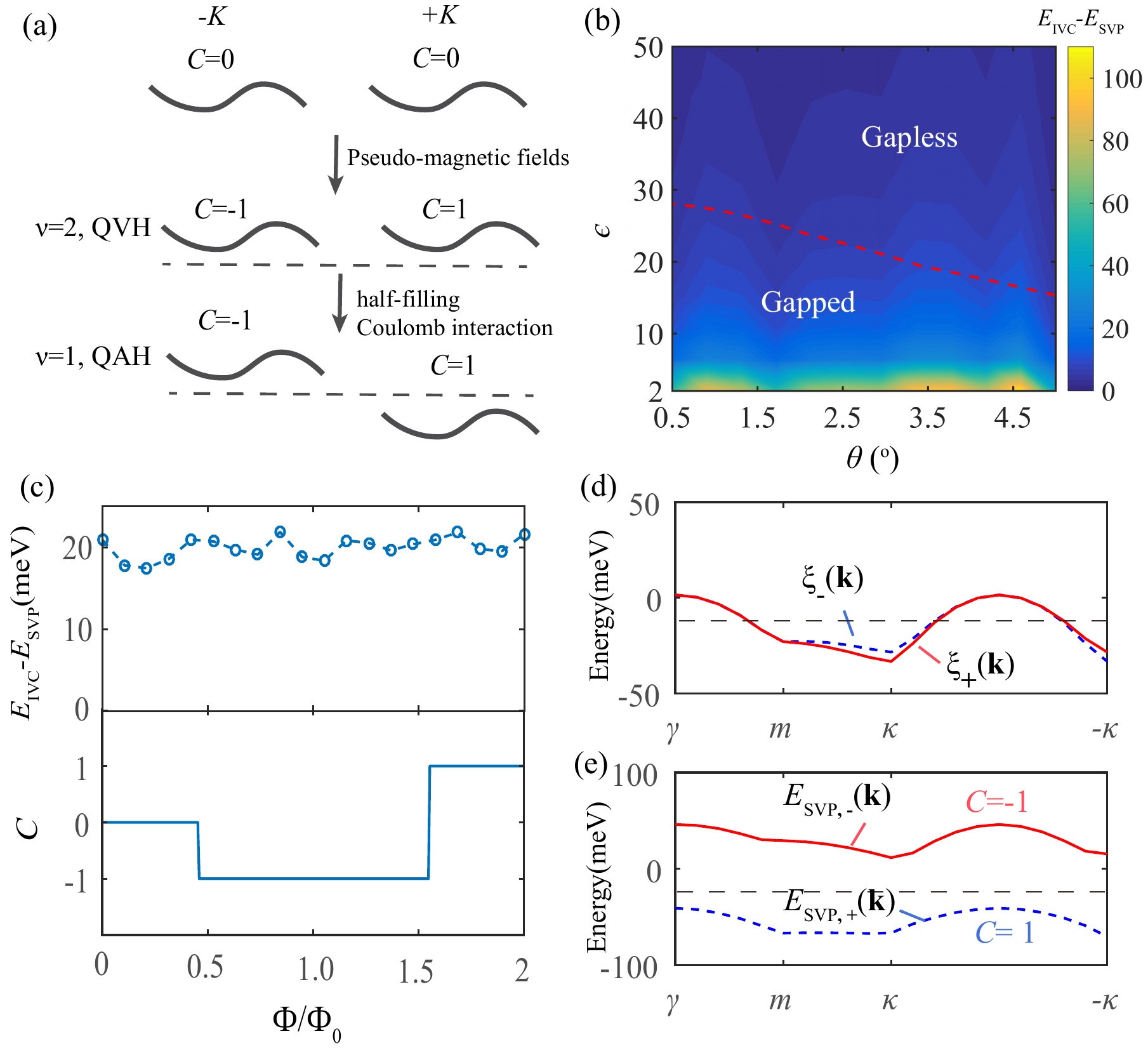}
	\caption{ (a) Schematic plot of the evolution of Chern number for top moir\'e bands at two valleys in the presence of pseudo-magnetic fields, different fillings and interaction with the emergence of quantum valley Hall (QVH) at filling $\nu=2$ and  valley-polarized quantum anomalous Hall (QAH) at filling $\nu=1$.  (b) The energy difference between the inter-valley coherent state  (IVC) and the spin-valley-polarized (SVP) state: $E_{IVC}-E_{SVP}$ (in units of meV) as a function of dielectric constant $\epsilon$ and twist angle $\theta$.     (c) shows the evolution of $E_{IVC}-E_{SVP}$ (top panel) and Chern number (lower panel) at finite flux. (d) and (e), respectively, show the moir\'e band structures $\xi_{\tau}(\bm{k})$ and the mean-field  band structures of  the SVP state $E_{SVP,\tau}(\bm{k})$ at $\Phi/\Phi_0=0.8$.  The parameters for moir\'e potential in (b) to (e)  are taken as $V_0=10$ meV $\phi=0.3\pi$, $\theta=0.53^{\circ}$. Here, $\epsilon=10$,  $\lambda^{-1}=10$ nm \cite{Wang2020,Fai2021,Jianpeng2021}  are adopted  for (c) to (e).}   
	\label{fig:fig3}
\end{figure}

{\emph {valley-polarized quantum anomalous Hall states at half-filling $\nu=1$.}}---After demonstrating the formation of  Chern bands in moir\'e   TMD heterobilayers with periodic pseudo-magnetic fields, we now study the interaction induced topological phases, as schematically depicted in Fig.~\ref{fig:fig3}(a).   We consider the case where the Chern number $C=\pm1$ at the $\pm K$ valley. Due to the spin-valley locking,  the moir\'e  TMD heterobilayer is  a quantum valley Hall insulating phase  at the full-filling $\nu=2$ (see Fig.~\ref{fig:fig3}(a)). When the chemical potential is tuned to half-filling $\nu=1$, as we will demonstrate later, the Coulomb interaction could lift the valley degeneracy and thus gives rise to a valley-polarized quantum anomalous Hall insulator. 



In moir\'e  TMD heterobilayer, due to the  spin-valley locking, the Coulomb interaction is simply
\begin{equation}
	H_{int}=\frac{1}{2S}\sum_{\bm{k},\bm{k'},\bm{q}}V(\bm{q})c^{\dagger}_{\tau}(\bm{k}+\bm{q})c^{\dagger}_{\tau'}(\bm{k'}-\bm{q})c_{\tau'}(\bm{k'})c_{\tau}(\bm{k}),
\end{equation}
where $S$ is the sample area, $V(\bm{q})=\frac{e^2}{2\epsilon\epsilon_0\sqrt{q^2+\lambda^2}}$  is the screened Coulomb interaction with $\epsilon$, $\epsilon_0$, $\lambda^{-1}$ denoting the dielectric constant, vacuum permittivity and a screened length. In practice,  the dielectric constant and screened length  are determined by the surrounding hBN and metallic gates \cite{Stepanov2020}. Within the Hartree-Fock mean-field analysis, we define the order parameter as  	$\braket{\psi_{G}|c_{\tau}^{\dagger}(\bm{k})c_{\tau'}(\bm{k'})|\psi_{G}}=\Delta_{\tau\tau'}(\bm{k})\delta_{\bm{k},\bm{k'}}$, where $\ket{\psi_{G}}$ denotes the ground state. 
Unlike  moir\'e superlattices of graphene \cite{Po2018,Yahui2019,Lee2019,XieMing2020,Jianpeng2021}, here due to the spin-valley locking,   the possible gapped correlated  ground  states for moir\'e  TMD heterobilayer can  be simply grouped into two categories   at half-filling: (i) the spin-valley-polarized (SVP) state  $\ket{\psi_{G}}=\Pi_{|\bm{k}|<k_F}c_{\tau}^{\dagger}(\bm{k})\ket{0}$, where  only $\tau$-valley is occupied; (ii) the spin-valley-locked intervalley coherent (IVC) state $\ket{\psi_{G}}=\Pi_{|\bm{k}|<k_F}[\sin\frac{\theta_{\bm{k}}}{2}e^{-i\frac{\varphi_{\bm{k}}}{2}}c_{+}^{\dagger}(\bm{k})+\cos\frac{\theta_{\bm{k}}}{2}e^{i\frac{\varphi_{\bm{k}}}{2}}c_{-}^{\dagger}(\bm{k})]\ket{0}$, where $\theta_{\bm{k}}=\pi-\theta_{-\bm{k}}$ and $\varphi_{\bm{k}}=\varphi_{-\bm{k}}$ to preserve the time-reversal symmetry. The SVP state breaks time-reversal symmetry, while the spin-valley-locked IVC state breaks the $U(1)$  valley-charge conservation. Importantly,  as a result of a single valley occupancy,  a SVP ground state at half-filling $\nu=1$ could lead to the valley-polarized quantum anomalous Hall insulating state. 


To study the stabilized ground state  at $\nu=1$, we performed the Hartree-Fock mean-field calculations.   In the calculations, we projected the interactions onto  the topmost moir\'e bands \cite{Po2018,Yahui2019,Lee2019}.  The details of the calculations can be found in SM Sec.~III and Sec.~IV \cite{Supp}.   Here, we summarize the numerical results in Fig.~\ref{fig:fig3}(b) to Fig.~\ref{fig:fig3}(f). Fig.~\ref{fig:fig3}(b) displays the energy difference of the  IVC and SVP state $E_{IVC}-E_{SVP}$ as a function of twist angle $\theta$ and dielectric constant $\epsilon$. It is shown that  in a wide parameter range, the SVP state exhibits  lower energy than the IVC state, being compatible  with previous results in moir\'e superlattices of graphene \cite{Yahui2019,Lee2019,Senthil2020_2}. Indeed, $E_{IVC}-E_{SVP}>0$ can be shown analytically in the long-wave limit $qL_{M}\ll 1$ as shown in SM Sec.~III \cite{Supp}. To obtain the observed insulating QAH state, a gapped SVP state is needed, which happens when the  strength of Coulomb interaction overcomes the band dispersion as highlighted in Fig.~\ref{fig:fig3}(b). 


We also performed the Hartree-Fock mean-field calculations with  finite  pseudo-magnetic fields ($\Phi=\frac{\sqrt{3}}{2}B _0L_M^2\neq 0$) which  enables  the moir\'e bands to carry finite Chern numbers. We found that the SVP state is still more stable than the IVC state in this case.  As shown in the upper panel of Fig.~\ref{fig:fig3}(c), the energy difference of $E_{IVC}-E_{SVP}$ is almost insensitive to the increase in the strength of pseudo-magnetic fields. This is because the corresponding  magnetic energy $\epsilon_B$ is much smaller than the Coulomb interacting strength ($\sim$ 100 meV). In contrast with the stability of the SVP state, the topology of the moir\'e bands is determined by specific pseudo-magnetic fields. As shown in the lower panel of Fig.~\ref{fig:fig3}(c), the SVP states acquired finite Chern numbers at some range of pseudo-magnetic fields. Therefore, by considering the effects of  pseudo-magnetic fields and Coulomb interactions, we demonstrated the degeneracy of the two moir\'e bands $\xi_{\pm}(\bm{k})$ can be lifted (see Fig. 3(d)) and a single  moir\'e band carrying a finite Chern number appears at half-filling (see Fig. 3(e)). As a result,  moir\'e   TMD heterobilayers can exhibit valley-polarized quantum anomalous Hall states.

{\emph {Discussion.}}--- It is important to note that another natural way to create nontrivial Chern bands is by reducing the energy offset of the valence bands of MoTe$_2$ and WSe$_2$ by applying a displacement field, such that  the moir\'e bands of the two TMD materials can hybridize to open a topologically non-trivial gap \cite{fengcheng2019, Fu_zhang2021}. However, it is not certain if the relatively small displacement field ($\sim$0.5 V/nm) used in the experiment \cite{Fai_ex2021} can hybridize the moir\'e bands from MoTe$_2$ and WSe$_2$, which are expected to have an energy offset of 300meV \cite{Fai2021,Fai_ex2021}. In the case of QAH effect observed at 3/4 filling in twisted bilayer graphene, the non-trivial Chern bands are generated by the coupling between the aligned graphene moir\'e superlattices and the boron nitride substrate \cite{David2019,Young2020,Chen2020,Yahui2019,Senthil2019,Zaletel2020}. In this work, we propose a new mechanism that pseudo-magnetic fields induced by lattice relaxation can cause topological band inversion for moir\'e bands originated from a single layer (such as the MoTe$_2$ layer). Our results in principle can be applicable to other TMD materials with pseudo-magnetic fields \cite{Falko2020,Mao2020,Yaowang}. Our model also provides a basis for the study of other strongly interacting phases such as fractional Chern insulating states \cite{Liuzhao2013,Liuzhao2020,Ashvin2020,Senthil2020} in heterobilayer TMDs.

In the SM Sec. V \cite{Supp}, we go beyond the pseudomagnetic field approximation and introduce an effective tight-binding model to describe the MoTe$_2$/WSe$_2$ heterobilayer with lattice relaxation. We demonstrate how lattice relaxation can cause gap closing and reopening and change the topology of the top moir\'e bands. Both the gap closing positions as well as the topology of bands of the effective tight-binding model are consistent with the results from the pseudomagnetic field description. 





{\emph {Acknowledgments.}}--- The authors thank the discussions with Liang Fu, Adrian Po, and Berthold J{\"a}ck. K.T.L. acknowledges the support of the Croucher Foundation and HKRGC through RFS2021-6S03, C6025-19G, AoE/P-701/20, 16310520, 16310219 and16309718. K.F.M. acknowledges the support of the Air Force Office of Scientific Research under award number FA9550-20-1-0219.
	
	\bibliographystyle{apsrev4-1} 
	\bibliography{Reference}

\begin{thebibliography}{72}%
\makeatletter
\providecommand \@ifxundefined [1]{%
 \@ifx{#1\undefined}
}%
\providecommand \@ifnum [1]{%
 \ifnum #1\expandafter \@firstoftwo
 \else \expandafter \@secondoftwo
 \fi
}%
\providecommand \@ifx [1]{%
 \ifx #1\expandafter \@firstoftwo
 \else \expandafter \@secondoftwo
 \fi
}%
\providecommand \natexlab [1]{#1}%
\providecommand \enquote  [1]{``#1''}%
\providecommand \bibnamefont  [1]{#1}%
\providecommand \bibfnamefont [1]{#1}%
\providecommand \citenamefont [1]{#1}%
\providecommand \href@noop [0]{\@secondoftwo}%
\providecommand \href [0]{\begingroup \@sanitize@url \@href}%
\providecommand \@href[1]{\@@startlink{#1}\@@href}%
\providecommand \@@href[1]{\endgroup#1\@@endlink}%
\providecommand \@sanitize@url [0]{\catcode `\\12\catcode `\$12\catcode
  `\&12\catcode `\#12\catcode `\^12\catcode `\_12\catcode `\%12\relax}%
\providecommand \@@startlink[1]{}%
\providecommand \@@endlink[0]{}%
\providecommand \url  [0]{\begingroup\@sanitize@url \@url }%
\providecommand \@url [1]{\endgroup\@href {#1}{\urlprefix }}%
\providecommand \urlprefix  [0]{URL }%
\providecommand \Eprint [0]{\href }%
\providecommand \doibase [0]{http://dx.doi.org/}%
\providecommand \selectlanguage [0]{\@gobble}%
\providecommand \bibinfo  [0]{\@secondoftwo}%
\providecommand \bibfield  [0]{\@secondoftwo}%
\providecommand \translation [1]{[#1]}%
\providecommand \BibitemOpen [0]{}%
\providecommand \bibitemStop [0]{}%
\providecommand \bibitemNoStop [0]{.\EOS\space}%
\providecommand \EOS [0]{\spacefactor3000\relax}%
\providecommand \BibitemShut  [1]{\csname bibitem#1\endcsname}%
\let\auto@bib@innerbib\@empty
\bibitem [{\citenamefont {Cao}\ \emph {et~al.}(2018{\natexlab{a}})\citenamefont
  {Cao}, \citenamefont {Fatemi}, \citenamefont {Demir}, \citenamefont {Fang},
  \citenamefont {Tomarken}, \citenamefont {Luo}, \citenamefont
  {Sanchez-Yamagishi}, \citenamefont {Watanabe}, \citenamefont {Taniguchi},
  \citenamefont {Kaxiras}, \citenamefont {Ashoori},\ and\ \citenamefont
  {Jarillo-Herrero}}]{Cao2018}%
  \BibitemOpen
  \bibfield  {author} {\bibinfo {author} {\bibfnamefont {Y.}~\bibnamefont
  {Cao}}, \bibinfo {author} {\bibfnamefont {V.}~\bibnamefont {Fatemi}},
  \bibinfo {author} {\bibfnamefont {A.}~\bibnamefont {Demir}}, \bibinfo
  {author} {\bibfnamefont {S.}~\bibnamefont {Fang}}, \bibinfo {author}
  {\bibfnamefont {S.~L.}\ \bibnamefont {Tomarken}}, \bibinfo {author}
  {\bibfnamefont {J.~Y.}\ \bibnamefont {Luo}}, \bibinfo {author} {\bibfnamefont
  {J.~D.}\ \bibnamefont {Sanchez-Yamagishi}}, \bibinfo {author} {\bibfnamefont
  {K.}~\bibnamefont {Watanabe}}, \bibinfo {author} {\bibfnamefont
  {T.}~\bibnamefont {Taniguchi}}, \bibinfo {author} {\bibfnamefont
  {E.}~\bibnamefont {Kaxiras}}, \bibinfo {author} {\bibfnamefont {R.~C.}\
  \bibnamefont {Ashoori}}, \ and\ \bibinfo {author} {\bibfnamefont
  {P.}~\bibnamefont {Jarillo-Herrero}},\ }\href {\doibase 10.1038/nature26154}
  {\bibfield  {journal} {\bibinfo  {journal} {Nature}\ }\textbf {\bibinfo
  {volume} {556}},\ \bibinfo {pages} {80} (\bibinfo {year}
  {2018}{\natexlab{a}})}\BibitemShut {NoStop}%
\bibitem [{\citenamefont {Cao}\ \emph {et~al.}(2018{\natexlab{b}})\citenamefont
  {Cao}, \citenamefont {Fatemi}, \citenamefont {Fang}, \citenamefont
  {Watanabe}, \citenamefont {Taniguchi}, \citenamefont {Kaxiras},\ and\
  \citenamefont {Jarillo-Herrero}}]{Cao2018sc}%
  \BibitemOpen
  \bibfield  {author} {\bibinfo {author} {\bibfnamefont {Y.}~\bibnamefont
  {Cao}}, \bibinfo {author} {\bibfnamefont {V.}~\bibnamefont {Fatemi}},
  \bibinfo {author} {\bibfnamefont {S.}~\bibnamefont {Fang}}, \bibinfo {author}
  {\bibfnamefont {K.}~\bibnamefont {Watanabe}}, \bibinfo {author}
  {\bibfnamefont {T.}~\bibnamefont {Taniguchi}}, \bibinfo {author}
  {\bibfnamefont {E.}~\bibnamefont {Kaxiras}}, \ and\ \bibinfo {author}
  {\bibfnamefont {P.}~\bibnamefont {Jarillo-Herrero}},\ }\href {\doibase
  10.1038/nature26160} {\bibfield  {journal} {\bibinfo  {journal} {Nature}\
  }\textbf {\bibinfo {volume} {556}},\ \bibinfo {pages} {43} (\bibinfo {year}
  {2018}{\natexlab{b}})}\BibitemShut {NoStop}%
\bibitem [{\citenamefont {Yankowitz}\ \emph {et~al.}(2019)\citenamefont
  {Yankowitz}, \citenamefont {Chen}, \citenamefont {Polshyn}, \citenamefont
  {Zhang}, \citenamefont {Watanabe}, \citenamefont {Taniguchi}, \citenamefont
  {Graf}, \citenamefont {Young},\ and\ \citenamefont {Dean}}]{Yankowitz2019}%
  \BibitemOpen
  \bibfield  {author} {\bibinfo {author} {\bibfnamefont {M.}~\bibnamefont
  {Yankowitz}}, \bibinfo {author} {\bibfnamefont {S.}~\bibnamefont {Chen}},
  \bibinfo {author} {\bibfnamefont {H.}~\bibnamefont {Polshyn}}, \bibinfo
  {author} {\bibfnamefont {Y.}~\bibnamefont {Zhang}}, \bibinfo {author}
  {\bibfnamefont {K.}~\bibnamefont {Watanabe}}, \bibinfo {author}
  {\bibfnamefont {T.}~\bibnamefont {Taniguchi}}, \bibinfo {author}
  {\bibfnamefont {D.}~\bibnamefont {Graf}}, \bibinfo {author} {\bibfnamefont
  {A.~F.}\ \bibnamefont {Young}}, \ and\ \bibinfo {author} {\bibfnamefont
  {C.~R.}\ \bibnamefont {Dean}},\ }\href {\doibase 10.1126/science.aav1910}
  {\bibfield  {journal} {\bibinfo  {journal} {Science}\ }\textbf {\bibinfo
  {volume} {363}},\ \bibinfo {pages} {1059} (\bibinfo {year}
  {2019})}\BibitemShut {NoStop}%
\bibitem [{\citenamefont {Sharpe}\ \emph {et~al.}(2019)\citenamefont {Sharpe},
  \citenamefont {Fox}, \citenamefont {Barnard}, \citenamefont {Finney},
  \citenamefont {Watanabe}, \citenamefont {Taniguchi}, \citenamefont
  {Kastner},\ and\ \citenamefont {Goldhaber-Gordon}}]{David2019}%
  \BibitemOpen
  \bibfield  {author} {\bibinfo {author} {\bibfnamefont {A.~L.}\ \bibnamefont
  {Sharpe}}, \bibinfo {author} {\bibfnamefont {E.~J.}\ \bibnamefont {Fox}},
  \bibinfo {author} {\bibfnamefont {A.~W.}\ \bibnamefont {Barnard}}, \bibinfo
  {author} {\bibfnamefont {J.}~\bibnamefont {Finney}}, \bibinfo {author}
  {\bibfnamefont {K.}~\bibnamefont {Watanabe}}, \bibinfo {author}
  {\bibfnamefont {T.}~\bibnamefont {Taniguchi}}, \bibinfo {author}
  {\bibfnamefont {M.~A.}\ \bibnamefont {Kastner}}, \ and\ \bibinfo {author}
  {\bibfnamefont {D.}~\bibnamefont {Goldhaber-Gordon}},\ }\href {\doibase
  10.1126/science.aaw3780} {\bibfield  {journal} {\bibinfo  {journal}
  {Science}\ }\textbf {\bibinfo {volume} {365}},\ \bibinfo {pages} {605}
  (\bibinfo {year} {2019})}\BibitemShut {NoStop}%
\bibitem [{\citenamefont {Kerelsky}\ \emph {et~al.}(2019)\citenamefont
  {Kerelsky}, \citenamefont {McGilly}, \citenamefont {Kennes}, \citenamefont
  {Xian}, \citenamefont {Yankowitz}, \citenamefont {Chen}, \citenamefont
  {Watanabe}, \citenamefont {Taniguchi}, \citenamefont {Hone}, \citenamefont
  {Dean}, \citenamefont {Rubio},\ and\ \citenamefont
  {Pasupathy}}]{Kerelsky2019}%
  \BibitemOpen
  \bibfield  {author} {\bibinfo {author} {\bibfnamefont {A.}~\bibnamefont
  {Kerelsky}}, \bibinfo {author} {\bibfnamefont {L.~J.}\ \bibnamefont
  {McGilly}}, \bibinfo {author} {\bibfnamefont {D.~M.}\ \bibnamefont {Kennes}},
  \bibinfo {author} {\bibfnamefont {L.}~\bibnamefont {Xian}}, \bibinfo {author}
  {\bibfnamefont {M.}~\bibnamefont {Yankowitz}}, \bibinfo {author}
  {\bibfnamefont {S.}~\bibnamefont {Chen}}, \bibinfo {author} {\bibfnamefont
  {K.}~\bibnamefont {Watanabe}}, \bibinfo {author} {\bibfnamefont
  {T.}~\bibnamefont {Taniguchi}}, \bibinfo {author} {\bibfnamefont
  {J.}~\bibnamefont {Hone}}, \bibinfo {author} {\bibfnamefont {C.}~\bibnamefont
  {Dean}}, \bibinfo {author} {\bibfnamefont {A.}~\bibnamefont {Rubio}}, \ and\
  \bibinfo {author} {\bibfnamefont {A.~N.}\ \bibnamefont {Pasupathy}},\ }\href
  {\doibase 10.1038/s41586-019-1431-9} {\bibfield  {journal} {\bibinfo
  {journal} {Nature}\ }\textbf {\bibinfo {volume} {572}},\ \bibinfo {pages}
  {95} (\bibinfo {year} {2019})}\BibitemShut {NoStop}%
\bibitem [{\citenamefont {Xie}\ \emph {et~al.}(2019)\citenamefont {Xie},
  \citenamefont {Lian}, \citenamefont {J{\"a}ck}, \citenamefont {Liu},
  \citenamefont {Chiu}, \citenamefont {Watanabe}, \citenamefont {Taniguchi},
  \citenamefont {Bernevig},\ and\ \citenamefont {Yazdani}}]{Xie2019}%
  \BibitemOpen
  \bibfield  {author} {\bibinfo {author} {\bibfnamefont {Y.}~\bibnamefont
  {Xie}}, \bibinfo {author} {\bibfnamefont {B.}~\bibnamefont {Lian}}, \bibinfo
  {author} {\bibfnamefont {B.}~\bibnamefont {J{\"a}ck}}, \bibinfo {author}
  {\bibfnamefont {X.}~\bibnamefont {Liu}}, \bibinfo {author} {\bibfnamefont
  {C.-L.}\ \bibnamefont {Chiu}}, \bibinfo {author} {\bibfnamefont
  {K.}~\bibnamefont {Watanabe}}, \bibinfo {author} {\bibfnamefont
  {T.}~\bibnamefont {Taniguchi}}, \bibinfo {author} {\bibfnamefont {B.~A.}\
  \bibnamefont {Bernevig}}, \ and\ \bibinfo {author} {\bibfnamefont
  {A.}~\bibnamefont {Yazdani}},\ }\href {\doibase 10.1038/s41586-019-1422-x}
  {\bibfield  {journal} {\bibinfo  {journal} {Nature}\ }\textbf {\bibinfo
  {volume} {572}},\ \bibinfo {pages} {101} (\bibinfo {year}
  {2019})}\BibitemShut {NoStop}%
\bibitem [{\citenamefont {Serlin}\ \emph {et~al.}(2020)\citenamefont {Serlin},
  \citenamefont {Tschirhart}, \citenamefont {Polshyn}, \citenamefont {Zhang},
  \citenamefont {Zhu}, \citenamefont {Watanabe}, \citenamefont {Taniguchi},
  \citenamefont {Balents},\ and\ \citenamefont {Young}}]{Young2020}%
  \BibitemOpen
  \bibfield  {author} {\bibinfo {author} {\bibfnamefont {M.}~\bibnamefont
  {Serlin}}, \bibinfo {author} {\bibfnamefont {C.~L.}\ \bibnamefont
  {Tschirhart}}, \bibinfo {author} {\bibfnamefont {H.}~\bibnamefont {Polshyn}},
  \bibinfo {author} {\bibfnamefont {Y.}~\bibnamefont {Zhang}}, \bibinfo
  {author} {\bibfnamefont {J.}~\bibnamefont {Zhu}}, \bibinfo {author}
  {\bibfnamefont {K.}~\bibnamefont {Watanabe}}, \bibinfo {author}
  {\bibfnamefont {T.}~\bibnamefont {Taniguchi}}, \bibinfo {author}
  {\bibfnamefont {L.}~\bibnamefont {Balents}}, \ and\ \bibinfo {author}
  {\bibfnamefont {A.~F.}\ \bibnamefont {Young}},\ }\href {\doibase
  10.1126/science.aay5533} {\bibfield  {journal} {\bibinfo  {journal}
  {Science}\ }\textbf {\bibinfo {volume} {367}},\ \bibinfo {pages} {900}
  (\bibinfo {year} {2020})}\BibitemShut {NoStop}%
\bibitem [{\citenamefont {Bistritzer}\ and\ \citenamefont
  {MacDonald}(2011)}]{Bistritzer2011}%
  \BibitemOpen
  \bibfield  {author} {\bibinfo {author} {\bibfnamefont {R.}~\bibnamefont
  {Bistritzer}}\ and\ \bibinfo {author} {\bibfnamefont {A.~H.}\ \bibnamefont
  {MacDonald}},\ }\href {\doibase 10.1073/pnas.1108174108} {\bibfield
  {journal} {\bibinfo  {journal} {Proceedings of the National Academy of
  Sciences}\ }\textbf {\bibinfo {volume} {108}},\ \bibinfo {pages} {12233}
  (\bibinfo {year} {2011})}\BibitemShut {NoStop}%
\bibitem [{\citenamefont {Po}\ \emph {et~al.}(2018)\citenamefont {Po},
  \citenamefont {Zou}, \citenamefont {Vishwanath},\ and\ \citenamefont
  {Senthil}}]{Po2018}%
  \BibitemOpen
  \bibfield  {author} {\bibinfo {author} {\bibfnamefont {H.~C.}\ \bibnamefont
  {Po}}, \bibinfo {author} {\bibfnamefont {L.}~\bibnamefont {Zou}}, \bibinfo
  {author} {\bibfnamefont {A.}~\bibnamefont {Vishwanath}}, \ and\ \bibinfo
  {author} {\bibfnamefont {T.}~\bibnamefont {Senthil}},\ }\href {\doibase
  10.1103/PhysRevX.8.031089} {\bibfield  {journal} {\bibinfo  {journal} {Phys.
  Rev. X}\ }\textbf {\bibinfo {volume} {8}},\ \bibinfo {pages} {031089}
  (\bibinfo {year} {2018})}\BibitemShut {NoStop}%
\bibitem [{\citenamefont {Koshino}\ \emph {et~al.}(2018)\citenamefont
  {Koshino}, \citenamefont {Yuan}, \citenamefont {Koretsune}, \citenamefont
  {Ochi}, \citenamefont {Kuroki},\ and\ \citenamefont {Fu}}]{Liang2018}%
  \BibitemOpen
  \bibfield  {author} {\bibinfo {author} {\bibfnamefont {M.}~\bibnamefont
  {Koshino}}, \bibinfo {author} {\bibfnamefont {N.~F.~Q.}\ \bibnamefont
  {Yuan}}, \bibinfo {author} {\bibfnamefont {T.}~\bibnamefont {Koretsune}},
  \bibinfo {author} {\bibfnamefont {M.}~\bibnamefont {Ochi}}, \bibinfo {author}
  {\bibfnamefont {K.}~\bibnamefont {Kuroki}}, \ and\ \bibinfo {author}
  {\bibfnamefont {L.}~\bibnamefont {Fu}},\ }\href {\doibase
  10.1103/PhysRevX.8.031087} {\bibfield  {journal} {\bibinfo  {journal} {Phys.
  Rev. X}\ }\textbf {\bibinfo {volume} {8}},\ \bibinfo {pages} {031087}
  (\bibinfo {year} {2018})}\BibitemShut {NoStop}%
\bibitem [{\citenamefont {Wang}\ \emph {et~al.}(2020)\citenamefont {Wang},
  \citenamefont {Shih}, \citenamefont {Ghiotto}, \citenamefont {Xian},
  \citenamefont {Rhodes}, \citenamefont {Tan}, \citenamefont {Claassen},
  \citenamefont {Kennes}, \citenamefont {Bai}, \citenamefont {Kim},
  \citenamefont {Watanabe}, \citenamefont {Taniguchi}, \citenamefont {Zhu},
  \citenamefont {Hone}, \citenamefont {Rubio}, \citenamefont {Pasupathy},\ and\
  \citenamefont {Dean}}]{Wang2020}%
  \BibitemOpen
  \bibfield  {author} {\bibinfo {author} {\bibfnamefont {L.}~\bibnamefont
  {Wang}}, \bibinfo {author} {\bibfnamefont {E.-M.}\ \bibnamefont {Shih}},
  \bibinfo {author} {\bibfnamefont {A.}~\bibnamefont {Ghiotto}}, \bibinfo
  {author} {\bibfnamefont {L.}~\bibnamefont {Xian}}, \bibinfo {author}
  {\bibfnamefont {D.~A.}\ \bibnamefont {Rhodes}}, \bibinfo {author}
  {\bibfnamefont {C.}~\bibnamefont {Tan}}, \bibinfo {author} {\bibfnamefont
  {M.}~\bibnamefont {Claassen}}, \bibinfo {author} {\bibfnamefont {D.~M.}\
  \bibnamefont {Kennes}}, \bibinfo {author} {\bibfnamefont {Y.}~\bibnamefont
  {Bai}}, \bibinfo {author} {\bibfnamefont {B.}~\bibnamefont {Kim}}, \bibinfo
  {author} {\bibfnamefont {K.}~\bibnamefont {Watanabe}}, \bibinfo {author}
  {\bibfnamefont {T.}~\bibnamefont {Taniguchi}}, \bibinfo {author}
  {\bibfnamefont {X.}~\bibnamefont {Zhu}}, \bibinfo {author} {\bibfnamefont
  {J.}~\bibnamefont {Hone}}, \bibinfo {author} {\bibfnamefont {A.}~\bibnamefont
  {Rubio}}, \bibinfo {author} {\bibfnamefont {A.~N.}\ \bibnamefont
  {Pasupathy}}, \ and\ \bibinfo {author} {\bibfnamefont {C.~R.}\ \bibnamefont
  {Dean}},\ }\href {\doibase 10.1038/s41563-020-0708-6} {\bibfield  {journal}
  {\bibinfo  {journal} {Nature Materials}\ }\textbf {\bibinfo {volume} {19}},\
  \bibinfo {pages} {861} (\bibinfo {year} {2020})}\BibitemShut {NoStop}%
\bibitem [{\citenamefont {Zhang}\ \emph
  {et~al.}(2020{\natexlab{a}})\citenamefont {Zhang}, \citenamefont {Wang},
  \citenamefont {Watanabe}, \citenamefont {Taniguchi}, \citenamefont {Ueno},
  \citenamefont {Tutuc},\ and\ \citenamefont {LeRoy}}]{Zhang2020}%
  \BibitemOpen
  \bibfield  {author} {\bibinfo {author} {\bibfnamefont {Z.}~\bibnamefont
  {Zhang}}, \bibinfo {author} {\bibfnamefont {Y.}~\bibnamefont {Wang}},
  \bibinfo {author} {\bibfnamefont {K.}~\bibnamefont {Watanabe}}, \bibinfo
  {author} {\bibfnamefont {T.}~\bibnamefont {Taniguchi}}, \bibinfo {author}
  {\bibfnamefont {K.}~\bibnamefont {Ueno}}, \bibinfo {author} {\bibfnamefont
  {E.}~\bibnamefont {Tutuc}}, \ and\ \bibinfo {author} {\bibfnamefont {B.~J.}\
  \bibnamefont {LeRoy}},\ }\href {\doibase 10.1038/s41567-020-0958-x}
  {\bibfield  {journal} {\bibinfo  {journal} {Nature Physics}\ }\textbf
  {\bibinfo {volume} {16}},\ \bibinfo {pages} {1093} (\bibinfo {year}
  {2020}{\natexlab{a}})}\BibitemShut {NoStop}%
\bibitem [{\citenamefont {Naik}\ and\ \citenamefont {Jain}(2018)}]{Jain2018}%
  \BibitemOpen
  \bibfield  {author} {\bibinfo {author} {\bibfnamefont {M.~H.}\ \bibnamefont
  {Naik}}\ and\ \bibinfo {author} {\bibfnamefont {M.}~\bibnamefont {Jain}},\
  }\href {\doibase 10.1103/PhysRevLett.121.266401} {\bibfield  {journal}
  {\bibinfo  {journal} {Phys. Rev. Lett.}\ }\textbf {\bibinfo {volume} {121}},\
  \bibinfo {pages} {266401} (\bibinfo {year} {2018})}\BibitemShut {NoStop}%
\bibitem [{\citenamefont {Wu}\ \emph {et~al.}(2019)\citenamefont {Wu},
  \citenamefont {Lovorn}, \citenamefont {Tutuc}, \citenamefont {Martin},\ and\
  \citenamefont {MacDonald}}]{fengcheng2019}%
  \BibitemOpen
  \bibfield  {author} {\bibinfo {author} {\bibfnamefont {F.}~\bibnamefont
  {Wu}}, \bibinfo {author} {\bibfnamefont {T.}~\bibnamefont {Lovorn}}, \bibinfo
  {author} {\bibfnamefont {E.}~\bibnamefont {Tutuc}}, \bibinfo {author}
  {\bibfnamefont {I.}~\bibnamefont {Martin}}, \ and\ \bibinfo {author}
  {\bibfnamefont {A.~H.}\ \bibnamefont {MacDonald}},\ }\href {\doibase
  10.1103/PhysRevLett.122.086402} {\bibfield  {journal} {\bibinfo  {journal}
  {Phys. Rev. Lett.}\ }\textbf {\bibinfo {volume} {122}},\ \bibinfo {pages}
  {086402} (\bibinfo {year} {2019})}\BibitemShut {NoStop}%
\bibitem [{\citenamefont {Bi}\ and\ \citenamefont {Fu}(2021)}]{Bi2021}%
  \BibitemOpen
  \bibfield  {author} {\bibinfo {author} {\bibfnamefont {Z.}~\bibnamefont
  {Bi}}\ and\ \bibinfo {author} {\bibfnamefont {L.}~\bibnamefont {Fu}},\ }\href
  {\doibase 10.1038/s41467-020-20802-z} {\bibfield  {journal} {\bibinfo
  {journal} {Nature Communications}\ }\textbf {\bibinfo {volume} {12}},\
  \bibinfo {pages} {642} (\bibinfo {year} {2021})}\BibitemShut {NoStop}%
\bibitem [{\citenamefont {Zhang}\ \emph {et~al.}(2017)\citenamefont {Zhang},
  \citenamefont {Chuu}, \citenamefont {Ren}, \citenamefont {Li}, \citenamefont
  {Li}, \citenamefont {Jin}, \citenamefont {Chou},\ and\ \citenamefont
  {Shih}}]{Zhange_hetero_2017}%
  \BibitemOpen
  \bibfield  {author} {\bibinfo {author} {\bibfnamefont {C.}~\bibnamefont
  {Zhang}}, \bibinfo {author} {\bibfnamefont {C.-P.}\ \bibnamefont {Chuu}},
  \bibinfo {author} {\bibfnamefont {X.}~\bibnamefont {Ren}}, \bibinfo {author}
  {\bibfnamefont {M.-Y.}\ \bibnamefont {Li}}, \bibinfo {author} {\bibfnamefont
  {L.-J.}\ \bibnamefont {Li}}, \bibinfo {author} {\bibfnamefont
  {C.}~\bibnamefont {Jin}}, \bibinfo {author} {\bibfnamefont {M.-Y.}\
  \bibnamefont {Chou}}, \ and\ \bibinfo {author} {\bibfnamefont {C.-K.}\
  \bibnamefont {Shih}},\ }\href {\doibase 10.1126/sciadv.1601459} {\bibfield
  {journal} {\bibinfo  {journal} {Science Advances}\ }\textbf {\bibinfo
  {volume} {3}} (\bibinfo {year} {2017}),\ 10.1126/sciadv.1601459}\BibitemShut
  {NoStop}%
\bibitem [{\citenamefont {Tong}\ \emph {et~al.}(2017)\citenamefont {Tong},
  \citenamefont {Yu}, \citenamefont {Zhu}, \citenamefont {Wang}, \citenamefont
  {Xu},\ and\ \citenamefont {Yao}}]{Yaowang2017}%
  \BibitemOpen
  \bibfield  {author} {\bibinfo {author} {\bibfnamefont {Q.}~\bibnamefont
  {Tong}}, \bibinfo {author} {\bibfnamefont {H.}~\bibnamefont {Yu}}, \bibinfo
  {author} {\bibfnamefont {Q.}~\bibnamefont {Zhu}}, \bibinfo {author}
  {\bibfnamefont {Y.}~\bibnamefont {Wang}}, \bibinfo {author} {\bibfnamefont
  {X.}~\bibnamefont {Xu}}, \ and\ \bibinfo {author} {\bibfnamefont
  {W.}~\bibnamefont {Yao}},\ }\href {\doibase 10.1038/nphys3968} {\bibfield
  {journal} {\bibinfo  {journal} {Nature Physics}\ }\textbf {\bibinfo {volume}
  {13}},\ \bibinfo {pages} {356} (\bibinfo {year} {2017})}\BibitemShut
  {NoStop}%
\bibitem [{\citenamefont {Wu}\ \emph {et~al.}(2018)\citenamefont {Wu},
  \citenamefont {Lovorn}, \citenamefont {Tutuc},\ and\ \citenamefont
  {MacDonald}}]{fengcheng2018}%
  \BibitemOpen
  \bibfield  {author} {\bibinfo {author} {\bibfnamefont {F.}~\bibnamefont
  {Wu}}, \bibinfo {author} {\bibfnamefont {T.}~\bibnamefont {Lovorn}}, \bibinfo
  {author} {\bibfnamefont {E.}~\bibnamefont {Tutuc}}, \ and\ \bibinfo {author}
  {\bibfnamefont {A.~H.}\ \bibnamefont {MacDonald}},\ }\href {\doibase
  10.1103/PhysRevLett.121.026402} {\bibfield  {journal} {\bibinfo  {journal}
  {Phys. Rev. Lett.}\ }\textbf {\bibinfo {volume} {121}},\ \bibinfo {pages}
  {026402} (\bibinfo {year} {2018})}\BibitemShut {NoStop}%
\bibitem [{\citenamefont {Jin}\ \emph {et~al.}(2019)\citenamefont {Jin},
  \citenamefont {Regan}, \citenamefont {Yan}, \citenamefont {Iqbal
  Bakti~Utama}, \citenamefont {Wang}, \citenamefont {Zhao}, \citenamefont
  {Qin}, \citenamefont {Yang}, \citenamefont {Zheng}, \citenamefont {Shi},
  \citenamefont {Watanabe}, \citenamefont {Taniguchi}, \citenamefont {Tongay},
  \citenamefont {Zettl},\ and\ \citenamefont {Wang}}]{Fengwang2019}%
  \BibitemOpen
  \bibfield  {author} {\bibinfo {author} {\bibfnamefont {C.}~\bibnamefont
  {Jin}}, \bibinfo {author} {\bibfnamefont {E.~C.}\ \bibnamefont {Regan}},
  \bibinfo {author} {\bibfnamefont {A.}~\bibnamefont {Yan}}, \bibinfo {author}
  {\bibfnamefont {M.}~\bibnamefont {Iqbal Bakti~Utama}}, \bibinfo {author}
  {\bibfnamefont {D.}~\bibnamefont {Wang}}, \bibinfo {author} {\bibfnamefont
  {S.}~\bibnamefont {Zhao}}, \bibinfo {author} {\bibfnamefont {Y.}~\bibnamefont
  {Qin}}, \bibinfo {author} {\bibfnamefont {S.}~\bibnamefont {Yang}}, \bibinfo
  {author} {\bibfnamefont {Z.}~\bibnamefont {Zheng}}, \bibinfo {author}
  {\bibfnamefont {S.}~\bibnamefont {Shi}}, \bibinfo {author} {\bibfnamefont
  {K.}~\bibnamefont {Watanabe}}, \bibinfo {author} {\bibfnamefont
  {T.}~\bibnamefont {Taniguchi}}, \bibinfo {author} {\bibfnamefont
  {S.}~\bibnamefont {Tongay}}, \bibinfo {author} {\bibfnamefont
  {A.}~\bibnamefont {Zettl}}, \ and\ \bibinfo {author} {\bibfnamefont
  {F.}~\bibnamefont {Wang}},\ }\href {\doibase 10.1038/s41586-019-0976-y}
  {\bibfield  {journal} {\bibinfo  {journal} {Nature}\ }\textbf {\bibinfo
  {volume} {567}},\ \bibinfo {pages} {76} (\bibinfo {year} {2019})}\BibitemShut
  {NoStop}%
\bibitem [{\citenamefont {Tran}\ \emph {et~al.}(2019)\citenamefont {Tran},
  \citenamefont {Moody}, \citenamefont {Wu}, \citenamefont {Lu}, \citenamefont
  {Choi}, \citenamefont {Kim}, \citenamefont {Rai}, \citenamefont {Sanchez},
  \citenamefont {Quan}, \citenamefont {Singh}, \citenamefont {Embley},
  \citenamefont {Zepeda}, \citenamefont {Campbell}, \citenamefont {Autry},
  \citenamefont {Taniguchi}, \citenamefont {Watanabe}, \citenamefont {Lu},
  \citenamefont {Banerjee}, \citenamefont {Silverman}, \citenamefont {Kim},
  \citenamefont {Tutuc}, \citenamefont {Yang}, \citenamefont {MacDonald},\ and\
  \citenamefont {Li}}]{Xiaoqin2019}%
  \BibitemOpen
  \bibfield  {author} {\bibinfo {author} {\bibfnamefont {K.}~\bibnamefont
  {Tran}}, \bibinfo {author} {\bibfnamefont {G.}~\bibnamefont {Moody}},
  \bibinfo {author} {\bibfnamefont {F.}~\bibnamefont {Wu}}, \bibinfo {author}
  {\bibfnamefont {X.}~\bibnamefont {Lu}}, \bibinfo {author} {\bibfnamefont
  {J.}~\bibnamefont {Choi}}, \bibinfo {author} {\bibfnamefont {K.}~\bibnamefont
  {Kim}}, \bibinfo {author} {\bibfnamefont {A.}~\bibnamefont {Rai}}, \bibinfo
  {author} {\bibfnamefont {D.~A.}\ \bibnamefont {Sanchez}}, \bibinfo {author}
  {\bibfnamefont {J.}~\bibnamefont {Quan}}, \bibinfo {author} {\bibfnamefont
  {A.}~\bibnamefont {Singh}}, \bibinfo {author} {\bibfnamefont
  {J.}~\bibnamefont {Embley}}, \bibinfo {author} {\bibfnamefont
  {A.}~\bibnamefont {Zepeda}}, \bibinfo {author} {\bibfnamefont
  {M.}~\bibnamefont {Campbell}}, \bibinfo {author} {\bibfnamefont
  {T.}~\bibnamefont {Autry}}, \bibinfo {author} {\bibfnamefont
  {T.}~\bibnamefont {Taniguchi}}, \bibinfo {author} {\bibfnamefont
  {K.}~\bibnamefont {Watanabe}}, \bibinfo {author} {\bibfnamefont
  {N.}~\bibnamefont {Lu}}, \bibinfo {author} {\bibfnamefont {S.~K.}\
  \bibnamefont {Banerjee}}, \bibinfo {author} {\bibfnamefont {K.~L.}\
  \bibnamefont {Silverman}}, \bibinfo {author} {\bibfnamefont {S.}~\bibnamefont
  {Kim}}, \bibinfo {author} {\bibfnamefont {E.}~\bibnamefont {Tutuc}}, \bibinfo
  {author} {\bibfnamefont {L.}~\bibnamefont {Yang}}, \bibinfo {author}
  {\bibfnamefont {A.~H.}\ \bibnamefont {MacDonald}}, \ and\ \bibinfo {author}
  {\bibfnamefont {X.}~\bibnamefont {Li}},\ }\href {\doibase
  10.1038/s41586-019-0975-z} {\bibfield  {journal} {\bibinfo  {journal}
  {Nature}\ }\textbf {\bibinfo {volume} {567}},\ \bibinfo {pages} {71}
  (\bibinfo {year} {2019})}\BibitemShut {NoStop}%
\bibitem [{\citenamefont {Seyler}\ \emph {et~al.}(2019)\citenamefont {Seyler},
  \citenamefont {Rivera}, \citenamefont {Yu}, \citenamefont {Wilson},
  \citenamefont {Ray}, \citenamefont {Mandrus}, \citenamefont {Yan},
  \citenamefont {Yao},\ and\ \citenamefont {Xu}}]{Xiaodong2019}%
  \BibitemOpen
  \bibfield  {author} {\bibinfo {author} {\bibfnamefont {K.~L.}\ \bibnamefont
  {Seyler}}, \bibinfo {author} {\bibfnamefont {P.}~\bibnamefont {Rivera}},
  \bibinfo {author} {\bibfnamefont {H.}~\bibnamefont {Yu}}, \bibinfo {author}
  {\bibfnamefont {N.~P.}\ \bibnamefont {Wilson}}, \bibinfo {author}
  {\bibfnamefont {E.~L.}\ \bibnamefont {Ray}}, \bibinfo {author} {\bibfnamefont
  {D.~G.}\ \bibnamefont {Mandrus}}, \bibinfo {author} {\bibfnamefont
  {J.}~\bibnamefont {Yan}}, \bibinfo {author} {\bibfnamefont {W.}~\bibnamefont
  {Yao}}, \ and\ \bibinfo {author} {\bibfnamefont {X.}~\bibnamefont {Xu}},\
  }\href {\doibase 10.1038/s41586-019-0957-1} {\bibfield  {journal} {\bibinfo
  {journal} {Nature}\ }\textbf {\bibinfo {volume} {567}},\ \bibinfo {pages}
  {66} (\bibinfo {year} {2019})}\BibitemShut {NoStop}%
\bibitem [{\citenamefont {Shimazaki}\ \emph {et~al.}(2020)\citenamefont
  {Shimazaki}, \citenamefont {Schwartz}, \citenamefont {Watanabe},
  \citenamefont {Taniguchi}, \citenamefont {Kroner},\ and\ \citenamefont
  {Imamo{\u{g}}lu}}]{Shimazaki2020}%
  \BibitemOpen
  \bibfield  {author} {\bibinfo {author} {\bibfnamefont {Y.}~\bibnamefont
  {Shimazaki}}, \bibinfo {author} {\bibfnamefont {I.}~\bibnamefont {Schwartz}},
  \bibinfo {author} {\bibfnamefont {K.}~\bibnamefont {Watanabe}}, \bibinfo
  {author} {\bibfnamefont {T.}~\bibnamefont {Taniguchi}}, \bibinfo {author}
  {\bibfnamefont {M.}~\bibnamefont {Kroner}}, \ and\ \bibinfo {author}
  {\bibfnamefont {A.}~\bibnamefont {Imamo{\u{g}}lu}},\ }\href {\doibase
  10.1038/s41586-020-2191-2} {\bibfield  {journal} {\bibinfo  {journal}
  {Nature}\ }\textbf {\bibinfo {volume} {580}},\ \bibinfo {pages} {472}
  (\bibinfo {year} {2020})}\BibitemShut {NoStop}%
\bibitem [{\citenamefont {Tang}\ \emph {et~al.}(2020)\citenamefont {Tang},
  \citenamefont {Li}, \citenamefont {Li}, \citenamefont {Xu}, \citenamefont
  {Liu}, \citenamefont {Barmak}, \citenamefont {Watanabe}, \citenamefont
  {Taniguchi}, \citenamefont {MacDonald}, \citenamefont {Shan},\ and\
  \citenamefont {Mak}}]{Fai_hubbard2020}%
  \BibitemOpen
  \bibfield  {author} {\bibinfo {author} {\bibfnamefont {Y.}~\bibnamefont
  {Tang}}, \bibinfo {author} {\bibfnamefont {L.}~\bibnamefont {Li}}, \bibinfo
  {author} {\bibfnamefont {T.}~\bibnamefont {Li}}, \bibinfo {author}
  {\bibfnamefont {Y.}~\bibnamefont {Xu}}, \bibinfo {author} {\bibfnamefont
  {S.}~\bibnamefont {Liu}}, \bibinfo {author} {\bibfnamefont {K.}~\bibnamefont
  {Barmak}}, \bibinfo {author} {\bibfnamefont {K.}~\bibnamefont {Watanabe}},
  \bibinfo {author} {\bibfnamefont {T.}~\bibnamefont {Taniguchi}}, \bibinfo
  {author} {\bibfnamefont {A.~H.}\ \bibnamefont {MacDonald}}, \bibinfo {author}
  {\bibfnamefont {J.}~\bibnamefont {Shan}}, \ and\ \bibinfo {author}
  {\bibfnamefont {K.~F.}\ \bibnamefont {Mak}},\ }\href {\doibase
  10.1038/s41586-020-2085-3} {\bibfield  {journal} {\bibinfo  {journal}
  {Nature}\ }\textbf {\bibinfo {volume} {579}},\ \bibinfo {pages} {353}
  (\bibinfo {year} {2020})}\BibitemShut {NoStop}%
\bibitem [{\citenamefont {Regan}\ \emph {et~al.}(2020)\citenamefont {Regan},
  \citenamefont {Wang}, \citenamefont {Jin}, \citenamefont {Bakti~Utama},
  \citenamefont {Gao}, \citenamefont {Wei}, \citenamefont {Zhao}, \citenamefont
  {Zhao}, \citenamefont {Zhang}, \citenamefont {Yumigeta}, \citenamefont
  {Blei}, \citenamefont {Carlstr{\"o}m}, \citenamefont {Watanabe},
  \citenamefont {Taniguchi}, \citenamefont {Tongay}, \citenamefont {Crommie},
  \citenamefont {Zettl},\ and\ \citenamefont {Wang}}]{Fengwang2020}%
  \BibitemOpen
  \bibfield  {author} {\bibinfo {author} {\bibfnamefont {E.~C.}\ \bibnamefont
  {Regan}}, \bibinfo {author} {\bibfnamefont {D.}~\bibnamefont {Wang}},
  \bibinfo {author} {\bibfnamefont {C.}~\bibnamefont {Jin}}, \bibinfo {author}
  {\bibfnamefont {M.~I.}\ \bibnamefont {Bakti~Utama}}, \bibinfo {author}
  {\bibfnamefont {B.}~\bibnamefont {Gao}}, \bibinfo {author} {\bibfnamefont
  {X.}~\bibnamefont {Wei}}, \bibinfo {author} {\bibfnamefont {S.}~\bibnamefont
  {Zhao}}, \bibinfo {author} {\bibfnamefont {W.}~\bibnamefont {Zhao}}, \bibinfo
  {author} {\bibfnamefont {Z.}~\bibnamefont {Zhang}}, \bibinfo {author}
  {\bibfnamefont {K.}~\bibnamefont {Yumigeta}}, \bibinfo {author}
  {\bibfnamefont {M.}~\bibnamefont {Blei}}, \bibinfo {author} {\bibfnamefont
  {J.~D.}\ \bibnamefont {Carlstr{\"o}m}}, \bibinfo {author} {\bibfnamefont
  {K.}~\bibnamefont {Watanabe}}, \bibinfo {author} {\bibfnamefont
  {T.}~\bibnamefont {Taniguchi}}, \bibinfo {author} {\bibfnamefont
  {S.}~\bibnamefont {Tongay}}, \bibinfo {author} {\bibfnamefont
  {M.}~\bibnamefont {Crommie}}, \bibinfo {author} {\bibfnamefont
  {A.}~\bibnamefont {Zettl}}, \ and\ \bibinfo {author} {\bibfnamefont
  {F.}~\bibnamefont {Wang}},\ }\href {\doibase 10.1038/s41586-020-2092-4}
  {\bibfield  {journal} {\bibinfo  {journal} {Nature}\ }\textbf {\bibinfo
  {volume} {579}},\ \bibinfo {pages} {359} (\bibinfo {year}
  {2020})}\BibitemShut {NoStop}%
\bibitem [{\citenamefont {Xu}\ \emph {et~al.}(2020)\citenamefont {Xu},
  \citenamefont {Liu}, \citenamefont {Rhodes}, \citenamefont {Watanabe},
  \citenamefont {Taniguchi}, \citenamefont {Hone}, \citenamefont {Elser},
  \citenamefont {Mak},\ and\ \citenamefont {Shan}}]{Fai_fractional2020}%
  \BibitemOpen
  \bibfield  {author} {\bibinfo {author} {\bibfnamefont {Y.}~\bibnamefont
  {Xu}}, \bibinfo {author} {\bibfnamefont {S.}~\bibnamefont {Liu}}, \bibinfo
  {author} {\bibfnamefont {D.~A.}\ \bibnamefont {Rhodes}}, \bibinfo {author}
  {\bibfnamefont {K.}~\bibnamefont {Watanabe}}, \bibinfo {author}
  {\bibfnamefont {T.}~\bibnamefont {Taniguchi}}, \bibinfo {author}
  {\bibfnamefont {J.}~\bibnamefont {Hone}}, \bibinfo {author} {\bibfnamefont
  {V.}~\bibnamefont {Elser}}, \bibinfo {author} {\bibfnamefont {K.~F.}\
  \bibnamefont {Mak}}, \ and\ \bibinfo {author} {\bibfnamefont
  {J.}~\bibnamefont {Shan}},\ }\href {\doibase 10.1038/s41586-020-2868-6}
  {\bibfield  {journal} {\bibinfo  {journal} {Nature}\ }\textbf {\bibinfo
  {volume} {587}},\ \bibinfo {pages} {214} (\bibinfo {year}
  {2020})}\BibitemShut {NoStop}%
\bibitem [{\citenamefont {Huang}\ \emph {et~al.}(2021)\citenamefont {Huang},
  \citenamefont {Wang}, \citenamefont {Miao}, \citenamefont {Wang},
  \citenamefont {Li}, \citenamefont {Lian}, \citenamefont {Taniguchi},
  \citenamefont {Watanabe}, \citenamefont {Okamoto}, \citenamefont {Xiao},
  \citenamefont {Shi},\ and\ \citenamefont {Cui}}]{Huang_fractional2021}%
  \BibitemOpen
  \bibfield  {author} {\bibinfo {author} {\bibfnamefont {X.}~\bibnamefont
  {Huang}}, \bibinfo {author} {\bibfnamefont {T.}~\bibnamefont {Wang}},
  \bibinfo {author} {\bibfnamefont {S.}~\bibnamefont {Miao}}, \bibinfo {author}
  {\bibfnamefont {C.}~\bibnamefont {Wang}}, \bibinfo {author} {\bibfnamefont
  {Z.}~\bibnamefont {Li}}, \bibinfo {author} {\bibfnamefont {Z.}~\bibnamefont
  {Lian}}, \bibinfo {author} {\bibfnamefont {T.}~\bibnamefont {Taniguchi}},
  \bibinfo {author} {\bibfnamefont {K.}~\bibnamefont {Watanabe}}, \bibinfo
  {author} {\bibfnamefont {S.}~\bibnamefont {Okamoto}}, \bibinfo {author}
  {\bibfnamefont {D.}~\bibnamefont {Xiao}}, \bibinfo {author} {\bibfnamefont
  {S.-F.}\ \bibnamefont {Shi}}, \ and\ \bibinfo {author} {\bibfnamefont
  {Y.-T.}\ \bibnamefont {Cui}},\ }\href {\doibase 10.1038/s41567-021-01171-w}
  {\bibfield  {journal} {\bibinfo  {journal} {Nature Physics}\ }\textbf
  {\bibinfo {volume} {17}},\ \bibinfo {pages} {715} (\bibinfo {year}
  {2021})}\BibitemShut {NoStop}%
\bibitem [{\citenamefont {Zhang}\ \emph
  {et~al.}(2020{\natexlab{b}})\citenamefont {Zhang}, \citenamefont {Yuan},\
  and\ \citenamefont {Fu}}]{Yangzhang2020}%
  \BibitemOpen
  \bibfield  {author} {\bibinfo {author} {\bibfnamefont {Y.}~\bibnamefont
  {Zhang}}, \bibinfo {author} {\bibfnamefont {N.~F.~Q.}\ \bibnamefont {Yuan}},
  \ and\ \bibinfo {author} {\bibfnamefont {L.}~\bibnamefont {Fu}},\ }\href
  {\doibase 10.1103/PhysRevB.102.201115} {\bibfield  {journal} {\bibinfo
  {journal} {Phys. Rev. B}\ }\textbf {\bibinfo {volume} {102}},\ \bibinfo
  {pages} {201115} (\bibinfo {year} {2020}{\natexlab{b}})}\BibitemShut
  {NoStop}%
\bibitem [{\citenamefont {Jin}\ \emph {et~al.}(2021)\citenamefont {Jin},
  \citenamefont {Tao}, \citenamefont {Li}, \citenamefont {Xu}, \citenamefont
  {Tang}, \citenamefont {Zhu}, \citenamefont {Liu}, \citenamefont {Watanabe},
  \citenamefont {Taniguchi}, \citenamefont {Hone}, \citenamefont {Fu},
  \citenamefont {Shan},\ and\ \citenamefont {Mak}}]{Fai_strip2021}%
  \BibitemOpen
  \bibfield  {author} {\bibinfo {author} {\bibfnamefont {C.}~\bibnamefont
  {Jin}}, \bibinfo {author} {\bibfnamefont {Z.}~\bibnamefont {Tao}}, \bibinfo
  {author} {\bibfnamefont {T.}~\bibnamefont {Li}}, \bibinfo {author}
  {\bibfnamefont {Y.}~\bibnamefont {Xu}}, \bibinfo {author} {\bibfnamefont
  {Y.}~\bibnamefont {Tang}}, \bibinfo {author} {\bibfnamefont {J.}~\bibnamefont
  {Zhu}}, \bibinfo {author} {\bibfnamefont {S.}~\bibnamefont {Liu}}, \bibinfo
  {author} {\bibfnamefont {K.}~\bibnamefont {Watanabe}}, \bibinfo {author}
  {\bibfnamefont {T.}~\bibnamefont {Taniguchi}}, \bibinfo {author}
  {\bibfnamefont {J.~C.}\ \bibnamefont {Hone}}, \bibinfo {author}
  {\bibfnamefont {L.}~\bibnamefont {Fu}}, \bibinfo {author} {\bibfnamefont
  {J.}~\bibnamefont {Shan}}, \ and\ \bibinfo {author} {\bibfnamefont {K.~F.}\
  \bibnamefont {Mak}},\ }\href {\doibase 10.1038/s41563-021-00959-8} {\bibfield
   {journal} {\bibinfo  {journal} {Nature Materials}\ } (\bibinfo {year}
  {2021}),\ 10.1038/s41563-021-00959-8}\BibitemShut {NoStop}%
\bibitem [{\citenamefont {Li}\ \emph {et~al.}(2021)\citenamefont {Li},
  \citenamefont {Jiang}, \citenamefont {Li}, \citenamefont {Zhang},
  \citenamefont {Kang}, \citenamefont {Zhu}, \citenamefont {Watanabe},
  \citenamefont {Taniguchi}, \citenamefont {Chowdhury}, \citenamefont {Fu},
  \citenamefont {Shan},\ and\ \citenamefont {Mak}}]{Fai2021}%
  \BibitemOpen
  \bibfield  {author} {\bibinfo {author} {\bibfnamefont {T.}~\bibnamefont
  {Li}}, \bibinfo {author} {\bibfnamefont {S.}~\bibnamefont {Jiang}}, \bibinfo
  {author} {\bibfnamefont {L.}~\bibnamefont {Li}}, \bibinfo {author}
  {\bibfnamefont {Y.}~\bibnamefont {Zhang}}, \bibinfo {author} {\bibfnamefont
  {K.}~\bibnamefont {Kang}}, \bibinfo {author} {\bibfnamefont {J.}~\bibnamefont
  {Zhu}}, \bibinfo {author} {\bibfnamefont {K.}~\bibnamefont {Watanabe}},
  \bibinfo {author} {\bibfnamefont {T.}~\bibnamefont {Taniguchi}}, \bibinfo
  {author} {\bibfnamefont {D.}~\bibnamefont {Chowdhury}}, \bibinfo {author}
  {\bibfnamefont {L.}~\bibnamefont {Fu}}, \bibinfo {author} {\bibfnamefont
  {J.}~\bibnamefont {Shan}}, \ and\ \bibinfo {author} {\bibfnamefont {K.~F.}\
  \bibnamefont {Mak}},\ }\href {\doibase 10.1038/s41586-021-03853-0} {\bibfield
   {journal} {\bibinfo  {journal} {Nature}\ }\textbf {\bibinfo {volume}
  {597}},\ \bibinfo {pages} {350} (\bibinfo {year} {2021})}\BibitemShut
  {NoStop}%
\bibitem [{\citenamefont {Morales-Dur\'an}\ \emph {et~al.}(2021)\citenamefont
  {Morales-Dur\'an}, \citenamefont {MacDonald},\ and\ \citenamefont
  {Potasz}}]{Macdonald2021}%
  \BibitemOpen
  \bibfield  {author} {\bibinfo {author} {\bibfnamefont {N.}~\bibnamefont
  {Morales-Dur\'an}}, \bibinfo {author} {\bibfnamefont {A.~H.}\ \bibnamefont
  {MacDonald}}, \ and\ \bibinfo {author} {\bibfnamefont {P.}~\bibnamefont
  {Potasz}},\ }\href {\doibase 10.1103/PhysRevB.103.L241110} {\bibfield
  {journal} {\bibinfo  {journal} {Phys. Rev. B}\ }\textbf {\bibinfo {volume}
  {103}},\ \bibinfo {pages} {L241110} (\bibinfo {year} {2021})}\BibitemShut
  {NoStop}%
\bibitem [{\citenamefont {Xiao}\ \emph {et~al.}(2012)\citenamefont {Xiao},
  \citenamefont {Liu}, \citenamefont {Feng}, \citenamefont {Xu},\ and\
  \citenamefont {Yao}}]{XiaoDi2012}%
  \BibitemOpen
  \bibfield  {author} {\bibinfo {author} {\bibfnamefont {D.}~\bibnamefont
  {Xiao}}, \bibinfo {author} {\bibfnamefont {G.-B.}\ \bibnamefont {Liu}},
  \bibinfo {author} {\bibfnamefont {W.}~\bibnamefont {Feng}}, \bibinfo {author}
  {\bibfnamefont {X.}~\bibnamefont {Xu}}, \ and\ \bibinfo {author}
  {\bibfnamefont {W.}~\bibnamefont {Yao}},\ }\href {\doibase
  10.1103/PhysRevLett.108.196802} {\bibfield  {journal} {\bibinfo  {journal}
  {Phys. Rev. Lett.}\ }\textbf {\bibinfo {volume} {108}},\ \bibinfo {pages}
  {196802} (\bibinfo {year} {2012})}\BibitemShut {NoStop}%
\bibitem [{\citenamefont {Xi}\ \emph {et~al.}(2016)\citenamefont {Xi},
  \citenamefont {Wang}, \citenamefont {Zhao}, \citenamefont {Park},
  \citenamefont {Law}, \citenamefont {Berger}, \citenamefont {Forr{\'o}},
  \citenamefont {Shan},\ and\ \citenamefont {Mak}}]{Xi2016}%
  \BibitemOpen
  \bibfield  {author} {\bibinfo {author} {\bibfnamefont {X.}~\bibnamefont
  {Xi}}, \bibinfo {author} {\bibfnamefont {Z.}~\bibnamefont {Wang}}, \bibinfo
  {author} {\bibfnamefont {W.}~\bibnamefont {Zhao}}, \bibinfo {author}
  {\bibfnamefont {J.-H.}\ \bibnamefont {Park}}, \bibinfo {author}
  {\bibfnamefont {K.~T.}\ \bibnamefont {Law}}, \bibinfo {author} {\bibfnamefont
  {H.}~\bibnamefont {Berger}}, \bibinfo {author} {\bibfnamefont
  {L.}~\bibnamefont {Forr{\'o}}}, \bibinfo {author} {\bibfnamefont
  {J.}~\bibnamefont {Shan}}, \ and\ \bibinfo {author} {\bibfnamefont {K.~F.}\
  \bibnamefont {Mak}},\ }\href {\doibase 10.1038/nphys3538} {\bibfield
  {journal} {\bibinfo  {journal} {Nature Physics}\ }\textbf {\bibinfo {volume}
  {12}},\ \bibinfo {pages} {139} (\bibinfo {year} {2016})}\BibitemShut
  {NoStop}%
\bibitem [{\citenamefont {Lu}\ \emph {et~al.}(2015)\citenamefont {Lu},
  \citenamefont {Zheliuk}, \citenamefont {Leermakers}, \citenamefont {Yuan},
  \citenamefont {Zeitler}, \citenamefont {Law},\ and\ \citenamefont
  {Ye}}]{Lu2015}%
  \BibitemOpen
  \bibfield  {author} {\bibinfo {author} {\bibfnamefont {J.~M.}\ \bibnamefont
  {Lu}}, \bibinfo {author} {\bibfnamefont {O.}~\bibnamefont {Zheliuk}},
  \bibinfo {author} {\bibfnamefont {I.}~\bibnamefont {Leermakers}}, \bibinfo
  {author} {\bibfnamefont {N.~F.~Q.}\ \bibnamefont {Yuan}}, \bibinfo {author}
  {\bibfnamefont {U.}~\bibnamefont {Zeitler}}, \bibinfo {author} {\bibfnamefont
  {K.~T.}\ \bibnamefont {Law}}, \ and\ \bibinfo {author} {\bibfnamefont
  {J.~T.}\ \bibnamefont {Ye}},\ }\href {\doibase 10.1126/science.aab2277}
  {\bibfield  {journal} {\bibinfo  {journal} {Science}\ }\textbf {\bibinfo
  {volume} {350}},\ \bibinfo {pages} {1353} (\bibinfo {year}
  {2015})}\BibitemShut {NoStop}%
\bibitem [{\citenamefont {{Li}}\ \emph {et~al.}(2021)\citenamefont {{Li}},
  \citenamefont {{Jiang}}, \citenamefont {{Shen}}, \citenamefont {{Zhang}},
  \citenamefont {{Li}}, \citenamefont {{Devakul}}, \citenamefont {{Watanabe}},
  \citenamefont {{Taniguchi}}, \citenamefont {{Fu}}, \citenamefont {{Shan}},\
  and\ \citenamefont {{Mak}}}]{Fai_ex2021}%
  \BibitemOpen
  \bibfield  {author} {\bibinfo {author} {\bibfnamefont {T.}~\bibnamefont
  {{Li}}}, \bibinfo {author} {\bibfnamefont {S.}~\bibnamefont {{Jiang}}},
  \bibinfo {author} {\bibfnamefont {B.}~\bibnamefont {{Shen}}}, \bibinfo
  {author} {\bibfnamefont {Y.}~\bibnamefont {{Zhang}}}, \bibinfo {author}
  {\bibfnamefont {L.}~\bibnamefont {{Li}}}, \bibinfo {author} {\bibfnamefont
  {T.}~\bibnamefont {{Devakul}}}, \bibinfo {author} {\bibfnamefont
  {K.}~\bibnamefont {{Watanabe}}}, \bibinfo {author} {\bibfnamefont
  {T.}~\bibnamefont {{Taniguchi}}}, \bibinfo {author} {\bibfnamefont
  {L.}~\bibnamefont {{Fu}}}, \bibinfo {author} {\bibfnamefont {J.}~\bibnamefont
  {{Shan}}}, \ and\ \bibinfo {author} {\bibfnamefont {K.~F.}\ \bibnamefont
  {{Mak}}},\ }\href@noop {} {\bibfield  {journal} {\bibinfo  {journal} {arXiv
  e-prints}\ ,\ \bibinfo {eid} {arXiv:2107.01796}} (\bibinfo {year} {2021})},\
  \Eprint {http://arxiv.org/abs/2107.01796} {arXiv:2107.01796
  [cond-mat.mes-hall]} \BibitemShut {NoStop}%
\bibitem [{\citenamefont {Kane}\ and\ \citenamefont {Mele}(2005)}]{Kane2005}%
  \BibitemOpen
  \bibfield  {author} {\bibinfo {author} {\bibfnamefont {C.~L.}\ \bibnamefont
  {Kane}}\ and\ \bibinfo {author} {\bibfnamefont {E.~J.}\ \bibnamefont
  {Mele}},\ }\href {\doibase 10.1103/PhysRevLett.95.226801} {\bibfield
  {journal} {\bibinfo  {journal} {Phys. Rev. Lett.}\ }\textbf {\bibinfo
  {volume} {95}},\ \bibinfo {pages} {226801} (\bibinfo {year}
  {2005})}\BibitemShut {NoStop}%
\bibitem [{\citenamefont {Nam}\ and\ \citenamefont
  {Koshino}(2017)}]{Koshino2017}%
  \BibitemOpen
  \bibfield  {author} {\bibinfo {author} {\bibfnamefont {N.~N.~T.}\
  \bibnamefont {Nam}}\ and\ \bibinfo {author} {\bibfnamefont {M.}~\bibnamefont
  {Koshino}},\ }\href {\doibase 10.1103/PhysRevB.96.075311} {\bibfield
  {journal} {\bibinfo  {journal} {Phys. Rev. B}\ }\textbf {\bibinfo {volume}
  {96}},\ \bibinfo {pages} {075311} (\bibinfo {year} {2017})}\BibitemShut
  {NoStop}%
\bibitem [{\citenamefont {Shi}\ \emph {et~al.}(2020)\citenamefont {Shi},
  \citenamefont {Zhan}, \citenamefont {Qi}, \citenamefont {Huang},
  \citenamefont {Veen}, \citenamefont {Silva-Guill{\'e}n}, \citenamefont
  {Zhang}, \citenamefont {Li}, \citenamefont {Xie}, \citenamefont {Ji},
  \citenamefont {Katsnelson}, \citenamefont {Yuan}, \citenamefont {Qin},\ and\
  \citenamefont {Zhang}}]{Shihaohao2020}%
  \BibitemOpen
  \bibfield  {author} {\bibinfo {author} {\bibfnamefont {H.}~\bibnamefont
  {Shi}}, \bibinfo {author} {\bibfnamefont {Z.}~\bibnamefont {Zhan}}, \bibinfo
  {author} {\bibfnamefont {Z.}~\bibnamefont {Qi}}, \bibinfo {author}
  {\bibfnamefont {K.}~\bibnamefont {Huang}}, \bibinfo {author} {\bibfnamefont
  {E.~v.}\ \bibnamefont {Veen}}, \bibinfo {author} {\bibfnamefont {J.~{\'A}.}\
  \bibnamefont {Silva-Guill{\'e}n}}, \bibinfo {author} {\bibfnamefont
  {R.}~\bibnamefont {Zhang}}, \bibinfo {author} {\bibfnamefont
  {P.}~\bibnamefont {Li}}, \bibinfo {author} {\bibfnamefont {K.}~\bibnamefont
  {Xie}}, \bibinfo {author} {\bibfnamefont {H.}~\bibnamefont {Ji}}, \bibinfo
  {author} {\bibfnamefont {M.~I.}\ \bibnamefont {Katsnelson}}, \bibinfo
  {author} {\bibfnamefont {S.}~\bibnamefont {Yuan}}, \bibinfo {author}
  {\bibfnamefont {S.}~\bibnamefont {Qin}}, \ and\ \bibinfo {author}
  {\bibfnamefont {Z.}~\bibnamefont {Zhang}},\ }\href {\doibase
  10.1038/s41467-019-14207-w} {\bibfield  {journal} {\bibinfo  {journal}
  {Nature Communications}\ }\textbf {\bibinfo {volume} {11}},\ \bibinfo {pages}
  {371} (\bibinfo {year} {2020})}\BibitemShut {NoStop}%
\bibitem [{\citenamefont {Enaldiev}\ \emph {et~al.}(2020)\citenamefont
  {Enaldiev}, \citenamefont {Z\'olyomi}, \citenamefont {Yelgel}, \citenamefont
  {Magorrian},\ and\ \citenamefont {Fal'ko}}]{Falko2020}%
  \BibitemOpen
  \bibfield  {author} {\bibinfo {author} {\bibfnamefont {V.~V.}\ \bibnamefont
  {Enaldiev}}, \bibinfo {author} {\bibfnamefont {V.}~\bibnamefont {Z\'olyomi}},
  \bibinfo {author} {\bibfnamefont {C.}~\bibnamefont {Yelgel}}, \bibinfo
  {author} {\bibfnamefont {S.~J.}\ \bibnamefont {Magorrian}}, \ and\ \bibinfo
  {author} {\bibfnamefont {V.~I.}\ \bibnamefont {Fal'ko}},\ }\href {\doibase
  10.1103/PhysRevLett.124.206101} {\bibfield  {journal} {\bibinfo  {journal}
  {Phys. Rev. Lett.}\ }\textbf {\bibinfo {volume} {124}},\ \bibinfo {pages}
  {206101} (\bibinfo {year} {2020})}\BibitemShut {NoStop}%
\bibitem [{\citenamefont {Maity}\ \emph {et~al.}(2021)\citenamefont {Maity},
  \citenamefont {Maiti}, \citenamefont {Krishnamurthy},\ and\ \citenamefont
  {Jain}}]{Jain2021}%
  \BibitemOpen
  \bibfield  {author} {\bibinfo {author} {\bibfnamefont {I.}~\bibnamefont
  {Maity}}, \bibinfo {author} {\bibfnamefont {P.~K.}\ \bibnamefont {Maiti}},
  \bibinfo {author} {\bibfnamefont {H.~R.}\ \bibnamefont {Krishnamurthy}}, \
  and\ \bibinfo {author} {\bibfnamefont {M.}~\bibnamefont {Jain}},\ }\href
  {\doibase 10.1103/PhysRevB.103.L121102} {\bibfield  {journal} {\bibinfo
  {journal} {Phys. Rev. B}\ }\textbf {\bibinfo {volume} {103}},\ \bibinfo
  {pages} {L121102} (\bibinfo {year} {2021})}\BibitemShut {NoStop}%
\bibitem [{\citenamefont {Magorrian}\ \emph {et~al.}(2021)\citenamefont
  {Magorrian}, \citenamefont {Enaldiev}, \citenamefont {Z\'olyomi},
  \citenamefont {Ferreira}, \citenamefont {Fal'ko},\ and\ \citenamefont
  {Ruiz-Tijerina}}]{Falko2021}%
  \BibitemOpen
  \bibfield  {author} {\bibinfo {author} {\bibfnamefont {S.~J.}\ \bibnamefont
  {Magorrian}}, \bibinfo {author} {\bibfnamefont {V.~V.}\ \bibnamefont
  {Enaldiev}}, \bibinfo {author} {\bibfnamefont {V.}~\bibnamefont {Z\'olyomi}},
  \bibinfo {author} {\bibfnamefont {F.}~\bibnamefont {Ferreira}}, \bibinfo
  {author} {\bibfnamefont {V.~I.}\ \bibnamefont {Fal'ko}}, \ and\ \bibinfo
  {author} {\bibfnamefont {D.~A.}\ \bibnamefont {Ruiz-Tijerina}},\ }\href
  {\doibase 10.1103/PhysRevB.104.125440} {\bibfield  {journal} {\bibinfo
  {journal} {Phys. Rev. B}\ }\textbf {\bibinfo {volume} {104}},\ \bibinfo
  {pages} {125440} (\bibinfo {year} {2021})}\BibitemShut {NoStop}%
\bibitem [{\citenamefont {Weston}\ \emph {et~al.}(2020)\citenamefont {Weston},
  \citenamefont {Zou}, \citenamefont {Enaldiev}, \citenamefont {Summerfield},
  \citenamefont {Clark}, \citenamefont {Z{\'o}lyomi}, \citenamefont {Graham},
  \citenamefont {Yelgel}, \citenamefont {Magorrian}, \citenamefont {Zhou},
  \citenamefont {Zultak}, \citenamefont {Hopkinson}, \citenamefont {Barinov},
  \citenamefont {Bointon}, \citenamefont {Kretinin}, \citenamefont {Wilson},
  \citenamefont {Beton}, \citenamefont {Fal'ko}, \citenamefont {Haigh},\ and\
  \citenamefont {Gorbachev}}]{Weston2020}%
  \BibitemOpen
  \bibfield  {author} {\bibinfo {author} {\bibfnamefont {A.}~\bibnamefont
  {Weston}}, \bibinfo {author} {\bibfnamefont {Y.}~\bibnamefont {Zou}},
  \bibinfo {author} {\bibfnamefont {V.}~\bibnamefont {Enaldiev}}, \bibinfo
  {author} {\bibfnamefont {A.}~\bibnamefont {Summerfield}}, \bibinfo {author}
  {\bibfnamefont {N.}~\bibnamefont {Clark}}, \bibinfo {author} {\bibfnamefont
  {V.}~\bibnamefont {Z{\'o}lyomi}}, \bibinfo {author} {\bibfnamefont
  {A.}~\bibnamefont {Graham}}, \bibinfo {author} {\bibfnamefont
  {C.}~\bibnamefont {Yelgel}}, \bibinfo {author} {\bibfnamefont
  {S.}~\bibnamefont {Magorrian}}, \bibinfo {author} {\bibfnamefont
  {M.}~\bibnamefont {Zhou}}, \bibinfo {author} {\bibfnamefont {J.}~\bibnamefont
  {Zultak}}, \bibinfo {author} {\bibfnamefont {D.}~\bibnamefont {Hopkinson}},
  \bibinfo {author} {\bibfnamefont {A.}~\bibnamefont {Barinov}}, \bibinfo
  {author} {\bibfnamefont {T.~H.}\ \bibnamefont {Bointon}}, \bibinfo {author}
  {\bibfnamefont {A.}~\bibnamefont {Kretinin}}, \bibinfo {author}
  {\bibfnamefont {N.~R.}\ \bibnamefont {Wilson}}, \bibinfo {author}
  {\bibfnamefont {P.~H.}\ \bibnamefont {Beton}}, \bibinfo {author}
  {\bibfnamefont {V.~I.}\ \bibnamefont {Fal'ko}}, \bibinfo {author}
  {\bibfnamefont {S.~J.}\ \bibnamefont {Haigh}}, \ and\ \bibinfo {author}
  {\bibfnamefont {R.}~\bibnamefont {Gorbachev}},\ }\href {\doibase
  10.1038/s41565-020-0682-9} {\bibfield  {journal} {\bibinfo  {journal} {Nature
  Nanotechnology}\ }\textbf {\bibinfo {volume} {15}},\ \bibinfo {pages} {592}
  (\bibinfo {year} {2020})}\BibitemShut {NoStop}%
\bibitem [{\citenamefont {Li}\ \emph {et~al.}(2021{\natexlab{a}})\citenamefont
  {Li}, \citenamefont {Hu}, \citenamefont {Feng}, \citenamefont {Zhou},
  \citenamefont {An}, \citenamefont {Law}, \citenamefont {Wang},\ and\
  \citenamefont {Lin}}]{Linnian2021}%
  \BibitemOpen
  \bibfield  {author} {\bibinfo {author} {\bibfnamefont {E.}~\bibnamefont
  {Li}}, \bibinfo {author} {\bibfnamefont {J.-X.}\ \bibnamefont {Hu}}, \bibinfo
  {author} {\bibfnamefont {X.}~\bibnamefont {Feng}}, \bibinfo {author}
  {\bibfnamefont {Z.}~\bibnamefont {Zhou}}, \bibinfo {author} {\bibfnamefont
  {L.}~\bibnamefont {An}}, \bibinfo {author} {\bibfnamefont {K.~T.}\
  \bibnamefont {Law}}, \bibinfo {author} {\bibfnamefont {N.}~\bibnamefont
  {Wang}}, \ and\ \bibinfo {author} {\bibfnamefont {N.}~\bibnamefont {Lin}},\
  }\href {https://doi.org/10.1038/s41467-021-25924-6} {\bibfield  {journal}
  {\bibinfo  {journal} {Nature Communications}\ }\textbf {\bibinfo {volume}
  {12}},\ \bibinfo {pages} {5601} (\bibinfo {year}
  {2021}{\natexlab{a}})}\BibitemShut {NoStop}%
\bibitem [{\citenamefont {Li}\ \emph {et~al.}(2021{\natexlab{b}})\citenamefont
  {Li}, \citenamefont {Li}, \citenamefont {Naik}, \citenamefont {Xie},
  \citenamefont {Li}, \citenamefont {Wang}, \citenamefont {Regan},
  \citenamefont {Wang}, \citenamefont {Zhao}, \citenamefont {Zhao},
  \citenamefont {Kahn}, \citenamefont {Yumigeta}, \citenamefont {Blei},
  \citenamefont {Taniguchi}, \citenamefont {Watanabe}, \citenamefont {Tongay},
  \citenamefont {Zettl}, \citenamefont {Louie}, \citenamefont {Wang},\ and\
  \citenamefont {Crommie}}]{Li_Crommie2021}%
  \BibitemOpen
  \bibfield  {author} {\bibinfo {author} {\bibfnamefont {H.}~\bibnamefont
  {Li}}, \bibinfo {author} {\bibfnamefont {S.}~\bibnamefont {Li}}, \bibinfo
  {author} {\bibfnamefont {M.~H.}\ \bibnamefont {Naik}}, \bibinfo {author}
  {\bibfnamefont {J.}~\bibnamefont {Xie}}, \bibinfo {author} {\bibfnamefont
  {X.}~\bibnamefont {Li}}, \bibinfo {author} {\bibfnamefont {J.}~\bibnamefont
  {Wang}}, \bibinfo {author} {\bibfnamefont {E.}~\bibnamefont {Regan}},
  \bibinfo {author} {\bibfnamefont {D.}~\bibnamefont {Wang}}, \bibinfo {author}
  {\bibfnamefont {W.}~\bibnamefont {Zhao}}, \bibinfo {author} {\bibfnamefont
  {S.}~\bibnamefont {Zhao}}, \bibinfo {author} {\bibfnamefont {S.}~\bibnamefont
  {Kahn}}, \bibinfo {author} {\bibfnamefont {K.}~\bibnamefont {Yumigeta}},
  \bibinfo {author} {\bibfnamefont {M.}~\bibnamefont {Blei}}, \bibinfo {author}
  {\bibfnamefont {T.}~\bibnamefont {Taniguchi}}, \bibinfo {author}
  {\bibfnamefont {K.}~\bibnamefont {Watanabe}}, \bibinfo {author}
  {\bibfnamefont {S.}~\bibnamefont {Tongay}}, \bibinfo {author} {\bibfnamefont
  {A.}~\bibnamefont {Zettl}}, \bibinfo {author} {\bibfnamefont {S.~G.}\
  \bibnamefont {Louie}}, \bibinfo {author} {\bibfnamefont {F.}~\bibnamefont
  {Wang}}, \ and\ \bibinfo {author} {\bibfnamefont {M.~F.}\ \bibnamefont
  {Crommie}},\ }\href {\doibase 10.1038/s41563-021-00923-6} {\bibfield
  {journal} {\bibinfo  {journal} {Nature Materials}\ }\textbf {\bibinfo
  {volume} {20}},\ \bibinfo {pages} {945} (\bibinfo {year}
  {2021}{\natexlab{b}})}\BibitemShut {NoStop}%
\bibitem [{\citenamefont {Mounet}\ \emph {et~al.}(2018)\citenamefont {Mounet},
  \citenamefont {Gibertini}, \citenamefont {Schwaller}, \citenamefont {Campi},
  \citenamefont {Merkys}, \citenamefont {Marrazzo}, \citenamefont {Sohier},
  \citenamefont {Castelli}, \citenamefont {Cepellotti}, \citenamefont {Pizzi},\
  and\ \citenamefont {Marzari}}]{Mounet2018}%
  \BibitemOpen
  \bibfield  {author} {\bibinfo {author} {\bibfnamefont {N.}~\bibnamefont
  {Mounet}}, \bibinfo {author} {\bibfnamefont {M.}~\bibnamefont {Gibertini}},
  \bibinfo {author} {\bibfnamefont {P.}~\bibnamefont {Schwaller}}, \bibinfo
  {author} {\bibfnamefont {D.}~\bibnamefont {Campi}}, \bibinfo {author}
  {\bibfnamefont {A.}~\bibnamefont {Merkys}}, \bibinfo {author} {\bibfnamefont
  {A.}~\bibnamefont {Marrazzo}}, \bibinfo {author} {\bibfnamefont
  {T.}~\bibnamefont {Sohier}}, \bibinfo {author} {\bibfnamefont {I.~E.}\
  \bibnamefont {Castelli}}, \bibinfo {author} {\bibfnamefont {A.}~\bibnamefont
  {Cepellotti}}, \bibinfo {author} {\bibfnamefont {G.}~\bibnamefont {Pizzi}}, \
  and\ \bibinfo {author} {\bibfnamefont {N.}~\bibnamefont {Marzari}},\ }\href
  {\doibase 10.1038/s41565-017-0035-5} {\bibfield  {journal} {\bibinfo
  {journal} {Nature Nanotechnology}\ }\textbf {\bibinfo {volume} {13}},\
  \bibinfo {pages} {246} (\bibinfo {year} {2018})}\BibitemShut {NoStop}%
\bibitem [{\citenamefont {Meckbach}\ \emph {et~al.}(2020)\citenamefont
  {Meckbach}, \citenamefont {Hader}, \citenamefont {Huttner}, \citenamefont
  {Neuhaus}, \citenamefont {Steiner}, \citenamefont {Stroucken}, \citenamefont
  {Moloney},\ and\ \citenamefont {Koch}}]{Meckbach2020}%
  \BibitemOpen
  \bibfield  {author} {\bibinfo {author} {\bibfnamefont {L.}~\bibnamefont
  {Meckbach}}, \bibinfo {author} {\bibfnamefont {J.}~\bibnamefont {Hader}},
  \bibinfo {author} {\bibfnamefont {U.}~\bibnamefont {Huttner}}, \bibinfo
  {author} {\bibfnamefont {J.}~\bibnamefont {Neuhaus}}, \bibinfo {author}
  {\bibfnamefont {J.~T.}\ \bibnamefont {Steiner}}, \bibinfo {author}
  {\bibfnamefont {T.}~\bibnamefont {Stroucken}}, \bibinfo {author}
  {\bibfnamefont {J.~V.}\ \bibnamefont {Moloney}}, \ and\ \bibinfo {author}
  {\bibfnamefont {S.~W.}\ \bibnamefont {Koch}},\ }\href {\doibase
  10.1103/PhysRevB.101.075401} {\bibfield  {journal} {\bibinfo  {journal}
  {Phys. Rev. B}\ }\textbf {\bibinfo {volume} {101}},\ \bibinfo {pages}
  {075401} (\bibinfo {year} {2020})}\BibitemShut {NoStop}%
\bibitem [{\citenamefont {Li}\ \emph {et~al.}(2018)\citenamefont {Li},
  \citenamefont {Bellus}, \citenamefont {Dai}, \citenamefont {Ma},
  \citenamefont {Li}, \citenamefont {Zhao},\ and\ \citenamefont
  {Zeng}}]{li2018type}%
  \BibitemOpen
  \bibfield  {author} {\bibinfo {author} {\bibfnamefont {M.}~\bibnamefont
  {Li}}, \bibinfo {author} {\bibfnamefont {M.~Z.}\ \bibnamefont {Bellus}},
  \bibinfo {author} {\bibfnamefont {J.}~\bibnamefont {Dai}}, \bibinfo {author}
  {\bibfnamefont {L.}~\bibnamefont {Ma}}, \bibinfo {author} {\bibfnamefont
  {X.}~\bibnamefont {Li}}, \bibinfo {author} {\bibfnamefont {H.}~\bibnamefont
  {Zhao}}, \ and\ \bibinfo {author} {\bibfnamefont {X.~C.}\ \bibnamefont
  {Zeng}},\ }\href@noop {} {\bibfield  {journal} {\bibinfo  {journal}
  {Nanotechnology}\ }\textbf {\bibinfo {volume} {29}},\ \bibinfo {pages}
  {335203} (\bibinfo {year} {2018})}\BibitemShut {NoStop}%
\bibitem [{Sup()}]{Supp}%
  \BibitemOpen
  \href@noop {} {}\bibinfo {note} {See supplementary Material for (1) Spinless
  time-reversal symmetry and the vanishing of Chern number; (2) Derivation of
  effective Hamiltonian through the perturbation theory; (3) Coulomb
  interaction and Hartree-Fock calculations; (4) Details for numerical
  calculations; (5) Effective tight-binding model calculation with lattice
  relaxation for moir\'e heterobilayer TMDs.}\BibitemShut {Stop}%
\bibitem [{\citenamefont {Mao}\ \emph {et~al.}(2020)\citenamefont {Mao},
  \citenamefont {Milovanovic}, \citenamefont {Andelkovic}, \citenamefont {Lai},
  \citenamefont {Cao}, \citenamefont {Watanabe}, \citenamefont {Taniguchi},
  \citenamefont {Covaci}, \citenamefont {Peeters}, \citenamefont {Geim},
  \citenamefont {Jiang},\ and\ \citenamefont {Andrei}}]{Mao2020}%
  \BibitemOpen
  \bibfield  {author} {\bibinfo {author} {\bibfnamefont {J.}~\bibnamefont
  {Mao}}, \bibinfo {author} {\bibfnamefont {S.~P.}\ \bibnamefont
  {Milovanovic}}, \bibinfo {author} {\bibfnamefont {M.}~\bibnamefont
  {Andelkovic}}, \bibinfo {author} {\bibfnamefont {X.}~\bibnamefont {Lai}},
  \bibinfo {author} {\bibfnamefont {Y.}~\bibnamefont {Cao}}, \bibinfo {author}
  {\bibfnamefont {K.}~\bibnamefont {Watanabe}}, \bibinfo {author}
  {\bibfnamefont {T.}~\bibnamefont {Taniguchi}}, \bibinfo {author}
  {\bibfnamefont {L.}~\bibnamefont {Covaci}}, \bibinfo {author} {\bibfnamefont
  {F.~M.}\ \bibnamefont {Peeters}}, \bibinfo {author} {\bibfnamefont {A.~K.}\
  \bibnamefont {Geim}}, \bibinfo {author} {\bibfnamefont {Y.}~\bibnamefont
  {Jiang}}, \ and\ \bibinfo {author} {\bibfnamefont {E.~Y.}\ \bibnamefont
  {Andrei}},\ }\href {\doibase 10.1038/s41586-020-2567-3} {\bibfield  {journal}
  {\bibinfo  {journal} {Nature}\ }\textbf {\bibinfo {volume} {584}},\ \bibinfo
  {pages} {215} (\bibinfo {year} {2020})}\BibitemShut {NoStop}%
\bibitem [{\citenamefont {Zhang}\ \emph {et~al.}(2021)\citenamefont {Zhang},
  \citenamefont {Devakul},\ and\ \citenamefont {Fu}}]{Fu_zhang2021}%
  \BibitemOpen
  \bibfield  {author} {\bibinfo {author} {\bibfnamefont {Y.}~\bibnamefont
  {Zhang}}, \bibinfo {author} {\bibfnamefont {T.}~\bibnamefont {Devakul}}, \
  and\ \bibinfo {author} {\bibfnamefont {L.}~\bibnamefont {Fu}},\ }\href
  {https://www.pnas.org/content/118/36/e2112673118} {\bibfield  {journal}
  {\bibinfo  {journal} {Proceedings of the National Academy of Sciences}\
  }\textbf {\bibinfo {volume} {118}} (\bibinfo {year} {2021})}\BibitemShut
  {NoStop}%
\bibitem [{\citenamefont {Haldane}(1988)}]{Haldane1988}%
  \BibitemOpen
  \bibfield  {author} {\bibinfo {author} {\bibfnamefont {F.~D.~M.}\
  \bibnamefont {Haldane}},\ }\href {\doibase 10.1103/PhysRevLett.61.2015}
  {\bibfield  {journal} {\bibinfo  {journal} {Phys. Rev. Lett.}\ }\textbf
  {\bibinfo {volume} {61}},\ \bibinfo {pages} {2015} (\bibinfo {year}
  {1988})}\BibitemShut {NoStop}%
\bibitem [{\citenamefont {Levy}\ \emph {et~al.}(2010)\citenamefont {Levy},
  \citenamefont {Burke}, \citenamefont {Meaker}, \citenamefont {Panlasigui},
  \citenamefont {Zettl}, \citenamefont {Guinea}, \citenamefont {Neto},\ and\
  \citenamefont {Crommie}}]{Levy2010}%
  \BibitemOpen
  \bibfield  {author} {\bibinfo {author} {\bibfnamefont {N.}~\bibnamefont
  {Levy}}, \bibinfo {author} {\bibfnamefont {S.~A.}\ \bibnamefont {Burke}},
  \bibinfo {author} {\bibfnamefont {K.~L.}\ \bibnamefont {Meaker}}, \bibinfo
  {author} {\bibfnamefont {M.}~\bibnamefont {Panlasigui}}, \bibinfo {author}
  {\bibfnamefont {A.}~\bibnamefont {Zettl}}, \bibinfo {author} {\bibfnamefont
  {F.}~\bibnamefont {Guinea}}, \bibinfo {author} {\bibfnamefont {A.~H.~C.}\
  \bibnamefont {Neto}}, \ and\ \bibinfo {author} {\bibfnamefont {M.~F.}\
  \bibnamefont {Crommie}},\ }\href {\doibase 10.1126/science.1191700}
  {\bibfield  {journal} {\bibinfo  {journal} {Science}\ }\textbf {\bibinfo
  {volume} {329}},\ \bibinfo {pages} {544} (\bibinfo {year}
  {2010})}\BibitemShut {NoStop}%
\bibitem [{\citenamefont {Guinea}\ \emph {et~al.}(2010)\citenamefont {Guinea},
  \citenamefont {Katsnelson},\ and\ \citenamefont {Geim}}]{Guinea2010}%
  \BibitemOpen
  \bibfield  {author} {\bibinfo {author} {\bibfnamefont {F.}~\bibnamefont
  {Guinea}}, \bibinfo {author} {\bibfnamefont {M.~I.}\ \bibnamefont
  {Katsnelson}}, \ and\ \bibinfo {author} {\bibfnamefont {A.~K.}\ \bibnamefont
  {Geim}},\ }\href {\doibase 10.1038/nphys1420} {\bibfield  {journal} {\bibinfo
   {journal} {Nature Physics}\ }\textbf {\bibinfo {volume} {6}},\ \bibinfo
  {pages} {30} (\bibinfo {year} {2010})}\BibitemShut {NoStop}%
\bibitem [{\citenamefont {Liu}\ and\ \citenamefont {Dai}(2021)}]{Jianpeng2021}%
  \BibitemOpen
  \bibfield  {author} {\bibinfo {author} {\bibfnamefont {J.}~\bibnamefont
  {Liu}}\ and\ \bibinfo {author} {\bibfnamefont {X.}~\bibnamefont {Dai}},\
  }\href {\doibase 10.1103/PhysRevB.103.035427} {\bibfield  {journal} {\bibinfo
   {journal} {Phys. Rev. B}\ }\textbf {\bibinfo {volume} {103}},\ \bibinfo
  {pages} {035427} (\bibinfo {year} {2021})}\BibitemShut {NoStop}%
\bibitem [{\citenamefont {Stepanov}\ \emph {et~al.}(2020)\citenamefont
  {Stepanov}, \citenamefont {Das}, \citenamefont {Lu}, \citenamefont
  {Fahimniya}, \citenamefont {Watanabe}, \citenamefont {Taniguchi},
  \citenamefont {Koppens}, \citenamefont {Lischner}, \citenamefont {Levitov},\
  and\ \citenamefont {Efetov}}]{Stepanov2020}%
  \BibitemOpen
  \bibfield  {author} {\bibinfo {author} {\bibfnamefont {P.}~\bibnamefont
  {Stepanov}}, \bibinfo {author} {\bibfnamefont {I.}~\bibnamefont {Das}},
  \bibinfo {author} {\bibfnamefont {X.}~\bibnamefont {Lu}}, \bibinfo {author}
  {\bibfnamefont {A.}~\bibnamefont {Fahimniya}}, \bibinfo {author}
  {\bibfnamefont {K.}~\bibnamefont {Watanabe}}, \bibinfo {author}
  {\bibfnamefont {T.}~\bibnamefont {Taniguchi}}, \bibinfo {author}
  {\bibfnamefont {F.~H.~L.}\ \bibnamefont {Koppens}}, \bibinfo {author}
  {\bibfnamefont {J.}~\bibnamefont {Lischner}}, \bibinfo {author}
  {\bibfnamefont {L.}~\bibnamefont {Levitov}}, \ and\ \bibinfo {author}
  {\bibfnamefont {D.~K.}\ \bibnamefont {Efetov}},\ }\href {\doibase
  10.1038/s41586-020-2459-6} {\bibfield  {journal} {\bibinfo  {journal}
  {Nature}\ }\textbf {\bibinfo {volume} {583}},\ \bibinfo {pages} {375}
  (\bibinfo {year} {2020})}\BibitemShut {NoStop}%
\bibitem [{\citenamefont {Zhang}\ \emph
  {et~al.}(2019{\natexlab{a}})\citenamefont {Zhang}, \citenamefont {Mao},
  \citenamefont {Cao}, \citenamefont {Jarillo-Herrero},\ and\ \citenamefont
  {Senthil}}]{Yahui2019}%
  \BibitemOpen
  \bibfield  {author} {\bibinfo {author} {\bibfnamefont {Y.-H.}\ \bibnamefont
  {Zhang}}, \bibinfo {author} {\bibfnamefont {D.}~\bibnamefont {Mao}}, \bibinfo
  {author} {\bibfnamefont {Y.}~\bibnamefont {Cao}}, \bibinfo {author}
  {\bibfnamefont {P.}~\bibnamefont {Jarillo-Herrero}}, \ and\ \bibinfo {author}
  {\bibfnamefont {T.}~\bibnamefont {Senthil}},\ }\href {\doibase
  10.1103/PhysRevB.99.075127} {\bibfield  {journal} {\bibinfo  {journal} {Phys.
  Rev. B}\ }\textbf {\bibinfo {volume} {99}},\ \bibinfo {pages} {075127}
  (\bibinfo {year} {2019}{\natexlab{a}})}\BibitemShut {NoStop}%
\bibitem [{\citenamefont {Lee}\ \emph {et~al.}(2019)\citenamefont {Lee},
  \citenamefont {Khalaf}, \citenamefont {Liu}, \citenamefont {Liu},
  \citenamefont {Hao}, \citenamefont {Kim},\ and\ \citenamefont
  {Vishwanath}}]{Lee2019}%
  \BibitemOpen
  \bibfield  {author} {\bibinfo {author} {\bibfnamefont {J.~Y.}\ \bibnamefont
  {Lee}}, \bibinfo {author} {\bibfnamefont {E.}~\bibnamefont {Khalaf}},
  \bibinfo {author} {\bibfnamefont {S.}~\bibnamefont {Liu}}, \bibinfo {author}
  {\bibfnamefont {X.}~\bibnamefont {Liu}}, \bibinfo {author} {\bibfnamefont
  {Z.}~\bibnamefont {Hao}}, \bibinfo {author} {\bibfnamefont {P.}~\bibnamefont
  {Kim}}, \ and\ \bibinfo {author} {\bibfnamefont {A.}~\bibnamefont
  {Vishwanath}},\ }\href {\doibase 10.1038/s41467-019-12981-1} {\bibfield
  {journal} {\bibinfo  {journal} {Nature Communications}\ }\textbf {\bibinfo
  {volume} {10}},\ \bibinfo {pages} {5333} (\bibinfo {year}
  {2019})}\BibitemShut {NoStop}%
\bibitem [{\citenamefont {Xie}\ and\ \citenamefont
  {MacDonald}(2020)}]{XieMing2020}%
  \BibitemOpen
  \bibfield  {author} {\bibinfo {author} {\bibfnamefont {M.}~\bibnamefont
  {Xie}}\ and\ \bibinfo {author} {\bibfnamefont {A.~H.}\ \bibnamefont
  {MacDonald}},\ }\href {\doibase 10.1103/PhysRevLett.124.097601} {\bibfield
  {journal} {\bibinfo  {journal} {Phys. Rev. Lett.}\ }\textbf {\bibinfo
  {volume} {124}},\ \bibinfo {pages} {097601} (\bibinfo {year}
  {2020})}\BibitemShut {NoStop}%
\bibitem [{\citenamefont {Repellin}\ \emph {et~al.}(2020)\citenamefont
  {Repellin}, \citenamefont {Dong}, \citenamefont {Zhang},\ and\ \citenamefont
  {Senthil}}]{Senthil2020_2}%
  \BibitemOpen
  \bibfield  {author} {\bibinfo {author} {\bibfnamefont {C.}~\bibnamefont
  {Repellin}}, \bibinfo {author} {\bibfnamefont {Z.}~\bibnamefont {Dong}},
  \bibinfo {author} {\bibfnamefont {Y.-H.}\ \bibnamefont {Zhang}}, \ and\
  \bibinfo {author} {\bibfnamefont {T.}~\bibnamefont {Senthil}},\ }\href
  {\doibase 10.1103/PhysRevLett.124.187601} {\bibfield  {journal} {\bibinfo
  {journal} {Phys. Rev. Lett.}\ }\textbf {\bibinfo {volume} {124}},\ \bibinfo
  {pages} {187601} (\bibinfo {year} {2020})}\BibitemShut {NoStop}%
\bibitem [{\citenamefont {Chen}\ \emph {et~al.}(2020)\citenamefont {Chen},
  \citenamefont {Sharpe}, \citenamefont {Fox}, \citenamefont {Zhang},
  \citenamefont {Wang}, \citenamefont {Jiang}, \citenamefont {Lyu},
  \citenamefont {Li}, \citenamefont {Watanabe}, \citenamefont {Taniguchi},
  \citenamefont {Shi}, \citenamefont {Senthil}, \citenamefont
  {Goldhaber-Gordon}, \citenamefont {Zhang},\ and\ \citenamefont
  {Wang}}]{Chen2020}%
  \BibitemOpen
  \bibfield  {author} {\bibinfo {author} {\bibfnamefont {G.}~\bibnamefont
  {Chen}}, \bibinfo {author} {\bibfnamefont {A.~L.}\ \bibnamefont {Sharpe}},
  \bibinfo {author} {\bibfnamefont {E.~J.}\ \bibnamefont {Fox}}, \bibinfo
  {author} {\bibfnamefont {Y.-H.}\ \bibnamefont {Zhang}}, \bibinfo {author}
  {\bibfnamefont {S.}~\bibnamefont {Wang}}, \bibinfo {author} {\bibfnamefont
  {L.}~\bibnamefont {Jiang}}, \bibinfo {author} {\bibfnamefont
  {B.}~\bibnamefont {Lyu}}, \bibinfo {author} {\bibfnamefont {H.}~\bibnamefont
  {Li}}, \bibinfo {author} {\bibfnamefont {K.}~\bibnamefont {Watanabe}},
  \bibinfo {author} {\bibfnamefont {T.}~\bibnamefont {Taniguchi}}, \bibinfo
  {author} {\bibfnamefont {Z.}~\bibnamefont {Shi}}, \bibinfo {author}
  {\bibfnamefont {T.}~\bibnamefont {Senthil}}, \bibinfo {author} {\bibfnamefont
  {D.}~\bibnamefont {Goldhaber-Gordon}}, \bibinfo {author} {\bibfnamefont
  {Y.}~\bibnamefont {Zhang}}, \ and\ \bibinfo {author} {\bibfnamefont
  {F.}~\bibnamefont {Wang}},\ }\href {\doibase 10.1038/s41586-020-2049-7}
  {\bibfield  {journal} {\bibinfo  {journal} {Nature}\ }\textbf {\bibinfo
  {volume} {579}},\ \bibinfo {pages} {56} (\bibinfo {year} {2020})}\BibitemShut
  {NoStop}%
\bibitem [{\citenamefont {Zhang}\ \emph
  {et~al.}(2019{\natexlab{b}})\citenamefont {Zhang}, \citenamefont {Mao},\ and\
  \citenamefont {Senthil}}]{Senthil2019}%
  \BibitemOpen
  \bibfield  {author} {\bibinfo {author} {\bibfnamefont {Y.-H.}\ \bibnamefont
  {Zhang}}, \bibinfo {author} {\bibfnamefont {D.}~\bibnamefont {Mao}}, \ and\
  \bibinfo {author} {\bibfnamefont {T.}~\bibnamefont {Senthil}},\ }\href
  {\doibase 10.1103/PhysRevResearch.1.033126} {\bibfield  {journal} {\bibinfo
  {journal} {Phys. Rev. Research}\ }\textbf {\bibinfo {volume} {1}},\ \bibinfo
  {pages} {033126} (\bibinfo {year} {2019}{\natexlab{b}})}\BibitemShut
  {NoStop}%
\bibitem [{\citenamefont {Bultinck}\ \emph {et~al.}(2020)\citenamefont
  {Bultinck}, \citenamefont {Chatterjee},\ and\ \citenamefont
  {Zaletel}}]{Zaletel2020}%
  \BibitemOpen
  \bibfield  {author} {\bibinfo {author} {\bibfnamefont {N.}~\bibnamefont
  {Bultinck}}, \bibinfo {author} {\bibfnamefont {S.}~\bibnamefont
  {Chatterjee}}, \ and\ \bibinfo {author} {\bibfnamefont {M.~P.}\ \bibnamefont
  {Zaletel}},\ }\href {\doibase 10.1103/PhysRevLett.124.166601} {\bibfield
  {journal} {\bibinfo  {journal} {Phys. Rev. Lett.}\ }\textbf {\bibinfo
  {volume} {124}},\ \bibinfo {pages} {166601} (\bibinfo {year}
  {2020})}\BibitemShut {NoStop}%
\bibitem [{\citenamefont {Zhai}\ and\ \citenamefont {Yao}(2020)}]{Yaowang}%
  \BibitemOpen
  \bibfield  {author} {\bibinfo {author} {\bibfnamefont {D.}~\bibnamefont
  {Zhai}}\ and\ \bibinfo {author} {\bibfnamefont {W.}~\bibnamefont {Yao}},\
  }\href {\doibase 10.1103/PhysRevMaterials.4.094002} {\bibfield  {journal}
  {\bibinfo  {journal} {Phys. Rev. Materials}\ }\textbf {\bibinfo {volume}
  {4}},\ \bibinfo {pages} {094002} (\bibinfo {year} {2020})}\BibitemShut
  {NoStop}%
\bibitem [{\citenamefont {Bergholtz}\ and\ \citenamefont
  {Liu}(2013)}]{Liuzhao2013}%
  \BibitemOpen
  \bibfield  {author} {\bibinfo {author} {\bibfnamefont {E.~J.}\ \bibnamefont
  {Bergholtz}}\ and\ \bibinfo {author} {\bibfnamefont {Z.}~\bibnamefont
  {Liu}},\ }\href {\doibase 10.1142/S021797921330017X} {\bibfield  {journal}
  {\bibinfo  {journal} {International Journal of Modern Physics B}\ }\textbf
  {\bibinfo {volume} {27}},\ \bibinfo {pages} {1330017} (\bibinfo {year}
  {2013})}\BibitemShut {NoStop}%
\bibitem [{\citenamefont {Abouelkomsan}\ \emph {et~al.}(2020)\citenamefont
  {Abouelkomsan}, \citenamefont {Liu},\ and\ \citenamefont
  {Bergholtz}}]{Liuzhao2020}%
  \BibitemOpen
  \bibfield  {author} {\bibinfo {author} {\bibfnamefont {A.}~\bibnamefont
  {Abouelkomsan}}, \bibinfo {author} {\bibfnamefont {Z.}~\bibnamefont {Liu}}, \
  and\ \bibinfo {author} {\bibfnamefont {E.~J.}\ \bibnamefont {Bergholtz}},\
  }\href {\doibase 10.1103/PhysRevLett.124.106803} {\bibfield  {journal}
  {\bibinfo  {journal} {Phys. Rev. Lett.}\ }\textbf {\bibinfo {volume} {124}},\
  \bibinfo {pages} {106803} (\bibinfo {year} {2020})}\BibitemShut {NoStop}%
\bibitem [{\citenamefont {Ledwith}\ \emph {et~al.}(2020)\citenamefont
  {Ledwith}, \citenamefont {Tarnopolsky}, \citenamefont {Khalaf},\ and\
  \citenamefont {Vishwanath}}]{Ashvin2020}%
  \BibitemOpen
  \bibfield  {author} {\bibinfo {author} {\bibfnamefont {P.~J.}\ \bibnamefont
  {Ledwith}}, \bibinfo {author} {\bibfnamefont {G.}~\bibnamefont
  {Tarnopolsky}}, \bibinfo {author} {\bibfnamefont {E.}~\bibnamefont {Khalaf}},
  \ and\ \bibinfo {author} {\bibfnamefont {A.}~\bibnamefont {Vishwanath}},\
  }\href {\doibase 10.1103/PhysRevResearch.2.023237} {\bibfield  {journal}
  {\bibinfo  {journal} {Phys. Rev. Research}\ }\textbf {\bibinfo {volume}
  {2}},\ \bibinfo {pages} {023237} (\bibinfo {year} {2020})}\BibitemShut
  {NoStop}%
\bibitem [{\citenamefont {Repellin}\ and\ \citenamefont
  {Senthil}(2020)}]{Senthil2020}%
  \BibitemOpen
  \bibfield  {author} {\bibinfo {author} {\bibfnamefont {C.}~\bibnamefont
  {Repellin}}\ and\ \bibinfo {author} {\bibfnamefont {T.}~\bibnamefont
  {Senthil}},\ }\href {\doibase 10.1103/PhysRevResearch.2.023238} {\bibfield
  {journal} {\bibinfo  {journal} {Phys. Rev. Research}\ }\textbf {\bibinfo
  {volume} {2}},\ \bibinfo {pages} {023238} (\bibinfo {year}
  {2020})}\BibitemShut {NoStop}%
\bibitem [{\citenamefont {Cazalilla}\ \emph {et~al.}(2014)\citenamefont
  {Cazalilla}, \citenamefont {Ochoa},\ and\ \citenamefont
  {Guinea}}]{Guinea2014}%
  \BibitemOpen
  \bibfield  {author} {\bibinfo {author} {\bibfnamefont {M.~A.}\ \bibnamefont
  {Cazalilla}}, \bibinfo {author} {\bibfnamefont {H.}~\bibnamefont {Ochoa}}, \
  and\ \bibinfo {author} {\bibfnamefont {F.}~\bibnamefont {Guinea}},\ }\href
  {\doibase 10.1103/PhysRevLett.113.077201} {\bibfield  {journal} {\bibinfo
  {journal} {Phys. Rev. Lett.}\ }\textbf {\bibinfo {volume} {113}},\ \bibinfo
  {pages} {077201} (\bibinfo {year} {2014})}\BibitemShut {NoStop}%
\bibitem [{\citenamefont {Liu}\ \emph {et~al.}(2013)\citenamefont {Liu},
  \citenamefont {Shan}, \citenamefont {Yao}, \citenamefont {Yao},\ and\
  \citenamefont {Xiao}}]{XiaoDi2013}%
  \BibitemOpen
  \bibfield  {author} {\bibinfo {author} {\bibfnamefont {G.-B.}\ \bibnamefont
  {Liu}}, \bibinfo {author} {\bibfnamefont {W.-Y.}\ \bibnamefont {Shan}},
  \bibinfo {author} {\bibfnamefont {Y.}~\bibnamefont {Yao}}, \bibinfo {author}
  {\bibfnamefont {W.}~\bibnamefont {Yao}}, \ and\ \bibinfo {author}
  {\bibfnamefont {D.}~\bibnamefont {Xiao}},\ }\href {\doibase
  10.1103/PhysRevB.88.085433} {\bibfield  {journal} {\bibinfo  {journal} {Phys.
  Rev. B}\ }\textbf {\bibinfo {volume} {88}},\ \bibinfo {pages} {085433}
  (\bibinfo {year} {2013})}\BibitemShut {NoStop}%
\bibitem [{\citenamefont {Fukui}\ \emph {et~al.}(2005)\citenamefont {Fukui},
  \citenamefont {Hatsugai},\ and\ \citenamefont {Suzuki}}]{Fukui2005}%
  \BibitemOpen
  \bibfield  {author} {\bibinfo {author} {\bibfnamefont {T.}~\bibnamefont
  {Fukui}}, \bibinfo {author} {\bibfnamefont {Y.}~\bibnamefont {Hatsugai}}, \
  and\ \bibinfo {author} {\bibfnamefont {H.}~\bibnamefont {Suzuki}},\ }\href
  {\doibase 10.1143/jpsj.74.1674} {\bibfield  {journal} {\bibinfo  {journal}
  {Journal of the Physical Society of Japan}\ }\textbf {\bibinfo {volume}
  {74}},\ \bibinfo {pages} {1674} (\bibinfo {year} {2005})}\BibitemShut
  {NoStop}%
\bibitem [{\citenamefont {Vozmediano}\ \emph {et~al.}(2010)\citenamefont
  {Vozmediano}, \citenamefont {Katsnelson},\ and\ \citenamefont
  {Guinea}}]{Guinea_review}%
  \BibitemOpen
  \bibfield  {author} {\bibinfo {author} {\bibfnamefont {M.}~\bibnamefont
  {Vozmediano}}, \bibinfo {author} {\bibfnamefont {M.}~\bibnamefont
  {Katsnelson}}, \ and\ \bibinfo {author} {\bibfnamefont {F.}~\bibnamefont
  {Guinea}},\ }\href {\doibase https://doi.org/10.1016/j.physrep.2010.07.003}
  {\bibfield  {journal} {\bibinfo  {journal} {Physics Reports}\ }\textbf
  {\bibinfo {volume} {496}},\ \bibinfo {pages} {109} (\bibinfo {year}
  {2010})}\BibitemShut {NoStop}%
\bibitem [{\citenamefont {Fang}\ \emph {et~al.}(2018)\citenamefont {Fang},
  \citenamefont {Carr}, \citenamefont {Cazalilla},\ and\ \citenamefont
  {Kaxiras}}]{Kaxiras2018}%
  \BibitemOpen
  \bibfield  {author} {\bibinfo {author} {\bibfnamefont {S.}~\bibnamefont
  {Fang}}, \bibinfo {author} {\bibfnamefont {S.}~\bibnamefont {Carr}}, \bibinfo
  {author} {\bibfnamefont {M.~A.}\ \bibnamefont {Cazalilla}}, \ and\ \bibinfo
  {author} {\bibfnamefont {E.}~\bibnamefont {Kaxiras}},\ }\href {\doibase
  10.1103/PhysRevB.98.075106} {\bibfield  {journal} {\bibinfo  {journal} {Phys.
  Rev. B}\ }\textbf {\bibinfo {volume} {98}},\ \bibinfo {pages} {075106}
  (\bibinfo {year} {2018})}\BibitemShut {NoStop}%
\bibitem [{\citenamefont {Zhou}\ \emph {et~al.}(2020)\citenamefont {Zhou},
  \citenamefont {Zhang},\ and\ \citenamefont {Law}}]{Benjamin2020}%
  \BibitemOpen
  \bibfield  {author} {\bibinfo {author} {\bibfnamefont {B.~T.}\ \bibnamefont
  {Zhou}}, \bibinfo {author} {\bibfnamefont {C.-P.}\ \bibnamefont {Zhang}}, \
  and\ \bibinfo {author} {\bibfnamefont {K.}~\bibnamefont {Law}},\ }\href
  {\doibase 10.1103/PhysRevApplied.13.024053} {\bibfield  {journal} {\bibinfo
  {journal} {Phys. Rev. Applied}\ }\textbf {\bibinfo {volume} {13}},\ \bibinfo
  {pages} {024053} (\bibinfo {year} {2020})}\BibitemShut {NoStop}%
\end{thebibliography}%
	
	%
	%

	%
	%

	%
	
	%
	%

		\clearpage
		\onecolumngrid
\begin{center}
		\textbf{\large Supplementary Material for  \lq\lq{}Valley-polarized Quantum Anomalous Hall State in Moir\'e MoTe$_2$/WSe$_2$ Heterobilayers\rq\rq{}}\\[.2cm]
		Ying-Ming Xie,$^{1}$ Cheng-Ping Zhang$^{1}$, Jin-Xin Hu$^1$, Kin Fai Mak$^{2,3,4}$, K. T. Law$^{1,*}$\\[.1cm]
		{\itshape ${}^1$Department of Physics, Hong Kong University of Science and Technology, Clear Water Bay, Hong Kong, China}\\
		{\itshape ${}^2$School of Applied and Engineering Physics, Cornell University, Ithaca, NY, USA}\\
				{\itshape ${}^3$ Kavli Institute at Cornell for Nanoscale Science, Ithaca, NY, USA}\\
				{\itshape${}^4$Laboratory of Atomic and Solid State Physics, Cornell University, Ithaca, NY, USA}\\[1cm]
\end{center}
	
	\maketitle

\setcounter{equation}{0}
\setcounter{section}{0}
\setcounter{figure}{0}
\setcounter{table}{0}
\setcounter{page}{1}
\renewcommand{\theequation}{S\arabic{equation}}
\renewcommand{\thesection}{ \Roman{section}}

\renewcommand{\thefigure}{S\arabic{figure}}
\renewcommand{\thetable}{\arabic{table}}
\renewcommand{\tablename}{Supplementary Table}

\renewcommand{\bibnumfmt}[1]{[S#1]}
\renewcommand{\citenumfont}[1]{#1}
\makeatletter

\maketitle
%
%

\section{Spinless time-reversal symmetry and the vanishing of Chern number}
If there is no pseudo-magnetic field, as we discussed in the main text,  there is an emergent spinless time-reversal symmetry $T'=\hat{K}$ with $T'\mathcal{H}_{\tau}(\bm{r})T'^{-1}=\mathcal{H}_{\tau}(\bm{r})$.  In this section, we show such spinless time-reversal symmetry would enforce the Chern number of a moir\'e band to be zero. 

We define the wavefunction as $\psi_{n,\tau,\bm{k}}(\bm{r})=u_{n,\tau,\bm{k}}(\bm{r})e^{i\bm{k}\cdot\bm{r}}$, where $n$ labels the band index, and $\tau$ is the valley index. The Berry curvature
\begin{equation}
	\Omega({\bm{k}})=-i(\braket{\partial_{k_x}u(\bm{k})|\partial_{k_y}u(\bm{k})}-\braket{\partial_{k_y}u(\bm{k})|\partial_{k_x}u(\bm{k})})
\end{equation}
and the Chern number $C=\int_{\bm{k}\in B.Z.}\frac{d\bm{k}}{(2\pi)^2}\Omega({\bm{k}})$. Under the spinless time-reversal symmetry $T': u_{n,\tau}(\bm{k})\mapsto u_{n,\tau}^*(-\bm{k})$ so that  $T':\Omega(\bm{k})\mapsto-i(\braket{\partial_{k_y}u(-\bm{k})|\partial_{k_x}u(-\bm{k})}-\braket{\partial_{k_x}u(-\bm{k})|\partial_{k_y}u(-\bm{k})})=-\Omega(-\bm{k})$. As the Hamiltonian is invariant under this spinless time-reversal symmetry, we thus obtain 
\begin{equation}
	\Omega(\bm{k})= -\Omega(-\bm{k}).
\end{equation}
Therefore, we have shown this emergent time-reversal symmetry $T'$ enforces the Berry curvature to be odd in the Brillouin zone. As a result, the Chern number of a moir\'e band that is equal to the integral of the Berry curvature over the whole Brillouin zone vanishes. 

\section{Derivation of Effective Hamiltonian through the perturbation theory}
\subsection{Three-band effective model Hamiltonian near  moir\'e Brillouin zone corners}
The moir\'e Hamiltonian is
\begin{equation}
	\mathcal{H}_{\tau}(\bm{r})=-\frac{(\bm{\hat{p}}+\tau e\bm{A})^2}{2m^*}+V(\bm{r}),
\end{equation}
where the moir\'e potential is
\begin{equation}
	V(\bm{r})=2V_0\sum_{j=1,3,5}\cos(\bm{G}_j\cdot\bm{r}+\phi)=\sum_{j} V(\bm{G}_j)e^{i\bm{G}_j\cdot \bm{r}},\label{S_moire}
\end{equation}
where $\bm{G}_{j}=\frac{4\pi}{\sqrt{3}L_{M}}(\cos(\frac{(j-1)\pi}{3}),\sin(\frac{(j-1)\pi}{3}))$. As discussed in the main text, we can take the corresponding gauge field of $B(\bm{r})$ as $\bm{A}(\bm{r})=A_0[\bm{a_2}\sin (\bm{G_1\cdot r})-\bm{a_1}\sin(\bm{G_3}\cdot\bm{r})-\bm{a_3}\sin(\bm{G_5}\cdot\bm{r})]$ , where we chose the Coulomb gauge $\nabla\cdot\bm{A}=0$, and define $\bm{a_1}=(\frac{\sqrt{3}}{2},\frac{1}{2})L_M$, $\bm{a_2}=(0,1)L_M$,  $\bm{a_3}=\bm{a_2}-\bm{a_1}$, and  $A_0=\sqrt{3}B_0/4\pi$. This gauge field can be generated by a strain displacement field $\bm{u}=u_0[\bm{G_1}\cos (\bm{G_1\cdot r})+\bm{G_3}\cos(\bm{G_3}\cdot\bm{r})+\bm{G_5}\cos(\bm{G_5}\cdot\bm{r})]$ with $u_0=-\frac{3A_0L_{M}^{3}}{16\pi^2\alpha}$. It gives rise to a strain field $u_{ij}(\bm{r})=(\partial_i u_j(\bm{r})+\partial_j u_i(\bm{r}))/2$, which leads to the gauge field  $\bm{A}=\alpha (2u_{xy}, u_{xx}-u_{yy})$ as desired. Note unlike the constant pseudo-magnetic fields considered in \cite{Guinea2014}, here the pseudo-magnetic fields we consider are periodic, which take the same periodicity as the moir\'e potential,  being motivated by a recent lattice relaxation calculation  for moir\'e transition metal dichalcogenides \cite{Falko2020}.

It can be found in the presence of gauge potential
\begin{equation}
	\mathcal{H}_{\tau}(\bm{r})=-\frac{(\hat{\bm{p}}+\tau\frac{\Phi}{\Phi_0}\tilde{\bm{A}})^2}{2m}+V(\bm{r})\approx -\frac{\hat{\bm{p}}^2}{2m}+\tau\gamma\tilde{\bm{A}}\cdot \bm{p}+V(\bm{r}),
\end{equation}
where the quantum flux $\Phi_0=\frac{h}{e}$, the flux $\Phi=\frac{\sqrt{3}}{2}B _0L_M^2$,  $\tilde{\bm{A}}=4\pi\bm{A}(\bm{r})/B_0\sqrt{3}L_M\text{and } \gamma=-\frac{\hbar\Phi}{mL_M\Phi_0}$, and we dropped $\bm{A}^2$ term here but it will be added in numerical calculations. In the following, we consider the moir\'e Hamiltonian at $+K$ valley: $\mathcal{H}_{+}(\bm{r})$, and the results for $-K$ valley is directly obtained with the time-reversal operation.

First, let us consider the case $\bm{A}=0$. 
Because the three corners of moir\'e Brillouin are connected by the superlattice reciprocal vectors, using the plane waves $\ket{\bm{k}}=e^{i\bm{k\cdot\bm{r}}}$, the effective Hamiltonian near $\pm \kappa$ is written as
\begin{equation}
	H_{\pm \kappa}(\bm{k})=\begin{pmatrix}
		\epsilon_{\pm \kappa}\pm v k_y&V_0e^{\pm i\phi}&V_0e^{\mp i\phi}\\
		V_0e^{\mp i\phi}&\epsilon_{\pm \kappa}\pm v(- \frac{\sqrt{3}}{2}k_x-\frac{1}{2}k_y)&V_0e^{\pm i\phi}\\
		V_0e^{\pm i\phi}&V_0e^{\mp i\phi}&\epsilon_{\pm \kappa}\pm v( \frac{\sqrt{3}}{2}k_x-\frac{1}{2}k_y)
	\end{pmatrix},
\end{equation}
where $\epsilon_{\pm \kappa}=\epsilon_0=-\kappa^2/2m^*$, $v=\kappa/m^*$.
At the Brillouin zone corners, the eigenenergies and eigenwavefunctions are
\begin{align}
	\epsilon_1= \epsilon_0+2V_0\cos \phi, \ket{\epsilon_1}=\frac{1}{\sqrt{3}}(\ket{\pm\kappa_1}+\ket{\pm\kappa_2}+\ket{\pm\kappa_3})\\
	\epsilon_2= \epsilon_0+2V_0\cos (\frac{2\pi}{3}+\phi), \ket{\epsilon_2}=\frac{e^{i\frac{\pi}{2}}}{\sqrt{3}}(\ket{\pm\kappa_1}+e^{\pm i \frac{2\pi}{3}}\ket{\pm\kappa_2}+e^{\mp i \frac{2\pi}{3}}\ket{\pm\kappa_3})\\
	\epsilon_3= \epsilon_0+2V_0\cos (\frac{4\pi}{3}+\phi), \ket{\epsilon_3}=\frac{e^{-i\frac{\pi}{2}}}{\sqrt{3}}(\ket{\pm\kappa_1}+e^{\mp i \frac{2\pi}{3}}\ket{\pm\kappa_2}+e^{\pm i \frac{2\pi}{3}}\ket{\pm\kappa_3})
\end{align}
When the gauge field is added, the gauge term $H'_{\pm \kappa}(\bm{A})=\gamma\tilde{\bm{A}}\cdot \bm{p}$ becomes
\begin{equation}
	H'_{\pm \kappa}(\tilde{\bm{A}})\approx\begin{pmatrix}
		0&\pm \gamma \tilde{\bm{A}}(\pm \bm{G}_5)\cdot \bm{\kappa}_2&\pm \gamma\tilde{\bm{A}}(\pm \bm{G}_6)\cdot \bm{\kappa}_3\\
		\pm \gamma \tilde{\bm{A}}(\pm \bm{G}_2)\cdot \bm{\kappa}_2&0&\pm \gamma\tilde{\bm{A}}(\pm \bm{G}_1)\cdot \bm{\kappa}_3\\
		\pm \gamma \tilde{\bm{A}}(\pm \bm{G}_3)\cdot \bm{\kappa}_3&\pm \gamma \tilde{\bm{A}}(\pm \bm{G}_4)\cdot \bm{\kappa}_3&0
	\end{pmatrix},
\end{equation}
where 
\begin{equation}
	\bm{\kappa}_1=\kappa(0,-1), \bm{\kappa}_2=\kappa(\frac{\sqrt{3}}{2},\frac{1}{2}), \bm{\kappa}_3=\kappa(-\frac{\sqrt{3}}{2},\frac{1}{2}),
\end{equation}
and 
\begin{equation}
	\tilde{\bm{A}}(\bm{G})=\frac{1}{S}\int d\bm{r} \tilde{\bm{A}}(\bm{r})e^{-i\bm{G}\cdot\bm{r}} \label{gauge}
\end{equation}with $\tilde{A}_x(\bm{r})=-\frac{\sqrt{3}}{2}\sin(\bm{G}_3\cdot\bm{r})+\frac{\sqrt{3}}{2}\sin(\bm{G}_5\cdot\bm{r})$,
$\tilde{A}_y(\bm{r})=\sin (\bm{G_1\cdot r})-\frac{1}{2}\sin(\bm{G_3}\cdot\bm{r})-\frac{1}{2}\sin(\bm{G_5}\cdot\bm{r})$, $S$ denoting the sample area.
As $\bm{A}(\bm{G})=\bm{A}^{*}(-\bm{G})$, $H'_{\pm\kappa}(\bm{A})$ would have the same form.  We can further obtain
\begin{equation}
	H'_{\pm \kappa}(\bm{A})=\begin{pmatrix}
		0&\frac{\kappa \gamma}{4i}&-\frac{\kappa \gamma}{4i}\\
		-\frac{\kappa \gamma}{4i}&0&\frac{\kappa \gamma}{4i}\\
		\frac{\kappa \gamma}{4i}&-\frac{\kappa \gamma}{4i}&0
	\end{pmatrix}
\end{equation}
After projecting $H'_{\pm \kappa}(\bm{A})$ into the basis formed by $(\ket{\epsilon_1},\ket{\epsilon_2},\ket{\epsilon_3})$,  we find
\begin{equation}
	H'_{\pm \kappa}(\bm{A})=\frac{\sqrt{3}\kappa\gamma}{4}\begin{pmatrix}
		0&0&0\\
		0&\pm 1&0\\
		0&0&\mp 1
	\end{pmatrix}
\end{equation}
Hence, in the basis spanned by $(\ket{\epsilon_1},\ket{\epsilon_2},\ket{\epsilon_3})$, we can write it as
\begin{equation}
	H_{\pm\kappa}^{eff}(\bm{k})=\begin{pmatrix}
		\epsilon_0+2V_0\cos \phi&\pm \frac{1}{2}v(k_x+ik_y)&\pm \frac{1}{2}v(k_x-ik_y)\\
		\pm\frac{1}{2}v(k_x-ik_y)&\epsilon_0+2V_0\cos (\frac{2\pi}{3}+\phi)\mp \epsilon_B&\pm\frac{i}{2}v(k_x+ik_y)\\
		\pm\frac{1}{2}v(k_x+ik_y)&\mp \frac{i}{2}v(k_x-ik_y)&\epsilon_0+2V_0\cos (\frac{4\pi}{3}+\phi)\pm \epsilon_B
	\end{pmatrix},
\end{equation}
\begin{figure}
	\centering
	\includegraphics[width=0.4\linewidth]{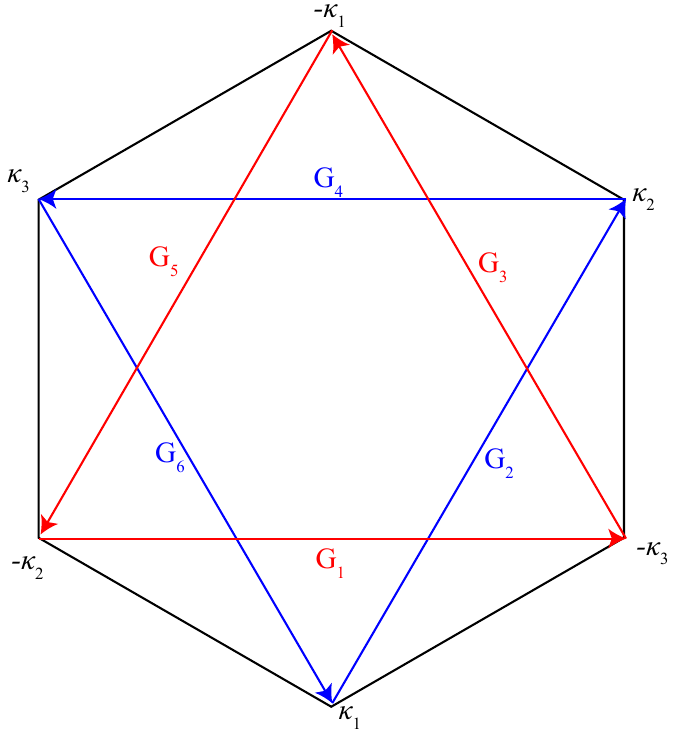}
	\caption{A sketch of the moir\'e Brillouin zone and how the moir\'e Brillouin corners $\kappa$ are connected by reciprocal lattice vectors $G_j$. }
	\label{FIGS1}
\end{figure}
where the  magnetic energy that gives rise to the Dirac mass gap is given by
\begin{equation}
	\epsilon_B=\frac{\sqrt{3}{v}}{4L_M}\frac{\Phi}{\Phi_0}=\frac{\hbar eB_0}{4m^*}
\end{equation}

In the main text, we determined the topological phase transition boundaries with above three-band model and compared with the topological phase diagram from numerical calculations. The details for numerical calculations, including the diagonalization of full moir\'e Hamiltonian and Chern number, are summarized in Sec. IV of Supplementary Material (SM). 
%

\subsection{Four-band effective model Hamiltonian near moir\'e Brillouin zone boundary}

\begin{figure}[h]
	\centering
	\includegraphics[width=0.8\linewidth]{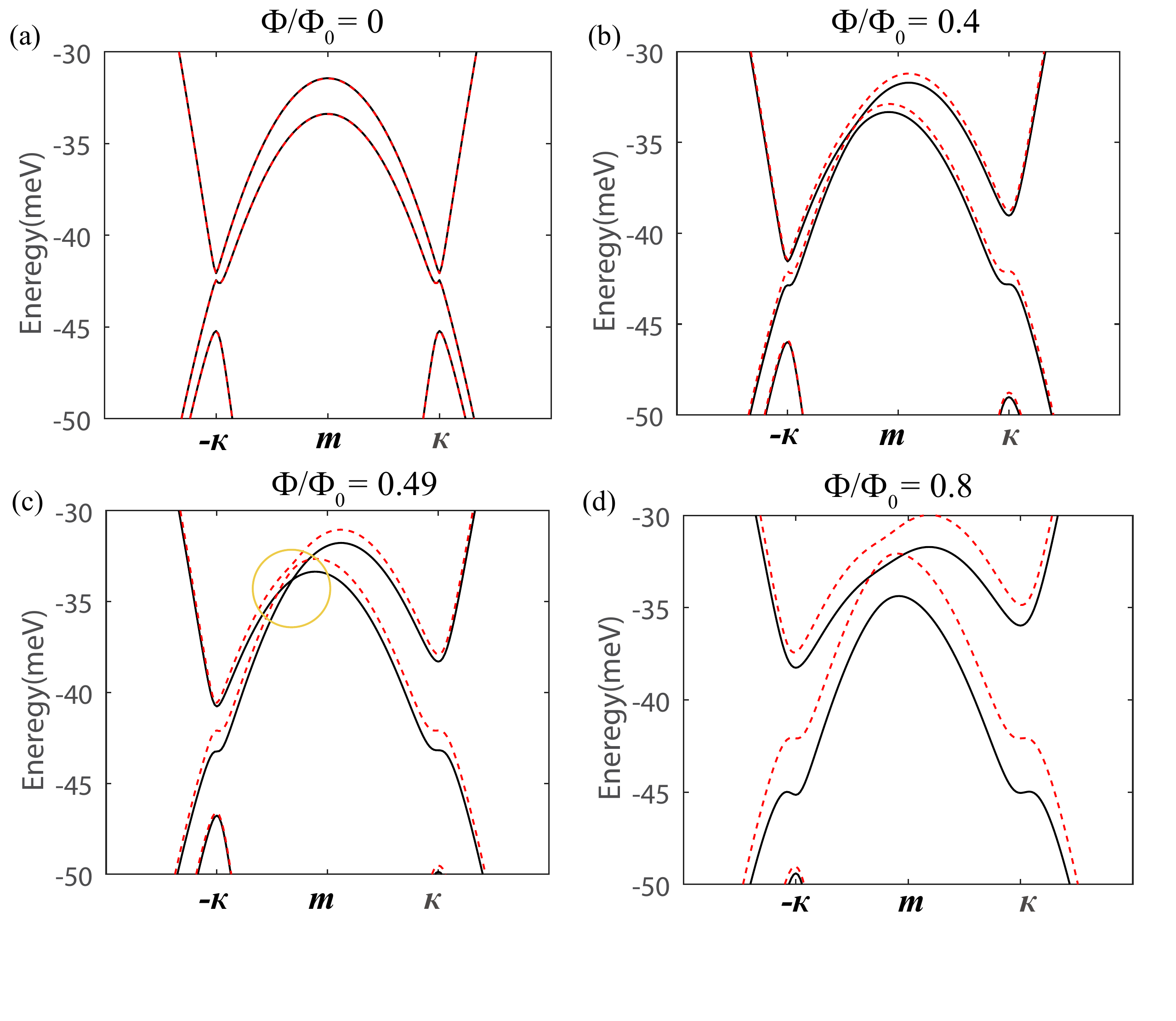}
	\caption{ (a) to (d) display the energy dispersion of top three moir\'e bands at moir\'e Brillouin zone boundary at $\Phi/\Phi_0=0, 0.4, 0.49, 0.8$ respectively. The red dashed lines are from the four-band effective model Hamiltonian Eq.~(\ref{Fourband}), while the black solid lines are from a direct diagonalization of the moir\'e Hamiltonian.  Here, the moir\'e potential parameters $V_0=1$ meV, $\phi=0.3\pi$ and angle $\theta=0.53^{\circ}$ (note $V_0$ is the same as Fig.~2(a) but different from Fig. 1). It shows the  gap closing point can happen at a general point of moir\'e Brillouin zone boundary beyond $m$ and $\kappa$ points. And this gives rise to the boundaries that separate $C=\pm 1$ phase and $C=\mp 2$ phase, where Chern number changes $\pm 3$, shown in the main text Fig. 2. }
	\label{fig:figs3}
\end{figure}

In the main text of Fig.~2, it can be seen the phase transition boundaries that separates $C=\pm 1$ phase and $C=\mp 2$ phase, where Chern number changes $\pm3$, are not captured by above three-band model. As shown in Fig.~\ref{fig:figs3}, where we calculated the moir\'e bands along moir\'e Brillouin zone boundary numerically (see black solid lines),  we found it is gap closings at general points of moir\'e Brillouin zone boundary that drives this topological phase transition. Due to $C_3$ symmetry, there are three gap closing points at three $m\kappa$ lines related by $C_3$ symmetry, which enables the change of Chern number to be $\pm 3$.

To capture all the topological transitions,  we derived an effective continuum model at the whole moir\'e boundary, denoting as $B_j$,  instead of near only $\pm \kappa$.
In the space $((\ket{\bm{k}_{B2}},\ket{\bm{k}_{B2}+\bm{G}_3}, \ket{\bm{k}_{B2}+\bm{G}_4},\ket{\bm{k}_{B2}+\bm{G}_5})$, after a similar derivation shown in previous part, we can obtain a four by four effective Hamiltonian for the whole moir\'e boundary:
\begin{equation}
	H_{L_2}^{eff}=\begin{pmatrix}
		\epsilon(\bm{k}_{B2})& V_0e^{i\phi}&V_0e^{-i\phi}&V_0e^{-i\phi}\\
		V_0e^{-i\phi}&\epsilon(\bm{k}_{B2})&V_0e^{i\phi}&V_0e^{i\phi}\\
		V_0e^{i\phi}&V_0e^{-i\phi}&\epsilon(\bm{k}_{B2}+\bm{G}_3)&0\\
		V_0e^{i\phi}&V_0e^{-i\phi}&0&\epsilon(\bm{k}_{B2}+\bm{G}_5)
	\end{pmatrix}+\begin{pmatrix}
		0&\frac{\gamma}{2i}k_y&\frac{\gamma}{4i}(k_y+\frac{3}{2}\kappa)&\frac{\gamma}{4i}(k_y-\frac{3}{2}\kappa)\\
		-\frac{\gamma}{2i}k_y&0&-\frac{\gamma}{4i}(k_y+\frac{3}{2}\kappa)&-\frac{\gamma}{4i}(k_y-\frac{3}{2}\kappa)\\
		-\frac{\gamma}{4i}(k_y+\frac{3}{2}\kappa)&\frac{\gamma}{4i}(k_y+\frac{3}{2}\kappa)&0&0\\
		-\frac{\gamma}{4i}(k_y-\frac{3}{2}\kappa)&\frac{\gamma}{4i}(k_y-\frac{3}{2}\kappa)&0&0
	\end{pmatrix}.\label{Fourband}
\end{equation}

The band dispersions given by this four-band continuum model are plotted  in Fig.~\ref{fig:figs3} as red dashed lines. It can be seen that it roughly  captures the topology of numerically calculated moir\'e bands, although the bands gradually deviates from directly numerically calculated moir\'e bands at large pseudo-magnetic fields such as the case $\Phi/\Phi_0=0.8$ shown in Fig.~\ref{fig:figs3}.   
\subsection{The comparison between the bands from the continuum model and DFT bands}
\begin{figure}
	\centering
	\includegraphics[width=1\linewidth]{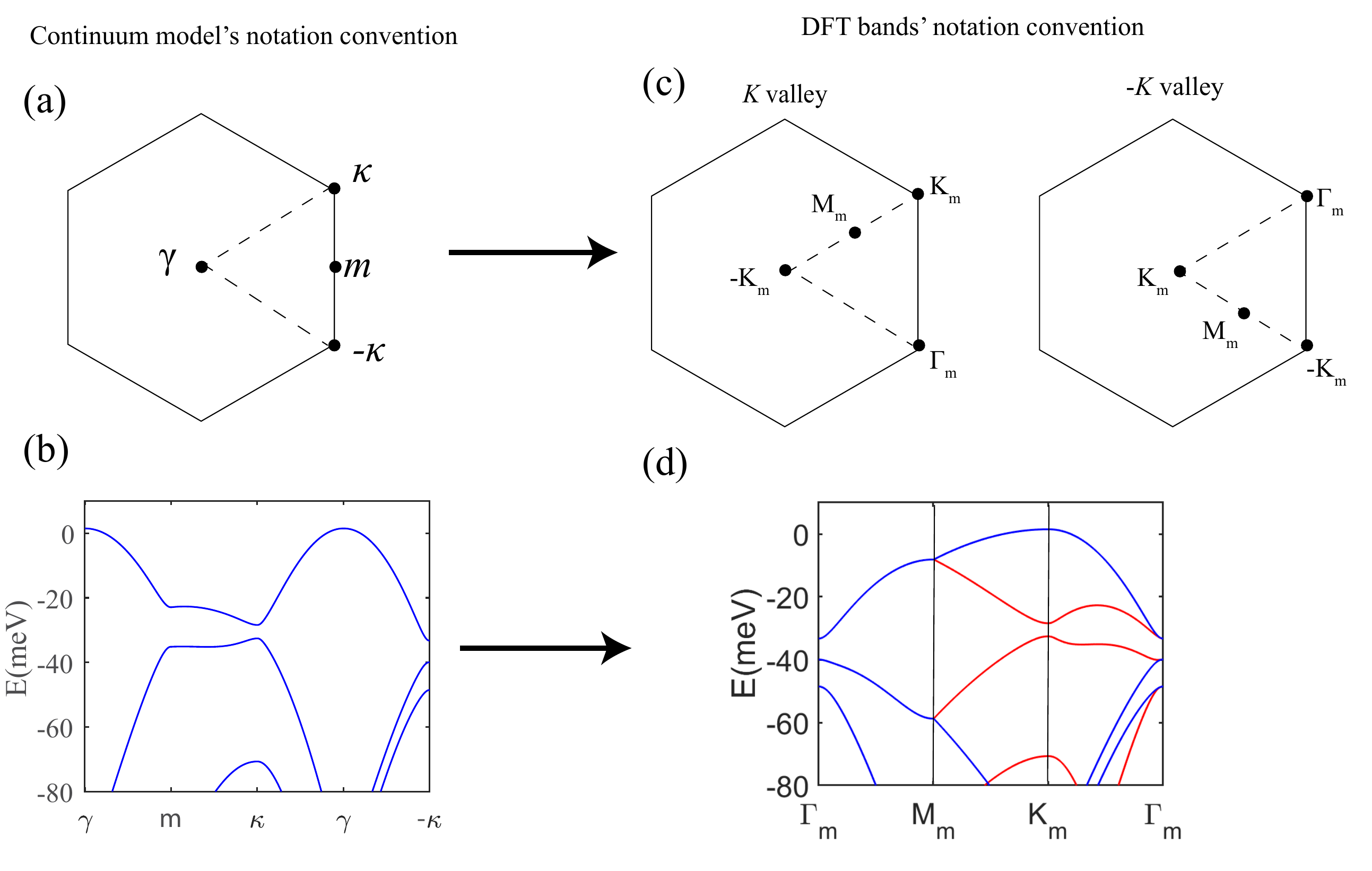}
	\caption{(a) and (c) respectively show the notation convention of continuum model, DFT bands \cite{Fai_ex2021} labelling the high symmetry points of moiré Brillouin zone.  (b) A copy of Fig. 1e in the main text. (d) is a plot of moiré bands of (b) using the DFT bands’ notation convention. Here the moiré bands from K valley are in blue and K’ valley are in red.  In this convention, the top moiré bands from two valleys exhibit different energy at $K_m$ which gives rise to a splitting. The features of top moiré bands are quite consistent with the ones in DFT bands (see Extended Data Figure 8 of ref.~\cite{Fai_ex2021}). }
	\label{fig:figs6}
\end{figure}

After we post this work, the DFT bands of AB-stacked MoTe$_2$/WSe$_2$ heterobilayer were presented in ref.~\cite{Fai_ex2021,Fu_zhang2021} recently. In this part, we present a comparison on the bands from our continuum model and  from Zhang and Fu's DFT calculations (see Extended Data Figure 8 of ref.~\cite{Fai_ex2021}). First, the valance band top of moir\'{e} bands in our work and Zhang and Fu's DFT bands are both originated from K/K' states of monolayer MoTe$_2$. In other words, although we labelled the valence band maximum as $\gamma $, it does not mean the valence bands are originated from $\Gamma$ states of monolayer MoTe$_2$.  It is important to note that the convention of labelling the high symmetry points of moir\'{e} Brillouin zone in our work and Zhang and Fu's DFT bands is different. 

Specifically, we followed the Wu and MacDonald's continuum model's notation convention \cite{fengcheng2018}, where the valence bands top of K/K' states after folding into moir\'{e} Brillouin zone is labelled as lowercase $\gamma $, as shown in Fig.~\ref{fig:figs6}(a). In contrast, the valence bands top becomes ${\mathrm{\pm }K}_m$ in the notation convention of Zhang and Fu's DFT bands  as plotted in Fig.~\ref{fig:figs6}(c), inferring from their theoretical paper \cite{Fu_zhang2021}. 

To compare the bands from our continuum model with their DFT bands more clearly, we replot the bands (Fig.~\ref{fig:figs6}(b)) following Zhang and Fu' convention. The resulting figure is Fig.~\ref{fig:figs6}(d), where the moir\'{e} bands from both original K/K' valleys are highlighted as red/blue colour.   Regardless of a constant shift, it can be seen that the feature of top moir\'{e} bands at two valleys in Fig.~\ref{fig:figs6}(d) is quite consistent with the DFT bands (Extended Data Figure 8 of ref.~\cite{Fai_ex2021}) .

Another slight difference in the DFT bands of ref.~\cite{Fai_ex2021} is that  there are flats that originate from $\Gamma$-valley states of monolayer intersecting the more dispersive K/K' bands. Note that due to a large separation of K/K' and $\Gamma$ valley in momentum space, the $\Gamma$-valley moir\'{e} bands and K/K' valley moir\'{e} bands are decoupled.  K valley is much higher than $\Gamma$-valley for the monolayer WSe$_2$ and MoTe$_2$ ($\mathrm{\sim}$300 meV difference for WSe$_2$ and 500 meV energy difference for MoTe$_2$ \cite{XiaoDi2013}), although the $\Gamma$-valley moir\'{e} bands are pushed up by the interlayer tunneling in this heterobilayer system. Nevertheless, the top moir\'{e} bands in DFT results are still contributed from K/K' valley, which give rise to the topological properties \cite{Fai_ex2021}. On the other hand, from the experimental side, there are no evidence that the $\Gamma$ valley moir\'{e} bands would come near Fermi energy and affect the quantum anomalous Hall state. Therefore, in our continuum model, we directly focused on the moir\'{e} bands folded from K/K' states of monolayer MoTe$_2$.  $\Gamma$-valley moir\'{e} bands are not relevant to our discussion concerning the topological properties from K/K' valley moir\'{e} bands.

It is worth noting that our continuum model captures both the AA and AB stacking, where MoTe$_2$ and WSe$_2$ are stacked in a parallel and antiparallel way respectively. The reason is that in both cases, it is the holes from MoTe$_2$ layer are moving in the superlattice potential generated by the WSe$_2$ layer. However, the strength of moir\'{e} potential and pseudo-magnetic fields for two distinct stackings should be  different. Importantly, in AB stacking due to the spin-valley locking at K/K'-valley, two layers would possess opposite spin in the same valley. As a result, the interlayer coupling is expected to be weaker in AB stacking, which results in a weaker moir\'{e} potential. As we showed in this work, to drive the system to be topological, the gap change caused by the pseudomagnetic fields should overcome the band gap opened by the periodic moir\'{e} potential.  Therefore, we would expect pseudo-magnetic fields are easier to create topological moir\'{e} bands in AB stacking than AA stacking case.  But from the symmetry point of view, it allows the topological moir\'{e} bands induced by pseudo-magnetic fields in AA-stacked MoTe$_2$/WSe$_2$ as well.

\section{Coulomb interaction and Hartree-Fock calculations}

The Coulomb interaction is written as
\begin{equation}
	H_{int}=\frac{1}{2}\int d\bm{r}\int d\bm{r'} c^{\dagger}_{\sigma}(\bm{r})c^{\dagger}_{\sigma'}(\bm{r'})V(\bm{r}-\bm{r'})c_{\sigma'}(\bm{r'})c_{\sigma}(\bm{r}),
\end{equation}
where $\sigma$ is the spin index. Here, the screened Coulomb interaction $V(\bm{r})=\frac{e^2}{4\pi\epsilon\epsilon_0}\frac{e^{-\lambda\bm{r}}}{|\bm{r}|}$ and  the electron creation operator 
\begin{equation}
	c_{\sigma}(\bm{r})=\sum_{n,\bm{k}\in\text{B.Z.}}\psi_{n,\sigma,\bm{k}}(\bm{r})c_{n,\sigma}(\bm{k}),
\end{equation}
with $\psi_{n,\sigma,\bm{k}}(\bm{r})=u_{n,\sigma,\bm{k}}(\bm{r})e^{i\bm{k}\cdot\bm{r}}=\frac{1}{\sqrt{S}}\sum_{\bm{G}}e^{i(\bm{G}+\bm{k})\cdot \bm{r}}u_{n,\sigma,\bm{k}}(\bm{G})$ as the Bloch wave function, $n$ is the band index, $\bm{k}$ is defined in the moir\'e Brillouin zone. Then Coulomb interaction in the momentum space reads 
\begin{equation}
	H_{int}=\frac{1}{2S}\sum_{\bm{k}_1,\bm{k}_2,\bm{k}_3,\bm{k}_4}V_{nn'mm'}(\bm{k}_1,\bm{k}_2,\bm{k}_3,\bm{k}_4)c^{\dagger}_{n,\sigma}(\bm{k}_1)c^{\dagger}_{m,\sigma'}(\bm{k}_2)c^{\dagger}_{m',\sigma'}(\bm{k}_3)c_{n',\sigma}(\bm{k}_4)
\end{equation}
and 
\begin{align}
	V_{nn'mm'}(\bm{k}_1,\bm{k}_2,\bm{k}_3,\bm{k}_4)&=\frac{1}{S}\int d\bm{r}\int d\bm{r'}\psi^*_{n,\sigma,\bm{k}_1}(\bm{r})\psi_{n',\sigma,\bm{k}_4}(\bm{r})V(\bm{r}-\bm{r'})\psi^*_{m,\sigma',\bm{k}_2}(\bm{r'})\psi_{m',\sigma,\bm{k}_3}(\bm{r'})\\
	&=\sum_{\bm{q},\bm{G}'_1,\bm{G}'_2,\bm{G}_3',\bm{G}_4'}V(\bm{q})\Lambda^{\sigma}_{nn'}(\bm{k}_1,\bm{k}_4)\Lambda^{\sigma}_{mm'}(\bm{k}_2,\bm{k}_3)\delta_{\bm{k_1}+\bm{G_1^{'}},\bm{k_4}+\bm{q}+\bm{G}_4^{'}}\delta_{\bm{k_2}+\bm{G}_2^{'},\bm{k_3}+\bm{G}_3^{'}-\bm{q}},\label{Coulomb}
\end{align}
where  $V(\bm{r})=\frac{1}{S}\sum_{\bm{q}}V(\bm{q})e^{i\bm{q}\cdot\bm{r}}$ with $V(\bm{q})=\frac{e^2}{2\epsilon\epsilon_0\sqrt{q^2+\lambda^2}}$, $\Lambda^{\sigma}_{nn'}(\bm{k}_i,\bm{k}_j)=u_{\bm{k}_i}^{\dagger}(\bm{G}'_i)u_{\bm{k}_j}(\bm{G}'_j)$. Due to giant Ising spin-orbit coupling (SOC $\sim$100 meV) near the valence band top of 2H-type transition metal dichalcogenide, the spin and valley are locked together, and thus we can replace the spin index $\sigma$ with valley index $\tau$. Then we obtain the Coulomb interaction as
\begin{equation}
	H_{int}=\frac{1}{2S}\sum_{\bm{k},\bm{k}',\bm{q}} V(\bm{q}) \Lambda^{\tau}_{nn'}(\bm{k}+\bm{q},\bm{k})\Lambda^{\tau'}_{mm'}(\bm{k'}-\bm{q},\bm{k'})c^{\dagger}_{n,\tau}(\bm{k}+\bm{q})c^{\dagger}_{m,\tau'}(\bm{k'}-\bm{q})c_{m',\tau'}(\bm{k'})c_{n',\tau}(\bm{k}).
\end{equation}

Note $\bm{q}$ is defined in $R^2$ so that $\bm{k}\pm \bm{q}$ can exceeds the first Brillouin and in the calculation, it needs to be projected back to the first Brillouin zone as $\bm{k}\pm\bm{q}=\bm{G}'(\bm{k}\pm \bm{q})+\bm{p}(\bm{k}\pm \bm{q})$, being equivalent to adjust $\bm{G}'_j$ in Eq.~(\ref{Coulomb}). Especially, after this projection, the form factor is given by

\begin{equation}
	\Lambda^{\tau}_{nn'}(\bm{k\pm\bm{q}},\bm{k})=\sum_{\bm{G}}u_{\bm{p}(\bm{k}\pm\bm{q}),\tau}^{\dagger}(\bm{G}+\bm{G'}(\bm{k}\pm\bm{q}))u_{\bm{k},\tau}(\bm{G})\equiv\braket{u_{n,\tau,\bm{k}\pm \bm{q}}|u_{n,\tau,\bm{k}}}.
\end{equation} 
It can be seen that the intervalley Hund's coupling \cite{Lee2019} is suppressed by the Ising SOC. From the definition, $\Lambda_{nn'}^{\tau}(\bm{k_1},\bm{k}_2)=(\Lambda_{n'n}^{\tau}(\bm{k_2},\bm{k}_1))^*$ and the time-reversal symmetry requires $\Lambda_{nn'}^{\tau}(\bm{k_1},\bm{k}_2)=\Lambda_{n'n}^{-\tau}(-\bm{k_2},-\bm{k}_1))$.

\subsection{Hartree-Fock mean-field approximation with a simple form of Coulomb interaction}
Next, let us perform the Hartree-Fock mean-field approximation. We first assume a simple form of interaction as
\begin{equation}
	H_{int}=\frac{g}{2N}\sum_{\bm{k},\bm{k'},\bm{q}}c^{\dagger}_{\tau}(\bm{k}+\bm{q})c^{\dagger}_{\tau'}(\bm{k'}-\bm{q})c_{\tau'}(\bm{k'})c_{\tau}(\bm{k}),
\end{equation}
where the interaction effects on the top moir\'e bands is studied and the the interaction strength is taken as $g$, $N$ is the number of moir\'e unit cells.   A more complete form will be discussed later. Then we define the expectation value 
\begin{equation}
	\Delta_{\tau\tau'}(\bm{k},\bm{k'})=\braket{c_{\tau}^{\dagger}(\bm{k})c_{\tau'}(\bm{k'})},
\end{equation}
which is assumed to be diagonal in momentum space, i.e., $\Delta_{\tau\tau'}(\bm{k},\bm{k'})=\Delta_{\tau\tau'}(\bm{k})\delta_{\bm{k},\bm{k'}}$.

The constraint for the order parameters are
\begin{eqnarray}
	filling: &&  
	\frac{1}{N}\sum_{\bm{k}}\text{tr}[\Delta(\bm{k})]=\nu.\\
	Symmetry:&& C_3: \Delta(\bm{k})\mapsto \Delta(C_3\bm{k})\\
	&&T: \Delta(\bm{k})\mapsto \tau_x\Delta^*(-\bm{k})\tau_x\label{time_reversal}
\end{eqnarray} 
where $C_3=\tau_0$ is the three-fold rotational symmetry, $T=\tau_xK$ is the time-reversal symmetry.	

We expand the $H_{int}$ in a  mean-field   manner:
\begin{align}
	&H^{MF}_{int}\approx\frac{g}{2N}\sum_{\bm{k},\bm{k'},\bm{q}} c^{\dagger}_{\tau}(\bm{k}+\bm{q})c_{\tau}(\bm{k})\braket{c^{\dagger}_{\tau'}(\bm{k'}-\bm{q})c_{\tau'}(\bm{k'})}+ \braket{c^{\dagger}_{\tau}(\bm{k}+\bm{q})c_{\tau}(\bm{k})}c^{\dagger}_{\tau'}(\bm{k'}-\bm{q})c_{\tau'}(\bm{k'})-\nonumber\\
	&\braket{c^{\dagger}_{\tau}(\bm{k}+\bm{q})c_{\tau}(\bm{k})}\braket{c^{\dagger}_{\tau'}(\bm{k'}-\bm{q})c_{\tau'}(\bm{k'})}-\braket{c^{\dagger}_{\tau}(\bm{k}+\bm{q})c_{\tau'}(\bm{k'})}c^{\dagger}_{\tau'}(\bm{k'}-\bm{q})c_{\tau}(\bm{k})-c^{\dagger}_{\tau}(\bm{k}+\bm{q})c_{\tau'}(\bm{k'})\braket{c^{\dagger}_{\tau'}(\bm{k'}-\bm{q})c_{\tau}(\bm{k})}+\nonumber\\
	&\braket{c^{\dagger}_{\tau}(\bm{k}+\bm{q})c_{\tau'}(\bm{k'})}\braket{c^{\dagger}_{\tau'}(\bm{k'}-\bm{q})c_{\tau}(\bm{k})}.
\end{align}

The first three terms are the Hartree contributions, which is assumed to be finite only at $\bm{q}=0$, and the last three terms are the Fock contributions. In a homogeneous electron gas, the Hartree contributions  are canceled by the direct interaction with the positive background, and such contributions are determined by the local density of electrons and should not be sensitive to the specific order. Thus,  the Fock terms are  usually kept for our purpose.
Then we obtain
\begin{equation}
	H^{MF}_{int}\approx -g\sum_{\bm{k}}c^{\dagger}_{\tau'}(\bm{k})\Delta^{T}_{\tau'\tau}c_{\tau}(\bm{k})+\frac{gN}{2}\text{tr}(\Delta^2).
\end{equation}

(1) the spin-valley-polarized (SVP) state. The time-reversal symmetry is broken.

The order parameter for the valley-polarized states are 
\begin{equation}
	\Delta(\bm{k})=\Delta_0(\bm{k})\tau_0+\Delta_z(\bm{k})\tau_z.
\end{equation} 
We can define the macroscopic mean-field order parameters as
\begin{equation}
	\Delta_{0}=\frac{1}{N}\sum_{\bm{k}}\Delta_0(\bm{k}), \Delta_{z}=\frac{1}{N}\sum_{\bm{k}}\Delta_z(\bm{k}).
\end{equation} As a result, $\Delta_0=1/2$ at half-filling. And we consider interaction is much larger than bandwidth so that only one band is filled in near half-filling. In this case, the self-consistent equation reads
\begin{align}
	\Delta_0+\Delta_z=\frac{1}{N}\sum_{\bm{k}}\Delta_{++}(\bm{k})=1\\
	\Delta_0-\Delta_z=\frac{1}{N}\sum_{\bm{k}}\Delta_{--}(\bm{k})=0.
\end{align}
This gives $\Delta_z=1/2$. The energy for the spin valley-polarized states are assumed to be 
\begin{equation}
	E_{SVP}=\sum_{\bm{k}}\xi_{+}(\bm{k})-gN/2.
\end{equation}

(2) the spin-valley-locked intervalley coherent  (IVC) state (we will simply refer it as IVC state in the later discussions).  The valley $U_{v}(1)$ symmetry is broken.

The order parameter for the valley-polarized states are assumed to be 
\begin{equation}
	\Delta(\bm{k})=\begin{pmatrix}
		\Delta_0(\bm{k})&\Delta_1(\bm{k})\\
		\Delta_1^*(\bm{k})&	\Delta_0(\bm{k})
	\end{pmatrix}.
\end{equation} 
Here, $\Delta_0(\bm{k})$ is real, while $\Delta_1(\bm{k})$ is complex. We define $\Delta_1=|\Delta_1|e^{i\varphi}=\frac{1}{N}\sum_{\bm{k}}\Delta_1(\bm{k})$.
The mean-field Hamiltonian becomes 
\begin{equation}
	H_{MF}=\sum_{\bm{k}}(\xi_{\tau}(\bm{k})-g\Delta_0)c^{\dagger}_{\tau}(\bm{k})c_{\tau}(\bm{k})-g\sum_{\bm{k}}(\Delta_1^*c^{\dagger}_{+}(\bm{k})c_{-}(\bm{k})+\Delta_1c^{\dagger}_{-}(\bm{k})c_{+}(\bm{k}))+gN(\Delta_0^2+|\Delta_1|^2).
\end{equation}
This mean-field Hamiltonian is diagonalized  with a unitary transformation:
\begin{eqnarray}
	c_{+}(\bm{k})&=&\cos\frac{\theta_{\bm{k}}}{2}e^{-i\frac{\varphi}{2}}\gamma_{+}(\bm{k})+\sin\frac{\theta_{\bm{k}}}{2}e^{-i\frac{\varphi}{2}}\gamma_{-}(\bm{k}),\nonumber\\
	c_{-}(\bm{k})&=&-\sin\frac{\theta_{\bm{k}}}{2}e^{i\frac{\varphi}{2}}\gamma_{+}(\bm{k})+\cos\frac{\theta_{\bm{k}}}{2}e^{i\frac{\varphi}{2}}\gamma_{-}(\bm{k}),\label{transform}
\end{eqnarray}
where 
\begin{equation}
	\cos\theta_{\bm{k}}=\frac{\xi_a(\bm{k})}{\sqrt{\xi^2_{a}(\bm{k})+g^2N^2|\Delta_1|^2}}, \sin\theta_{\bm{k}}=\frac{gN|\Delta_1|}{\sqrt{\xi^2_{a}(\bm{k})+g^2N^2|\Delta_1|^2}},
\end{equation}
and we denote $\xi_{s(a)}(\bm{k})=(\xi_{+}(\bm{k})\pm \xi_{-}(\bm{k}))/2$. After this transformation, the mean field Hamiltonian becomes
\begin{equation}
	H_{MF}=\sum_{\bm{k}}E_{+}(\bm{k})\gamma^{\dagger}_{+}(\bm{k})\gamma_{+}(\bm{k})+E_{-}(\bm{k})\gamma^{\dagger}_{-}(\bm{k})\gamma_{-}(\bm{k})++gN(\Delta_0^2+|\Delta_1|^2).
\end{equation}
Here,	the eigenenergies $E_{\pm}(\bm{k})$ are  given by
\begin{equation}
	E_{\pm}(\bm{k})=\xi_{s}(\bm{k})-gN\Delta_0\pm \sqrt{(\xi_{a}(\bm{k}))^2+g^2N^2|\Delta_1|^2}.
\end{equation}
The self-consistent equation is
\begin{equation}
	\Delta_1=\frac{1}{N}\sum_{\bm{k}}\braket{c^{\dagger}_{+}(\bm{k})c_{-}(\bm{k})}=\frac{\Delta_1}{2N}\sum_{\bm{k}}\sin\theta_{\bm{k}}\braket{-\gamma^{\dagger}_{+}(\bm{k})\gamma_{+}(\bm{k})+\gamma^{\dagger}_{-}(\bm{k})\gamma_{-}(\bm{k})}
\end{equation}	
At half-filling, only the $E_{-}(\bm{k})$ bands are filled. The self-consistent equation is simplified as
\begin{equation}
	|\Delta_1|=\frac{1}{2}\sum_{\bm{k}}\frac{g|\Delta_1|}{\sqrt{\xi^2_{a}(\bm{k})+g^2N^2|\Delta_1|^2}}.
\end{equation}
When the interaction is much larger than bandwidth, we can obtain $|\Delta_1|\approx 1/2$. The filling constraint further gives $\Delta_0=1/2$. Hence, the energy for the IVC state is approximated as
\begin{equation}
	E_{IVC}=\sum_{\bm{k}}\xi_s(\bm{k})-\sqrt{\xi^2_a(\bm{k})+g^2N^2/4}.
\end{equation}
The energy difference between the SVP states and the IVC states is
\begin{equation}
	\delta E=E_{SVP}-E_{IVC}=\sum_{\bm{k}}\sqrt{\xi_a^2(\bm{k})+g^2N^2/4}-gN/2>0
\end{equation}
Therefore, without considering the form factor, which encodes the information of the nontrivial wave-function, the states are tend to  form the IVC states instead of SVP state. 
\subsection{Hartree-Fock mean-field approximation with a more general form of Coulomb interaction}

Next, let us  consider a more general form of Coulomb interaction:
\begin{equation}
	H_{int}=\frac{V_0}{2N}\sum_{\bm{k},\bm{k'},\bm{q}}v_{\bm{q}}\Lambda^{\tau}(\bm{k}+\bm{q},\bm{k})\Lambda^{\tau'}(\bm{k'}-\bm{q},\bm{k'})c^{\dagger}_{\tau}(\bm{k}+\bm{q})c^{\dagger}_{\tau'}(\bm{k'}-\bm{q})c_{\tau'}(\bm{k'})c_{\tau}(\bm{k}),
\end{equation}
where  $V_0=\frac{e^2}{2\epsilon\epsilon_0|\kappa|\Omega}$, the dimensionless screened Coulomb interaction is $v_{\bm{q}}=|\kappa|/\sqrt{\bm{q}^2+\lambda^2}$, and only the top moir\'e band is considered. 
In a mean-field manner, the interaction is expanded as
\begin{align}
	H_{int}^{MF}&\approx \frac{V_0}{2N}\sum_{\bm{k},\bm{k'},\bm{q}}v_{\bm{q}}\Lambda^{\tau}(\bm{k}+\bm{q},\bm{k})\Lambda^{\tau'}(\bm{k'}-\bm{q},\bm{k'})[c^{\dagger}_{\tau}(\bm{k}+\bm{q})c_{\tau}(\bm{k})\braket{c^{\dagger}_{\tau'}(\bm{k'}-\bm{q})c_{\tau'}(\bm{k'})}+ \braket{c^{\dagger}_{\tau}(\bm{k}+\bm{q})c_{\tau}(\bm{k})}c^{\dagger}_{\tau'}(\bm{k'}-\bm{q})c_{\tau'}(\bm{k'})-\nonumber\\
	&\braket{c^{\dagger}_{\tau}(\bm{k}+\bm{q})c_{\tau}(\bm{k})}\braket{c^{\dagger}_{\tau'}(\bm{k'}-\bm{q})c_{\tau'}(\bm{k'})}-\braket{c^{\dagger}_{\tau}(\bm{k}+\bm{q})c_{\tau'}(\bm{k'})}c^{\dagger}_{\tau'}(\bm{k'}-\bm{q})c_{\tau}(\bm{k})-c^{\dagger}_{\tau}(\bm{k}+\bm{q})c_{\tau'}(\bm{k'})\braket{c^{\dagger}_{\tau'}(\bm{k'}-\bm{q})c_{\tau}(\bm{k})}+\nonumber\\
	&\braket{c^{\dagger}_{\tau}(\bm{k}+\bm{q})c_{\tau'}(\bm{k'})}\braket{c^{\dagger}_{\tau'}(\bm{k'}-\bm{q})c_{\tau}(\bm{k})}].
\end{align}
To make the Coulomb interaction Hamiltonian more compact, let us define the Hartree  and Fock order parameters:
\begin{eqnarray}
	\Delta^H_{\tau\tau'}(\bm{G})&&=\frac{1}{N}\sum_{\bm{k'}}v_{\bm{G}}\Lambda^\tau(\bm{k'}+ \bm{G},\bm{k'})\braket{c^{\dagger}_{\tau}(\bm{k'}+\bm{G})c_{\tau'}(\bm{k'})}\delta_{\tau\tau'},\\
	\Delta_{\tau\tau'}^{F}(\bm{k},\bm{G})&&=\frac{1}{N}\sum_{\bm{k'}}v_{\bm{k'}-\bm{k}+\bm{G}}\Lambda^{\tau}(\bm{k'}+\bm{G},\bm{k})\Lambda^{\tau'}(\bm{k}-\bm{G},\bm{k'})\braket{c^{\dagger}_{\tau}(\bm{k'}+\bm{G})c_{\tau'}(\bm{k'})}
\end{eqnarray} 
With these definitions, we rewrite the first three Hartree terms as
\begin{equation}
	H^{H}_{MF}\approx V_0 \sum_{\bm{k}}\sum_{\bm{G}} \text{tr}[\Delta^H(\bm{G})]\Lambda^{\tau}(\bm{k}-\bm{G},\bm{k}) c^{\dagger}_{\tau}(\bm{k}-\bm{G})c_{\tau}(\bm{k})-\frac{NV_0}{2}\sum_{\bm{G}}\frac{\text{tr}[\Delta^{H}(\bm{G})]\text{tr}[\Delta^{H}(-\bm{G})]}{v_{\bm{G}}}.
\end{equation}
Note  the Hartree terms at $\bm{G}=0$ are still considered to be canceled by some positive charge background.
The next three are Fock terms:
\begin{align}
	H^{F}_{MF}&\approx-V_0\sum_{\bm{k}}\sum_{\bm{G}}c^{\dagger}_{\tau}(\bm{k}-\bm{G})[\Delta^{F}(\bm{k},\bm{G})]^{T}_{\tau\tau'}c_{\tau'}(\bm{k})\\
	&+\frac{V_0}{2N}\sum_{\bm{k},\bm{k'},\bm{G}}v_{\bm{k'}-\bm{k}+\bm{G}}\Lambda^{\tau}(\bm{k'}+\bm{G},\bm{k})\Lambda^{\tau'}(\bm{k}-\bm{G},\bm{k'})\braket{c^{\dagger}_{\tau}(\bm{k'}+\bm{G})c_{\tau'}(\bm{k'})}\braket{c^{\dagger}_{\tau'}(\bm{k}-\bm{G})c_{\tau}(\bm{k})}.
\end{align}
Since we always do the calculation in the first Brillouin zone, $c^{\dagger}(\bm{k}\pm \bm{G})$ needs to be projected back to  the first Brillouin zone as mentioned previously. After this projection, we arrive at a mean-field Hamiltonian as
\begin{align}
	H_{MF}&\approx\sum_{\bm{k}}c^{\dagger}_{\tau}(\bm{k})(\xi_{\tau}(\bm{k})+V_0\sum_{\bm{G}}\text{tr}[\Delta^{H}(\bm{G})]\Lambda^{\tau}(\bm{k}-\bm{G},\bm{k}))c_{\tau}(\bm{k})-V_0\sum_{\bm{k}}\sum_{\bm{G}}c^{\dagger}_{\tau}(\bm{k})[\Delta^{F}(\bm{k},\bm{G})]^{T}_{\tau\tau'}c_{\tau'}(\bm{k})\nonumber\\
	&-\frac{NV_0}{2}\sum_{\bm{G}}\frac{\text{tr}[\Delta^{H}(\bm{G})]\text{tr}[\Delta^{H}(-\bm{G})]}{v_{\bm{G}}}+\frac{V_0}{2N}\sum_{\bm{k},\bm{k'},\bm{G}}v_{\bm{k'}-\bm{k}+\bm{G}}\text{tr}[\Lambda(\bm{k'}+\bm{G},\bm{k})\Delta(\bm{k'})\Lambda(\bm{k}-\bm{G},\bm{k'})\Delta(\bm{k})],\label{Full_mean}
\end{align}
where
\begin{eqnarray}
	\Delta^H_{\tau\tau'}(\bm{G})&&=\frac{1}{N}\sum_{\bm{k'}}v_{\bm{G}}\Lambda^\tau(\bm{k'}+ \bm{G},\bm{k'})\braket{c^{\dagger}_{\tau}(\bm{k'}+\bm{G})c_{\tau'}(\bm{k'})}\delta_{\tau\tau'},\\
	\Delta_{\tau\tau'}^{F}(\bm{k},\bm{G})&&=\frac{1}{N}\sum_{\bm{k'}}v_{\bm{k'}-\bm{k}+\bm{G}}\Lambda^{\tau}(\bm{k'}+\bm{G},\bm{k})\Lambda^{\tau'}(\bm{k}-\bm{G},\bm{k'})\braket{c^{\dagger}_{\tau}(\bm{k'}+\bm{G})c_{\tau'}(\bm{k'})}
\end{eqnarray}
and the form factor is given by
\begin{eqnarray}
	\Lambda^{\tau}(\bm{k}\pm \bm{G},\bm{k})=\sum_{\bm{G'}}u^{\dagger}_{\bm{k},\tau}(\bm{G'}\pm \bm{G})u_{\bm{k},\tau}(\bm{G'})\\
	\Lambda^{\tau}(\bm{k'}\pm \bm{G},\bm{k})=\sum_{\bm{G'}}u^{\dagger}_{\bm{k'},\tau}(\bm{G'}\pm \bm{G})u_{\bm{k},\tau}(\bm{G'})
\end{eqnarray}

Let us consider the long-wave limit $qL_M\ll 1$ so that only $\bm{G}=0$ in the sum needs to be considered. A more general case will be evaluated numerically as we will present later.  In the long-wave limit case,
\begin{align}
	H_{MF}\approx \sum_{\bm{k}}c^{\dagger}_{\tau}(\bm{k})\xi_{\tau}(\bm{k})c_{\tau}(\bm{k})-V_0\sum_{\bm{k}}c_{\tau}^{\dagger}(\bm{k})[\Delta^{F}(\bm{k})]_{\tau\tau'}^{T}c_{\tau'}(\bm{k})+\frac{V_0}{2N}\sum_{\bm{k},\bm{q}}v_{\bm{q}}\text{tr}[\Lambda(\bm{k}+\bm{q},\bm{k})\Delta(\bm{k}+\bm{q})\Lambda(\bm{k},\bm{k}+\bm{q})\Delta(\bm{k})],\label{mean_79}
\end{align} 
where
\begin{equation}
	\Delta^{F}_{\tau\tau'}(\bm{k})=\frac{1}{N}\sum_{\bm{q}}v_{\bm{q}}\Lambda^{\tau}(\bm{k}+\bm{q},\bm{k})\Delta_{\tau\tau'}(\bm{k}+\bm{q})\Lambda^{\tau'}(\bm{k},\bm{k}+\bm{q}).
\end{equation}

For the SVP states, only one valley is occupied. Without loss of generality, we assume the $+$ valley is occupied, which gives the mean-field order parameter: $\Delta(\bm{k})=1/2(1+\tau_z)$.
The total energy of this SVP state is obtained as
\begin{equation}
	E_{SVP}=\sum_{\bm{k}}[\xi_{+}(\bm{k})-\frac{V_0}{2N}\sum_{\bm{q}}v_{\bm{q}}|\Lambda^+(\bm{k}+\bm{q},\bm{k})|^2].
\end{equation}

Let us further consider the IVC states.  In this case, the valley is not a good index. In this case, we  take a general form of the order parameter:
\begin{equation}
	\Delta(\bm{k})=\begin{pmatrix}
		\braket{c^{\dagger}_{+}(\bm{k})c_{+}(\bm{k})}&\braket{c^{\dagger}_{+}(\bm{k})c_{-}(\bm{k})}\\
		\braket{c^{\dagger}_{-}(\bm{k})c_{+}(\bm{k})}&\braket{c^{\dagger}_{-}(\bm{k})c_{-}(\bm{k})}
	\end{pmatrix}=\begin{pmatrix}
		\Delta_{++}(\bm{k})&\Delta_{+-}(\bm{k})\\
		\Delta_{-+}(\bm{k})&\Delta_{--}(\bm{k})
	\end{pmatrix}.\label{mean_order}
\end{equation}
The mean-field Hamiltonian Eq.~(\ref{mean_79}) becomes
\begin{equation}
	H(\bm{k})=\begin{pmatrix}
		\xi_{+}(\bm{k})-\frac{V_0}{N}\sum_{\bm{q}}v_{\bm{q}}|\Lambda^+(\bm{k}+\bm{q},\bm{k})|^2\Delta_{++}(\bm{k}+\bm{q})&-\frac{V_0}{N}\sum_{\bm{q}}v_{\bm{q}}\Lambda^-(\bm{k}+\bm{q},\bm{k})\Delta_{-+}(\bm{k}+\bm{q})\Lambda^+(\bm{k},\bm{k}+\bm{q})\\
		-\frac{V_0}{N}\sum_{\bm{q}}v_{\bm{q}}\Lambda^+(\bm{k}+\bm{q},\bm{k})\Delta_{+-}(\bm{k}+\bm{q})\Lambda^-(\bm{k},\bm{k}+\bm{q})&\xi_{-}(\bm{k})-\frac{V_0}{N}\sum_{\bm{q}}v_{\bm{q}}|\Lambda^{-}(\bm{k}+\bm{q},\bm{k})|^2\Delta_{--}(\bm{k}+\bm{q})
	\end{pmatrix}.
\end{equation}
Here, the basis is $(c_+(\bm{k}),c_{-}(\bm{k}))^{T}$. The last term in Eq.~(\ref{mean_79}) is a potential energy which will be added later. We can further parameterize $H(\bm{k})$ as
\begin{equation}
	H(\bm{k})=\begin{pmatrix}
		h_0(\bm{k})+h_1(\bm{k})&-h_2^*(\bm{k})\\
		-h_2(\bm{k})&h_0(\bm{k})-h_1(\bm{k})
	\end{pmatrix}.\label{mean_form}
\end{equation}
According to Eq.~(\ref{transform}), we can take  the following transform to diagonalize $H(\bm{k})$:
\begin{eqnarray}
	c_{+}(\bm{k})&=&\cos\frac{\theta_{\bm{k}}}{2}e^{-i\frac{\varphi_{\bm{k}}}{2}}\gamma_{+}(\bm{k})+\sin\frac{\theta_{\bm{k}}}{2}e^{-i\frac{\varphi_{\bm{k}}}{2}}\gamma_{-}(\bm{k}),\nonumber\\
	c_{-}(\bm{k})&=&-\sin\frac{\theta_{\bm{k}}}{2}e^{i\frac{\varphi_{\bm{k}}}{2}}\gamma_{+}(\bm{k})+\cos\frac{\theta_{\bm{k}}}{2}e^{i\frac{\varphi_{\bm{k}}}{2}}\gamma_{-}(\bm{k}).\label{unitary}
\end{eqnarray}
After the unitary transform, we obtain the mean-field Hamiltonian
\begin{equation}
	H=\sum_{\bm{k}}E_{\pm}(\bm{k})\gamma^{\dagger}_{\pm}(\bm{k})\gamma_{\pm}(\bm{k}).
\end{equation}
The eigenenergies $E_{\pm}(\bm{k})=h_0(\bm{k})\pm \sqrt{h_1(\bm{k})^2+|h_2(\bm{k})|^2}$.
In the half-filling, only $E_{-}(\bm{k})$ is occupied. This also means $\braket{\gamma^{\dagger}_{-}(\bm{k}) \gamma_{-}(\bm{k})}=1$ and $\braket{\gamma^{\dagger}_{+}(\bm{k}) \gamma_{+}(\bm{k})}=0$. The parameters $\theta_{\bm{k}}$ and $\varphi_{\bm{k}}$ are determined by the following equations
\begin{equation}
	\sin\theta_{\bm{k}}=\frac{|h_2(\bm{k})|}{\sqrt{h^2_1(\bm{k})+|h_2(\bm{k})|^2}},  \cos\theta_{\bm{k}}=\frac{h_1(\bm{k})}{\sqrt{h^2_1(\bm{k})+|h_2(\bm{k})|^2}}\label{self_cons}, \tan \varphi_{\bm{k}}=\frac{\text{Im}(h_2(\bm{k}))}{\text{Re}(h_2(\bm{k}))}
\end{equation}
Here, 
\begin{eqnarray}
	h_0(\bm{k})&&=\frac{1}{2}(\xi_+(\bm{k})+\xi_{-}(\bm{k}))-\frac{V_0}{2N}\sum_{\bm{q}}v_{\bm{q}}[|\Lambda^+(\bm{k}+\bm{q},\bm{k})|^2\sin^2\frac{\theta_{\bm{k}+\bm{q}}}{2}+|\Lambda^-(\bm{k}+\bm{q},\bm{k})|^2\cos^2\frac{\theta_{\bm{k}+\bm{q}}}{2}]\label{h0}\\
	h_1(\bm{k})&&=\frac{1}{2}(\xi_+(\bm{k})-\xi_{-}(\bm{k}))-\frac{V_0}{2N}\sum_{\bm{q}}v_{\bm{q}}[|\Lambda^+(\bm{k}+\bm{q},\bm{k})|^2\sin^2\frac{\theta_{\bm{k}+\bm{q}}}{2}-|\Lambda^-(\bm{k}+\bm{q},\bm{k})|^2\cos^2\frac{\theta_{\bm{k}+\bm{q}}}{2}]\\
	h_2(\bm{k})&&=\frac{V_0}{2N}\sum_{\bm{q}}v_{\bm{q}}\sin\theta_{\bm{k}+\bm{q}}\Lambda^+(\bm{k}+\bm{q},\bm{k})\Lambda^{-}(\bm{k},\bm{k}+\bm{q})e^{i\varphi_{\bm{k}+\bm{q}}}.
\end{eqnarray}
Note  we have replaced $\Delta(\bm{k})$ in the Hamiltonian as 
\begin{equation}
	\Delta(\bm{k})=\begin{pmatrix}
		\sin^2\frac{\theta_{\bm{k}}}{2}&\frac{1}{2}\sin\theta_{\bm{k}}e^{i\varphi_{\bm{k}}}\\
		\frac{1}{2}\sin\theta_{\bm{k}}e^{-i\varphi_{\bm{k}}}&  \cos^2\frac{\theta_{\bm{k}}}{2}\label{IVCorder}
	\end{pmatrix},
\end{equation}
which is obtained by substituting Eq.~(\ref{unitary}) in Eq.~(\ref{mean_order}), and    $\theta_{\bm{k}}=\pi-\theta_{-\bm{k}}$, $\varphi_{\bm{k}}=\varphi_{-\bm{k}}$ due to the the constraint of time reversal symmetry $T=\tau_xK$ given in Eq.~\ref{time_reversal}.  It can be seen that the half-filling constraint $\frac{1}{N}\sum_{\bm{k}}\text{tr}[\Delta(\bm{k})]=1$ is satisfied.   
Moreover, there exhibits a gauge degree of freedom: $c_{\pm}(\bm{k})\rightarrow  c_{\pm}(\bm{k})e^{\mp i\frac{\phi_{\bm{k}}}{2}}$. Under this gauge transform, $\Lambda^{\pm}(\bm{k}+\bm{q},\bm{k})\rightarrow \Lambda^{\pm}(\bm{k}+\bm{q},\bm{k})e^{\pm i(\phi_{\bm{k}}-\phi_{\bm{k}+\bm{q}})/2}$. This gauge phase will affect the $h_2(\bm{k})$, but will not affect the total energy.

If we approximate $\theta_{\bm{k}}\approx \theta_{\bm{k}+\bm{q}}$ considering $qL_{M}\ll 1$,    the form of $E_{\pm}(\bm{k})$ can be simplified with the first  equation in Eq.~(\ref{self_cons}), which is rewritten as
\begin{equation}
	\sqrt{h_1^2(\bm{k})+|h_2(\bm{k})|^2}=\frac{V_0}{2N}\sum_{\bm{q}}v_{\bm{q}}\Lambda^+(\bm{k}+\bm{q},\bm{k})\Lambda^{-}(\bm{k},\bm{k}+\bm{q})e^{i(\varphi_{\bm{k}+\bm{q}}-\varphi_{\bm{k}})}.
\end{equation}
As only the band with energy $E_{-}(\bm{k})$ is filled in, we can obtain the total energy for the IVC states as
\begin{align}
	E_{IVC}&=\sum_{\bm{k}} h_0(\bm{k})-\frac{V_0}{2N}\sum_{\bm{k}}\sum_{\bm{q}}v_{\bm{q}}\Lambda^+(\bm{k}+\bm{q},\bm{k})\Lambda^{-}(\bm{k},\bm{k}+\bm{q})e^{i(\varphi_{\bm{k}+\bm{q}}-\varphi_{\bm{k}})}+ \frac{V_0}{2N}\sum_{\bm{k}}\sum_{\bm{q}}v_{\bm{q}}[\sin^2(\frac{\theta_{\bm{k}+\bm{q}}}{2})\sin^2(\frac{\theta_{\bm{k}}}{2})\nonumber\\&|\Lambda^{+}(\bm{k}+\bm{q},\bm{k})|^2+\cos^2(\frac{\theta_{\bm{k}+\bm{q}}}{2})\cos^2(\frac{\theta_{\bm{k}}}{2})|\Lambda^{-}(\bm{k}+\bm{q},\bm{k})|^2+\frac{1}{2}\sin(\theta_{\bm{k}})\sin(\theta_{\bm{k}+\bm{q}})\Lambda^{+}(\bm{k}+\bm{q},\bm{k})\Lambda^{-}(\bm{k},\bm{k}+\bm{q})e^{i(\varphi_{\bm{k}+\bm{q}}-\varphi_{\bm{k}})}].
\end{align}
The last potential term in Eq.~(\ref{mean_79}) is also added. Using the aforementioned approximation $\theta_{\bm{k}}\approx  \theta_{\bm{k}+\bm{q}}$, we obtain
\begin{align}
	E_{IVC}&=\sum_{\bm{k}} \xi_{+}(\bm{k})-\frac{V_0}{2N}\sum_{\bm{k},\bm{q}}\sin^2(\frac{\theta_{\bm{k}}}{2})\cos^2(\frac{\theta_{\bm{k}}}{2})[|\Lambda^{+}(\bm{k}+\bm{q},\bm{k})|^2+|\Lambda^{-}(\bm{k}+\bm{q},\bm{k})|^2]\\
	&-\frac{V_0}{2N}\sum_{\bm{k}}\sum_{\bm{q}}v_{\bm{q}}(1-\frac{1}{2}\sin^2(\theta_{\bm{k}}))\Lambda^+(\bm{k}+\bm{q},\bm{k})\Lambda^{-}(\bm{k},\bm{k}+\bm{q})e^{i(\varphi_{\bm{k}+\bm{q}}-\varphi_{\bm{k}})}
\end{align} 
The lowest value of $E_{IVC}$ is obtained as

\begin{align}
	E^{(0)}_{IVC}&=\sum_{\bm{k}} \xi_{+}(\bm{k})-\frac{V_0}{2N}\sum_{\bm{k},\bm{q}}\sin^2(\frac{\theta_{\bm{k}}}{2})\cos^2(\frac{\theta_{\bm{k}}}{2})[|\Lambda^{+}(\bm{k}+\bm{q},\bm{k})|^2+|\Lambda^{-}(\bm{k}+\bm{q},\bm{k})|^2]\\
	&-\frac{V_0}{2N}\sum_{\bm{k}}\sum_{\bm{q}}v_{\bm{q}}(1-\frac{1}{2}\sin^2(\theta_{\bm{k}}))|\Lambda^+(\bm{k}+\bm{q},\bm{k})||\Lambda^{-}(\bm{k},\bm{k}+\bm{q})|.
\end{align}
Therefore, the smallest energy difference between the SVP state and IVC state is 
\begin{equation}
	E^{(0)}_{IVC}-E_{SVP}\approx\frac{V_0}{4N} \sum_{\bm{k},\bm{q}}(1-\frac{1}{2}\sin^2\theta_{\bm{k}})(|\Lambda^+(\bm{k}+\bm{q},\bm{k})|-|\Lambda^{-}(\bm{k},\bm{k}+\bm{q})|)^2>0,
\end{equation}
where $\sum_{\bm{k}}\sum_{\bm{q}}|\Lambda^+(\bm{k}+\bm{q},\bm{k})|^2=\sum_{\bm{k}}\sum_{\bm{q}}|\Lambda^-(\bm{k}+\bm{q},\bm{k})|^2$ is used. It is found that the SVP states are favorable in this case, which is compatible with the result given in \cite{Yahui2019,Lee2019}. Actually, we will present a more general formalism with finite $\bm{G}$  later and we found the SVP state is still more stable than the IVC state. 



\section{Details for numerical calculations}
\subsection{ Diagonalization of the moir\'e Hamiltonian}
The moir\'e Hamiltonian $\mathcal{H}_{\tau}(\bm{r})=-\frac{(\bm{\hat{p}}+\tau e\bm{A})^2}{2m^*}+V(\bm{r})$ is diagonalized with the plane wave bases $\{\ket{\bm{k}+\bm{G}}\}$, where $\bm{k}$ is defined within the  moir\'e Brillouin zone,  $\bm{G}=m\bm{G}_2+n\bm{G}_3$ are the reciprocal lattice vectors for the moir\'e pattern, $m,n$ are integer numbers. In this basis, the representation of the Hamiltonian is
\begin{equation}
	\hat{H}_{\bm{k}}(\bm{G'},\bm{G})=\braket{\bm{k}+\bm{G'}|\mathcal{H}_{\tau}(\bm{r})|\bm{k}+\bm{G}}.
\end{equation}
There are four different terms in the moir\'e Hamiltonian, i.e., $\mathcal{H}_{\tau}(\bm{r})=-\hat{\bm{p}}^2/2m^*-\frac{\tau e}{m^*}\bm{A}\cdot\bm{\hat{p}}-e^2\bm{A}^2/2m^*+V(\bm{r})\equiv\mathcal{H}_{\bm{\hat{p}}^2}+\mathcal{H}_{\bm{A}\cdot\bm{\hat{p}}}+\mathcal{H}_{\bm{A}^2}+\mathcal{H}_{V}$. It is straightforward to obtain
\begin{eqnarray}
	\braket{\bm{k}+\bm{G'}|\mathcal{H}_{\hat{\bm{p}}^2}(\bm{r})|\bm{k}+\bm{G}}&&=\frac{-(\bm{k}+\bm{G})^2}{2m^*}\delta_{\bm{G},\bm{G'}},\\
	\braket{\bm{k}+\bm{G'}|\mathcal{H}_V|\bm{k}+\bm{G}}&&=V(\bm{G'}-\bm{G}),\\
	\braket{\bm{k}+\bm{G'}|\mathcal{H}_{\bm{A}\cdot\hat{\bm{p}}}(\bm{r})|\bm{k}+\bm{G}}&&=\tau\gamma \tilde{\bm{A}}(\bm{G'}-\bm{G})\cdot(\bm{k}+\bm{G}),\\
	\braket{\bm{k}+\bm{G'}|\mathcal{H}_{\bm{A}^2}|\bm{k}+\bm{G}}&&=-\frac{\hbar^2}{2m^*L_M^2}\frac{\Phi^2}{\Phi_0^2}\tilde{\bm{A}}^2(\bm{G'}-\bm{G}).
\end{eqnarray}
where  the Fourier components $V(\bm{G'}-\bm{G})$ and  $\tilde{\bm{A}}(\bm{G'}-\bm{G})$ are obtained straightforwardly according to Eq.~(\ref{moire}) and Eq.~(\ref{gauge}), while $\tilde{\bm{A}}^2(\bm{G'}-\bm{G})=\frac{1}{S}\int d\bm{r} [\tilde{A}_x^2(\bm{r})+\tilde{A}_y^2(\bm{r})]e^{-i\bm{G}\cdot\bm{r}}$. Specifically, we can obtain
\begin{align}
	&\braket{\bm{k}+\bm{G'}|\mathcal{H}_V|\bm{k}+\bm{G}}=V_0e^{i\phi}\sum_{j=1,3,5}\delta_{\bm{G'}-\bm{G},\bm{G_j}}+V_0e^{-i\phi}\sum_{j=2,4,6}\delta_{\bm{G'}-\bm{G},\bm{G_j}};\\
	&\braket{\bm{k}+\bm{G'}|\mathcal{H}_{\bm{A}\cdot\hat{\bm{p}}}(\bm{r})|\bm{k}+\bm{G}}=\frac{\sqrt{3}\tau\gamma}{4i}(-\delta_{\bm{G'}-\bm{G},\bm{G}_3}-\delta_{\bm{G'}-\bm{G},\bm{G}_2}+\delta_{\bm{G'}-\bm{G},\bm{G}_5}+\delta_{\bm{G'}-\bm{G},\bm{G}_6})(\bm{k}+\bm{G})_x\nonumber\\
	&+\frac{\tau\gamma}{4i}(2\delta_{\bm{G'}-\bm{G},\bm{G}_1}-2\delta_{\bm{G'}-\bm{G},\bm{G}_4}-\delta_{\bm{G'}-\bm{G},\bm{G}_3}+\delta_{\bm{G'}-\bm{G},\bm{G}_6}-\delta_{\bm{G'}-\bm{G},\bm{G}_5}+\delta_{\bm{G'}-\bm{G},\bm{G}_2})(\bm{k}+\bm{G})_y;\\
	&\braket{\bm{k}+\bm{G'}|\mathcal{H}_{\bm{A}^2}|\bm{k}+\bm{G}}=-\frac{\hbar^2}{2m^*L_M^2}\frac{\Phi^2}{\Phi_0^2}\{\frac{3}{2}\delta_{\bm{G},\bm{G'}}-\frac{1}{4}\sum_{\pm}(\delta_{\bm{G}-\bm{G'},\pm 2\bm{G}_1}+\delta_{\bm{G}-\bm{G'},\pm 2\bm{G}_2}+\delta_{\bm{G}-\bm{G'},\pm 2\bm{G}_3})\\
	&-\frac{1}{4}\sum_{\pm}(\delta_{\bm{G}-\bm{G'},\pm(\bm{G}_1+\bm{G}_2)}-\delta_{\bm{G}-\bm{G'},\pm(\bm{G}_1-\bm{G}_2)}+\delta_{\bm{G}-\bm{G'},\pm(\bm{G}_2+\bm{G}_3)}-\delta_{\bm{G}-\bm{G'},\pm(\bm{G}_2-\bm{G}_3)}-\delta_{\bm{G}-\bm{G'},\pm(\bm{G}_1+\bm{G}_3)}+\delta_{\bm{G}-\bm{G'},\pm(\bm{G}_1-\bm{G}_3)})\}.
\end{align}

Using above relations,  we can obtain the matrix representation of $\hat{H}_{\bm{k}}(\bm{G'},\bm{G})$ with $\bm{G}=m\bm{G}_2+n\bm{G}_3$.	The moir\'e bands are calculated by diagonalizing $\hat{H}_{\bm{k}}(\bm{G'},\bm{G})$ numerically with a momentum cut-off of  $-N\le m,n\le N$.

\subsection{ The Chern number of moir\'e bands}
\begin{figure}
	\centering
	\includegraphics[width=0.5\linewidth]{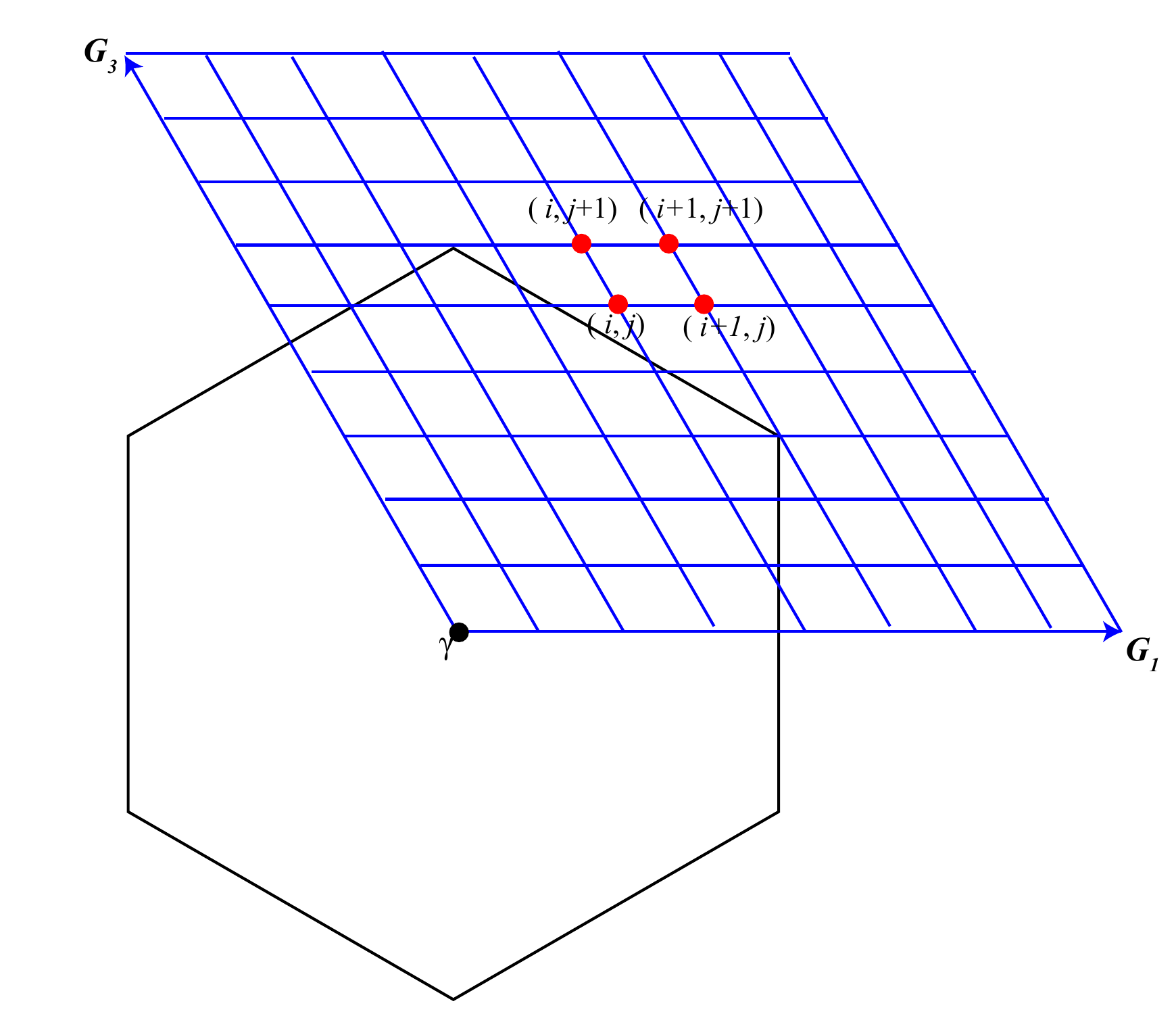}
	\caption{A schematic plot of the  Brillouin Zone formed with $\bm{G}_1$ and $\bm{G}_3$. We discretized this Brillouin Zone  to calculate the Chern number of the moir\'e bands. }
	\label{fig:figs2}
\end{figure}

In the Fig.2 of  main text, the Chern numbers of moir\'e bands were evaluated with various of phase $\phi$  and pseudo-magnetic field strength. To make the calculation more efficient,   we used the method proposed in Ref.~\cite{Fukui2005} to evaluate the Chern number, where the Brillouin zone  is discretized.  To be convenient, as shown in  Fig.~\ref{fig:figs2}, we discretized the Brillouin zone formed with reciprocal lattice vector $\bm{G}_1$ and $\bm{G}_3$, where the $\bm{k}$ is spanned as discretized $\bm{k}$-points $\bm{k}_{i,j}=\frac{i}{N_t}\bm{G}_1+ \frac{j}{N_t}\bm{G}_3$. The Chern number of $n-$th band is given by
\begin{equation}
	C=\sum_{i=1}^{N_t-1}\sum_{j=1}^{N_t-1}\ln[U_n(\bm{k}_{i,j},\bm{k}_{i+1,j})U_n(\bm{k}_{i+1,j},\bm{k}_{i+1,j+1})U_n(\bm{k}_{i+1,j+1},\bm{k}_{i,j+1})U_n(\bm{k}_{i,j+1},\bm{k}_{i,j})],
\end{equation}
where $U_n(\bm{k}_{i,j},\bm{k}_{i',j'})=\braket{\psi_n(\bm{k}_{i,j})|\psi_n(\bm{k}_{i',j'})}/|\braket{\psi_n(\bm{k}_{i,j})|\psi_n(\bm{k}_{i',j'})}|$ and  $\psi_n(\bm{k}_{i,j})$ denote the eigen wavefunction obtained by diagonalizing the moir\'e Hamiltonian at momentum $\bm{k}=\bm{k}_{i,j}$. In the calculation, we took $N_t=21$.

\subsection{ Hartree-Fock  mean-field calculations}

In the main text, we have presented  the numerical results of Hartree-Fock  mean-field calculations. In this section, we sketch the essential formalisms and processes to evaluate the energies of the SVP state and IVC states numerically. Here we consider a more general case, where the full mean-field Hamiltonian is given in Eq.~(\ref{Full_mean}). As discussed, the order parameter is $\Delta(\bm{k})=1/2(1+\tau_z)$ for the SVP state. The energy for the SVP state can thus be straightforwardly obtained as
\begin{equation}
	E_{SVP}=\sum_{\bm{k}}\xi_{+}(\bm{k})+\frac{V_0}{2N}\sum_{\bm{G}}v_{\bm{G}}|\sum_{\bm{k'}}\Lambda^+(\bm{k'}+\bm{G},\bm{k'})|^2-\frac{V_0}{2N}\sum_{\bm{k},\bm{k'},\bm{G}}v_{\bm{k'}-\bm{k}+\bm{G}}|\Lambda^+(\bm{k'}+\bm{G},\bm{k})|^2.
\end{equation}
For the IVC state, by using the IVC order parameter given in Eq.~(\ref{IVCorder}), we can obtain a similar mean-field Hamiltonian $H(\bm{k})$ as Eq.~(\ref{mean_form}) with
\begin{align}
	h_0(\bm{k})&=\frac{1}{2}(\xi_+(\bm{k})+\xi_{-}(\bm{k}))+\frac{V_0}{2}\sum_{\bm{G}}\text{tr}[\Delta^{H}(\bm{G})](\Lambda^{+}(\bm{k}-\bm{G},\bm{k})+\Lambda^{-}(\bm{k}-\bm{G},\bm{k}))-\frac{V_0}{2}\sum_{\bm{G}}(\Delta^{F}_{++}(\bm{k},\bm{G})+\Delta^{F}_{--}(\bm{k},\bm{G}))\\
	h_1(\bm{k})&=\frac{1}{2}(\xi_+(\bm{k})-\xi_{-}(\bm{k}))+\frac{V_0}{2}\sum_{\bm{G}}\text{tr}[\Delta^{H}(\bm{G})](\Lambda^{+}(\bm{k}-\bm{G},\bm{k})-\Lambda^{-}(\bm{k}-\bm{G},\bm{k}))-\frac{V_0}{2}\sum_{\bm{G}}(\Delta^{F}_{++}(\bm{k},\bm{G})-\Delta^{F}_{--}(\bm{k},\bm{G}))\\
	h_2(\bm{k})&=-V_0\sum_{\bm{G}}\Delta^F_{+-}(\bm{k},\bm{G}).
\end{align} 
Here, the Hartree order parameter is given by
\begin{equation}
	\Delta^{H}(\bm{G})=\frac{v_{\bm{G}}}{N}\sum_{\bm{k'}}\begin{pmatrix}
		\Lambda^+(\bm{k'}+\bm{G},\bm{k'})\sin^2\frac{\theta_{\bm{k'}}}{2}&0\\
		0&\Lambda^-(\bm{k'}+\bm{G},\bm{k'})\cos^2\frac{\theta_{\bm{k'}}}{2}
	\end{pmatrix}
\end{equation}
and Fock order parameter is given by
\begin{equation}
	\Delta^{F}(\bm{k},\bm{G})=\frac{1}{N}\sum_{\bm{k'}}v_{\bm{k'}-\bm{k}+\bm{G}}\begin{pmatrix}
		|\Lambda^+(\bm{k'}+\bm{G},\bm{k})|^2\sin^2\frac{\theta_{\bm{k'}}}{2}&\frac{1}{2}\Lambda^+(\bm{k'}+\bm{G},\bm{k})\Lambda^-(\bm{k}-\bm{G},\bm{k'})\sin\theta_{\bm{k'}}e^{i\varphi_{\bm{k'}}}\\
		\frac{1}{2}\Lambda^-(\bm{k'}+\bm{G},\bm{k})\Lambda^+(\bm{k}-\bm{G},\bm{k'})\sin\theta_{\bm{k'}}e^{-i\varphi_{\bm{k'}}}&|\Lambda^-(\bm{k'}+\bm{G},\bm{k})|^2\cos^2\frac{\theta_{\bm{k'}}}{2}
	\end{pmatrix}.
\end{equation}

The self-consistent equation reads
\begin{equation}
	\cos\theta_{\bm{k}}=\frac{h_1(\bm{k})}{\sqrt{h^2_1(\bm{k})+|h_2(\bm{k})|^2}}, \tan \varphi_{\bm{k}}=\frac{\text{Im}(h_2(\bm{k}))}{\text{Re}(h_2(\bm{k}))}\label{self_consistent}
\end{equation}
This self-consistent equation is solved iteratively. Specifically, we chose discrete $\bm{k}$ points with $N\times N$ grid in the Brillouin Zone and set  some initial values for $\theta_{\bm{k}}$ and $\varphi_{\bm{k}}$ with $\theta_{\bm{k}}=\pi-\theta_{-\bm{k}} $ and $\varphi_{\bm{k}}=\varphi_{-\bm{k}}$. In the Fig.~3 of  main text, we set $N=11$. Then we can evaluate  $\theta_{\bm{k}}$ and  $\varphi_{\bm{k}}$ with Eq.~(\ref{self_consistent}). This can be done iteratively until the difference $\sum_{\bm{k}}|\theta^{m+1}_{\bm{k}}-\theta^{m}_{\bm{k}}|$ is smaller than a critical value such as $10^{-3}$, where $m$ labels the $m-$th iterative step.

After adding the potential term back, the total energy for the IVC state is obtained as
\begin{equation}
	E_{IVC}=\sum_{\bm{k}}E_{-}(\bm{k})-\frac{NV_0}{2}\sum_{\bm{G}}\frac{\text{tr}[\Delta^{H}(\bm{G})]\text{tr}[\Delta^{H}(-\bm{G})]}{v_{\bm{G}}}+\frac{V_0}{2N}\sum_{\bm{k},\bm{k'},\bm{G}}v_{\bm{k'}-\bm{k}+\bm{G}}\text{tr}[\Lambda(\bm{k'}+\bm{G},\bm{k})\Delta(\bm{k'})\Lambda(\bm{k}-\bm{G},\bm{k'})\Delta(\bm{k})].
\end{equation}

Note we need to further subtract a   large density-like Hartree term from $\Delta^{H}(\bm{G}=0)$, i.e., $E^{H}(\bm{G}=0)=\frac{NV_0}{2}v_{\bm{G}=0}$, which does not affect the order but  gives a large charge background and should be canceled with positive ion background.  In other words, the total energy for the SVP state and IVC state are $\tilde{E}_{SVP}=E_{SVP}-E^{H}(\bm{G}=0)$ and  $\tilde{E}_{IVC}=E_{IVC}-E^{H}(\bm{G}=0)$. The gap of the SVP states are defined as $\Delta_{SVP}=\text{min}[\tilde{E}_{SVP,-}(\bm{k})]-\text{max}[\tilde{E}_{SVP,+}(\bm{k})]$ and  $\tilde{E}_{SVP,\tau}(\bm{k})$ represents the mean-field energy dispersion of the SVP state at momentum $\bm{k}$.

\section{Effective tight-binding model calculation with lattice relaxation for moir\'e heterobilayer TMDs}

\subsection{Lattice relaxation in moir\'e TMDs}

\begin{figure}
	\centering
	\includegraphics[width=7in]{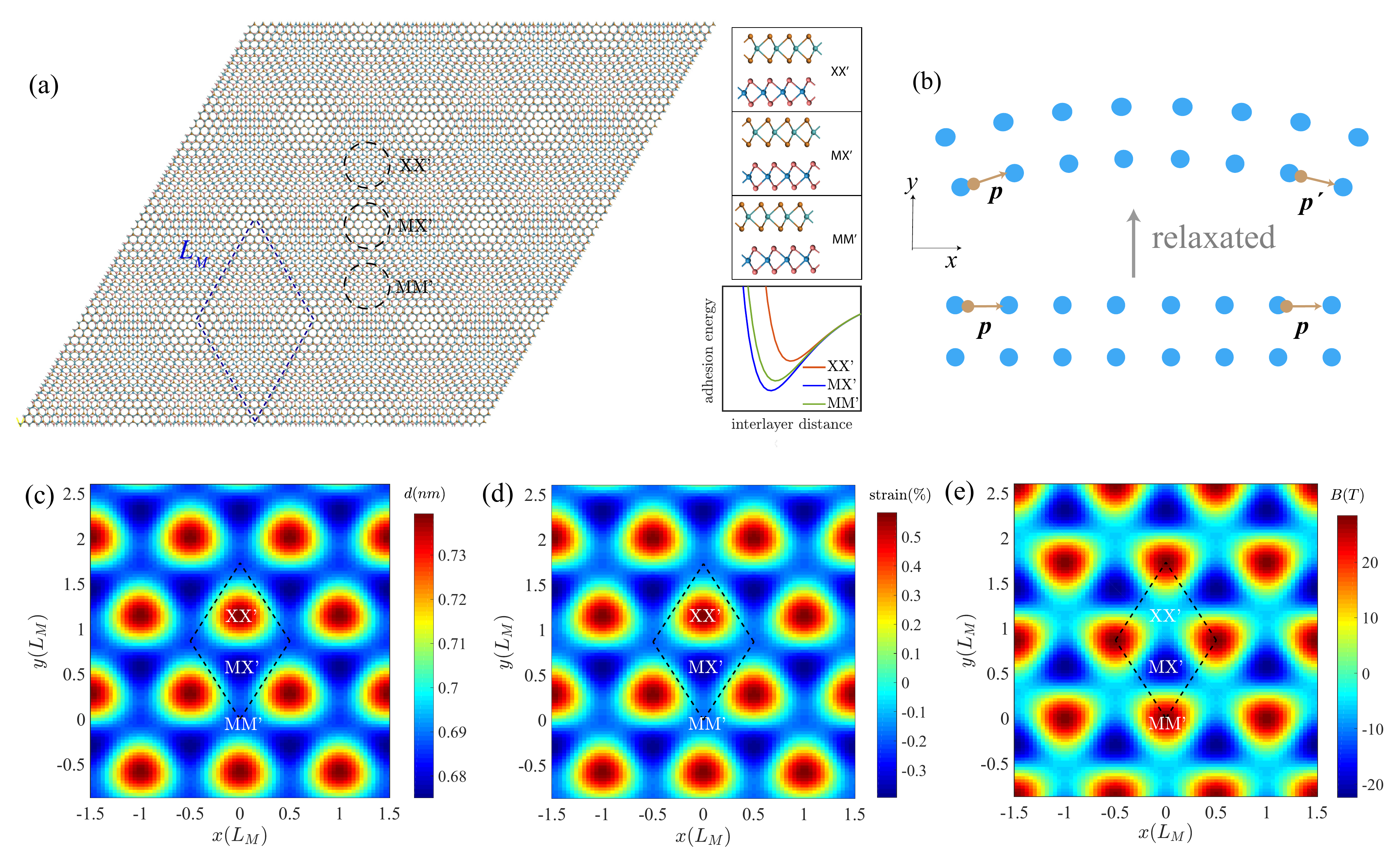}
	\caption{(a) Moir\'e supercell marked with blue dashed line, and the local MM$'$, MX$'$, XX$'$ stacking configurations highlighted with black dashed circles of heterobilayer MoTe$_2$/WSe$_2$ in the zero twist angle limit. The side views of different stacking areas are magnified in the top right panel. The adhesion energy with respect to interlayer distance for the three stacking configurations is schematically shown in the bottom right panel \cite{Falko2020}. (b) schematically shows a uniform lattice and a lattice with some in-plane distortions. The $\bm{p}$ represents  the momentum with a direction along the the arrow. (c) Inter-layer distance which shows strong out-of-plane corrugation. (d) In-plane strain field (trace of the strain tensor) in the MoTe$_2$ layer. (e) Pseudo-magnetic field in the MoTe$_2$ layer. We have adopted the following parameters in accordance with the DFT results~\cite{Fu_zhang2021,Li_Crommie2021}: $\bm{u}^{(2)}_{1,0}=(0,-0.6 + 0.2i)\times 10^{-3} \; \rm{nm}$, $\Delta h_{1,0}=-(2.4 + 6.2 i) \times 10^{-3} \; \rm{nm}$, $d_{0}=0.7 \; \rm{nm}$ and $\beta=2$.  Note that we set the $x$-component of $\bm{u}_{1,0}$ to be zero due to the mirror symmetry $M_x$ in the untwisted heterobilayer TMDs.}
	\label{fig:FIGS4}
\end{figure}

Let us first illustrate the underlying mechanism of  the lattice relaxation in moir\'e bilayer transition metal dichalcogenides(TMDs). As shown in Fig.~\ref{fig:FIGS4}(a), the  lattice structure of a moir\'e bilayer TMD  locally resembles the regular stacking,  such as MX', MM' and XX' regions. Here, the M, X respectively denotes the transition metal atom and chalcogenide atom. To simplify the discussion,  we can focus on the MX' and XX' stacking which change the adhesion energy most prominently.  Specifically, at the MX' stacking region, the M and X' atoms are aligned with each other. In this case,  the attraction between the transition metal atoms and  chalcogenide atoms will locally lower the adhesion energy between the two layers. The MX' stacking configuration thus has the lowest energy as shown in the bottom right panel of Fig.~\ref{fig:FIGS4}(a), and actually corresponds to  the stacking configuration of bulk 2H-structure TMDs. On the contrary, at the XX' stacking regions, the X and X' atoms are aligned with each other. Due to the repulsion between the chalcogenide atoms, the adhesion energy  would be locally increased. Consequently, to save the adhesion energy~\cite{Falko2020},  the heterobilayer TMDs lattice will relax itself to enlarge the MX'-stacking area and reduce the XX'-stacking area. If the intra-layer elastic energy caused by lattice relaxation can be compensated by lowering adhesion energy, the lattice relaxation will be energy favored. Indeed, lattice relaxation and even lattice reconstructions are commonly observed in various moir\'e materials \cite{Weston2020, Linnian2021,Li_Crommie2021}.

%



The lattice relaxation can be characterized by the displacement vectors of atoms positions between the relaxed and unrelaxed lattice. Let us first look at the   in-plane displacement vectors, which we denote as $\bm{u}^{(l)}(\bm{r})$. The layer index $l=1$ for the WSe$_2$ layer, and $l=2$ for the MoTe$_2$ layer. The lattice deformation is expected to  keep the original superlattice period and respects the $C_3$ symmetry. As a result, we can express the in-plane displacement $\bm{u}^{(l)}(\bm{r})$  with the Fourier components $\bm{u}^{(l)}_{\bm{q}}$ as
\begin{equation}
	\bm{u}^{(l)}({\bm{r}})=\sum_{\bm{q}}\bm{u}^{(l)}_{\bm{q}}e^{i\bm{q}\cdot\bm{r}}, \label{inplane}
\end{equation}
with $\bm{q}=m\bm{G}_1+n\bm{G}_3$ with $\bm{G}_j=\frac{4\pi}{\sqrt{3}L_M}(\sin\frac{(j-1)\pi}{3},\cos\frac{(j-1)\pi}{3})$. Note that in this section, we rotate the coordinate convention by ninety degree  compared to the main text for the sake of convenience. For simplicity, we only keep the leading-order contributions from $\bm{u}^{(l)}_{\bm{q}}$ with $|\bm{q}|=G$, where $G=|\bm{G}_{j}|=\sqrt{3}\kappa$. To be more specific, we will also use $\bm{u}^{(l)}_{m,n}$ to label $\bm{u}^{(l)}_{\bm{q}}$ sometimes.  Note that $\bm{u}^{(l)}_{0,0}$ is forced to vanish by the $C_3$ symmetry. Thus, we consider the Fourier components from six $\bm{q}$ points in total. The point group symmetry requires $\bm{u}^{(l)}_{\hat{g}{\bm{q}}}=\hat{g}\bm{u}^{(l)}_{\bm{q}}$. Moreover, the displacement $\bm{u}^{(l)}({\bm{r}})$ should be real, which requires $\bm{u}^{(l)}_{\bm{-q}}=\bm{u}^{(l)*}_{\bm{q}}$. Therefore, there is only one independent $\bm{u}^{(l)}_{\bm{q}}$ for each layer, and we will use the $\bm{u}^{(l)}_{1,0}$ component. As we will discuss  later, the in-plane displacement can be chosen according to the strength of build-in strain found by the DFT results \cite{Fu_zhang2021,Li_Crommie2021}.


Apart from in-plane relaxation, the heterobilayers will also exhibit out-of-plane corrugation. The interlayer distance which minimizes the adhesion energy can be inferred from the bottom right panel of Fig.~\ref{fig:FIGS4}(a). The lattice tends to exhibit a minimum inter-layer spacing at the MX'-stacking regions due to the attraction between the transition metal atoms and chalcogenide atoms, while has a maximum inter-layer distance at the XX'-stacking region, which gives rise to the corrugation effect.

To capture this corrugation effect, we define the out-of-plane displacement of the two layers as $h^{(l)}(\bm{r})$.  The inter-layer spacing can thus be written as $d(\bm{r})=d_{0}+\Delta h(\bm{r})$ with $\Delta h(\bm{r})=h^{(2)}(\bm{r})-h^{(1)}(\bm{r})$, where $d_{0}$ is the average spacing between the two layers. The Fourier components $h^{(l)}_{\bm{q}}$ for the out-of-plane corrugation is written as
\begin{equation}
	h^{(l)}({\bm{r}})=\sum_{\bm{q}}h^{(l)}_{\bm{q}}e^{i\bm{q}\cdot\bm{r}}.\label{outplane}
\end{equation}
Still, only  the leading-order contributions from $h^{(l)}_{\bm{q}}$ with $|\bm{q}|=G$ is kept. The point group symmetry gives $h^{(l)}_{\hat{g}\bm{q}}=h^{(l)}_{\bm{q}}$, and a real displacement $h^{(l)}({\bm{r}})$ requires $h^{(l)}_{\bm{-q}}=h^{(l)*}_{\bm{q}}$.   To show the interlayer distance given by Eq.~\ref{outplane} in moir\'e heterobilayer MoTe$_2$/WSe$_2$, we plotted  the inter-layer distance as shown in Fig.~\ref{fig:FIGS4}(c), where we set $\Delta h_{1,0}=-(2.4 + 6.2 i) \times 10^{-3} \; \rm{nm}$, $d_{0}=0.7 \; \rm{nm}$ that fit the DFT results \cite{Fu_zhang2021,Li_Crommie2021}. As expected, the interlayer spacing is maximum at XX' stacking region  and minimum at the MX'-stacking region.

\subsection{ Pseudomagnetic fields emerged from the atomic lattice relaxation in moir\'e TMDs}

After discussing the lattice relaxation in moir\'e TMDs, we next introduce how the pseudomagnetic fields can be generated by such periodic lattice displacement. To give an intuitive physical picture why the  pseudomagnetic fields can arise from lattice relaxation in general, we depicted Fig.~\ref{fig:FIGS4} which shows  a uniform lattice  and  a distorted lattice after the lattice relaxation. Phenomenologically, the momentum of an electron moves along the lattice is preserved in the uniform case, while  the momentum of an electron moves along a distorted lattice has to be gradually changed (see Fig.~\ref{fig:FIGS4}(b)). The scenario in the distorted lattice mimics an electron circulating an out-of-plane magnetic fields by the Lorentz force classically. From this simple picture, it is understandable that the lattice relaxation could give rise to the effects of pseudo-magnetic fields.

It was known that the pseudo-magnetic fields from the lattice relaxation can be directly obtained from 
the strain tensor, which is defined as the spatial gradients of the displacement field \cite{Guinea2010,Guinea_review,Guinea2014}
\begin{equation}
	u^{(l)}_{ij}=\frac{1}{2}(\partial_i u^{(l)}_j + \partial_j u^{(l)}_i + \partial_i h^{(l)}  \partial_j h^{(l)}).
\end{equation}
The in-plane displacement $u_{i}^{l}$ and out-of-plan displacement $h^{l}$ are defined in Eq.~(\ref{inplane}) and Eq.~(\ref{outplane}), respectively. The last term represents the in-plane strain induced by the out-of-plane corrugation, which we  find to be a second-order effect and will be neglected  in the following discussion for simplicity. Then, the strain tensor can  be expressed  with the Fourier components as
\begin{equation}
	u^{(l)}_{ij}({\bm{r}})=\frac{i}{2}\sum_{\bm{q}}(q_i u^{(l)}_{\bm{q}, j}+ q_j u^{(l)}_{\bm{q}, i}) e^{i\bm{q}\cdot\bm{r}}.\label{strain_tensor}
\end{equation}

To be specific, we consider an in-plane displacement field with Fourier component $\bm{u}^{(2)}_{1,0}=(0,-0.6 + 0.2i)\times 10^{-3} \; \rm{nm}$, and 
the calculated in-plane strain field (trace of the strain tensor) of the MoTe$_2$ layer using Eq.~(\ref{strain_tensor})  is plotted in Fig.~\ref{fig:FIGS4}(d). Note that the  Fourier component of in-plane displacement is set to give rise to roughly $0.5\%$ built-in strain strength revealed by previews DFT calculations for moir\'e TMDs~\cite{Fu_zhang2021,Li_Crommie2021}. As shown in Fig.~\ref{fig:FIGS4}(d),   the strain at the MX'-stacking region is compressive, while the strain at XX'-stacking region is tensile. As the MoTe$_2$ layer possesses larger lattice constant than WSe$_2$ layer, such relaxation could effectively increase the MX'-stacking region and reduce the XX'-stacking region to save energy.

As we presented in the main text, the strain tensor will induce an effective gauge field $\bm{A}$.  We thus can map out pseudomagnetic fields from lattice relaxation as $\bm{B}=\nabla \times \bm{A}$. Specifically, the  effective gauge field from the strain is given by 
\begin{eqnarray}
	\bm{A}^{(l)} = \frac{\sqrt{3}\beta\hbar}{2ea}(u_{xx}^{(l)}-u_{yy}^{(l)}, -2u_{xy}^{(l)})= \frac{\sqrt{3}\beta\Phi_{0}}{4 \pi a} i \sum_{\bm{q}} (q_{x}u^{(l)}_{\bm{q},x}-q_{y}u^{(l)}_{\bm{q},y}, \; -q_{x}u^{(l)}_{\bm{q},y}-q_{y}u^{(l)}_{\bm{q},x}) e^{i\bm{q}\cdot\bm{r}}, \label{guage}
\end{eqnarray}
where the strain tensor form from Eq.~(\ref{strain_tensor}) is  inserted,  the value of $\beta$ depends on the specific material or model, and $\Phi_{0}=\frac{h}{e}$ is the flux quantum. This results in an out-of-plane pseudomagnetic field~\cite{Guinea2010,Guinea_review}
\begin{eqnarray}
	B^{(l)}_{z} =\partial_x A^{(l)}_y - \partial_y A^{(l)}_x = \frac{\sqrt{3}\beta\Phi_{0}}{4 \pi a} \sum_{\bm{q}} [(q_{x}^{2}-q_{y}^{2})u^{(l)}_{\bm{q},y}+2q_{x}q_{y}u^{(l)}_{\bm{q},x}]e^{i\bm{q}\cdot\bm{r}}.\label{B_field}
\end{eqnarray}

We plot the distribution of the pseudomagnetic field in the MoTe$_2$ layer in Fig.~\ref{fig:FIGS4}(e) with $\beta=2$. A periodic pseudomagnetic field  can be achieved in the presence of the lattice relaxation, which is expected for moir\'e TMD materials \cite{Falko2020}. Note that it is mainly the spatial gradient of strain fields rather than the strength of strain only determines the magnitude of  pseudomagnetic fields. For example, the XX' stacking region in Fig.~\ref{fig:FIGS4}(d) displays a sizable tensile strain ($\sim$ 0.5\%), while the corresponding pseudomagnetic field is small (see Fig.~\ref{fig:FIGS4}(e)). In contrast, the strain configuration given in Fig.~\ref{fig:FIGS4}(d) exhibits  largest spatial gradients near  MM' regions, which results in largest pseudomagnetic fields (see Fig.~\ref{fig:FIGS4}(e)).

\subsection{Effective strained tight-binding model for MoTe$_2$/WSe$_2$ heterobilayers}

\begin{figure}[h]
	\centering
	\includegraphics[width=1\linewidth]{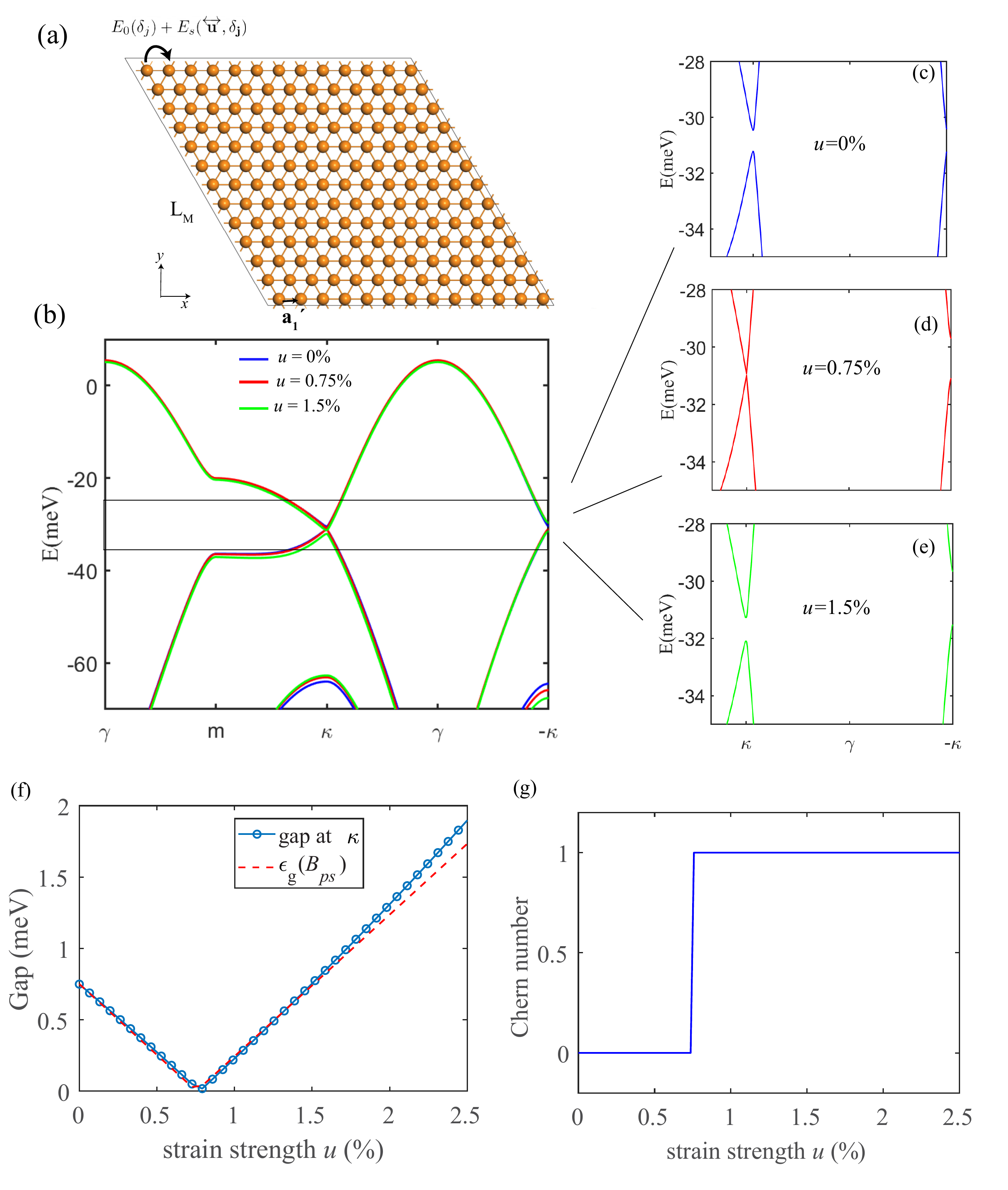}
	\caption{(a) A schematic plot of  $13\times 13$ Mo atoms in each moir\'e unit cell, where the modified nearest-neighbor hopping under lattice relaxation, i.e., Eq.~(\ref{hop_change}), is highlighted. (b)shows the moir\'e band  structures without strain (blue) and with strain strength $u=0.75\%$ (red) and $u=1.5\%$ (green). The strain strength $u$ is defined as the maximal value of $u_{xy}(\bm{r})$ within the moir\'e unit cell.  (c), (d) ,(e) are enlarged from (b) near [-28,35] meV, which clearly indicates a band inversion at $\kappa$ valley   when the strain strength exceeds $0.75\%$. (f) shows the separation between the first and the second moir\'e bands at $\kappa$ valley,  as a function of strain strength $u$. The red dashed line is plotted with $\epsilon_{g}(B_{ps})\approx |\epsilon_g(0)-0.05B_{ps}|$ (see Eq.~\ref{Eq_gap}), where $B_{ps}$ denotes the magnitude of pseudomagnetic fields induced by the lattice relaxation. (g) shows the Chern number of the top moir\'e band  as a function of strain strength $u$.  }
	\label{fig:FIGS5}
\end{figure}

To calculate the band structure and study its topology in the presence of lattice relaxation, we further write a tight-binding model for the MoTe$_2$ monolayer, and treat the effect of the WSe$_2$ layer as a moir\'e potential. The moir\'e supercell of MoTe$_2$/WSe$_2$ heterobilayers in zero twist angle limit contains $13 \times 13$ MoTe$_2$ unit cells and $14 \times 14$ WSe$_2$ unit cells. We will construct the tight-binding Hamiltonian using the six d-orbitals $\{\ket{d_{z^2, s}},\ket{d_{xy, s}},\ket{d_{x^2-y^2, s}}\}$ with $s=\uparrow/\downarrow$ from the transition metal atoms \cite{XiaoDi2013}. As there are $13 \times 13$ Mo atoms in each moir\'e unit cell (see Fig.~\ref{fig:FIGS5}(a)), the dimension of our tight-binding Hamiltonian is $13 \times 13 \times 6 = 1014$.

%
%

First of all, the unstrained tight-binding Hamiltonian up to nearest-neighbor hopping is written as \cite{XiaoDi2013}
\begin{eqnarray}
	H_{0}=\sum_{\bm{k}, \bm{R}_i}\sum_{\bm{\delta}_{j}} c^{\dagger}_{\bm{k},\bm{R}_{i}}E_{0}(\bm{\delta}_{j})c_{\bm{k},\bm{R}_{i}+\bm{\delta}_{j}},
\end{eqnarray}
where $\bm{R_i}$ labels the positions of the $13\times 13$ Mo atoms, $c^{\dagger}_{\bm{k},\bm{R}_{i}}$ denotes the creation operator of  the Bloch states with momentum $\bm{k}$ (defined in moir\'e Brillouin zone), $\bm{\delta}_{j}$ are the lattice vectors connecting the sites, and $E^{\alpha\alpha'}_{0}(\bm{\delta}_{j})=\bra{\phi^{\alpha}(\bm{r})}\hat{H}\ket{\phi^{\alpha'}(\bm{r}-\bm{\delta}_{j})}$ is the hopping integral. The on-stie energy and the nearest-neighbor hopping term take the form:
\begin{eqnarray}
	E_{0}(\bm{\delta}=0) &=& \sigma_{0} \otimes
	\begin{pmatrix}
		\epsilon_{1} & 0 & 0 \\ 
		0 & \epsilon_{2} & 0 \\ 
		0 & 0 & \epsilon_{2}
	\end{pmatrix} - \mu I_{6\times6}, \\
	E_{0}(\bm{\delta}=\bm{a}_{1}') &=& \sigma_{0} \otimes
	\begin{pmatrix}
		t_{0} & t_{1} & t_{2} \\ 
		- t_{1} & t_{11} & t_{12} \\ 
		t_{2} & - t_{12} & t_{22}
	\end{pmatrix},
\end{eqnarray}
and the hopping terms along other directions can be determined by symmetry as:
\begin{eqnarray}
	E_{0}(\hat{g}\bm{\delta}_{j})=D(\hat{g})E_{0}(\bm{\delta}_{j})D^{\dagger}(\hat{g}),
\end{eqnarray}
where $D(\hat{g})$ is representation of the symmetry operation $\hat{g}$ under the six-orbital basis.

The SOC term takes the form of
\begin{eqnarray}
	H_{SOC}=\sum_{\bm{k},\bm{R}_i} c^{\dagger}_{\bm{k},\bm{R}_{i}}(\frac{\lambda}{2}\sigma_z \otimes L_z)c_{\bm{k},\bm{R}_{i}},
\end{eqnarray}
where $L_z =
\begin{pmatrix}
	0 & 0 & 0 \\ 
	0 & 0 & -2i \\ 
	0 & 2i & 0
\end{pmatrix}$
is the $z$-component of the orbital angular momentum. It should be noted the coordinated convention here is rotated by $90^\circ$ compared to the main text. In the following presentation, we fixed  the coordinated convention to be the same as this tight-binding model.

\begin{center}
	\begin{table}[ht]
		\caption{Parameters for unstrained tight-binding Hamiltonian of monolayer MoTe$_2$ adapted from Refs.\cite{XiaoDi2013}. All parameters set in units of eV.} 
		\centering 
		\begin{ruledtabular}
			\begin{tabular}{c c c c c c c c c} 
				$\epsilon_1$ & $\epsilon_2$ & $t_0$ & $t_1$ & $t_2$ & $t_{11}$ & $t_{12}$ & $t_{22}$ & $\lambda$\\ [0.5ex]\hline 
				0.605 & 1.972 & -0.169 & 0.228 & 0.390 & 0.207 & 0.239 & 0.252 & -0.107 \\ 
			\end{tabular}
		\end{ruledtabular}
		\label{table:01} 
	\end{table}
\end{center}

As we mentioned, due to the moir\'e potential, the unit cell is extended as the moir\'e unit cell which contains $13\times 13$ Mo atoms. The hopping terms are captured by the above tight-binding Hamiltonian. Next, let us add  the moir\'e potential introduced by the WSe$_2$ layer and the effects from the periodic strain fields $\bm{u}(\bm{r})$ that contains the pseudomagnetic fields.  To be consistent with the main text and reduce the number of parameters, we will focus on these effects on the valence band only, where the top moir\'e bands are originated from. In other words, the strain and moir\'e terms will be added in the $(\ket{d_{xy,s}},\ket{d_{x^2-y^2,s}})$ subspace. Note that such simplification neglect the physics from the inter-orbital  mixing from the conduction and valence band, which is not the focus of this work and we will leave this as a future study.

Let us discuss how the tight-binding Hamiltonian is changed in the presence of the strain fields first. Due to the displacement of atoms, 
when the strain $\bm{u}$  is present, the hopping terms between the atomic orbitals will be modified:


\begin{eqnarray}
	E_{0}(\bm{\delta}_{j}) \rightarrow E_{0}(\bm{\delta}_{j}) + E_{s}(\overleftrightarrow{\bm{u}},\bm{\delta}_{j}),\label{hop_change}
\end{eqnarray} 
which gives an extra contribution to the Hamiltonian
\begin{eqnarray}
	H_{s}=\sum_{\bm{k}, \bm{R}_i}\sum_{\bm{\delta}_{j}} c^{\dagger}_{\bm{k},\bm{R}_{i}}E_{s}[\overleftrightarrow{\bm{u}}(\bm{R}_{i}),\bm{\delta}_{j}] c_{\bm{k},\bm{R}_{i}+\bm{\delta}_{j}}.
\end{eqnarray}
Here,  the spatial dependence of $\bm{u}(\bm{R}_{i})$ in each moir\'e unit cell is obtained by submitting the positions of transition  metal atoms into Eq.~(\ref{strain_tensor}), and as we will show, the strain tensor  $\overleftrightarrow{\bm{u}}$ would be decomposed as $u_0=u_{xx}+u_{yy}$ and $(u_1,u_2)=(u_{xx}-u_{yy},-2u_{xy})$. The former transforms as a scalar, while the latter transforms as a vector, i.e., $E$-representation, under $C_3$ operation.

Based on the symmetry analysis\cite{Kaxiras2018, Benjamin2020}, to linear order of the strain tensor,  the contribution to the on-site energy is~

\begin{eqnarray}\label{eq:InvariantForm}\nonumber
	E_{s}(\overleftrightarrow{\bm{u}},\bm{\delta}_{j} = 0) &=& \sigma_{0} \otimes \bigg[ (u_{xx} + u_{yy}) 
	\begin{pmatrix}
		E_2^{S} & 0\\
		0 & E_2^{S}
	\end{pmatrix} + 2 u_{xy}
	\begin{pmatrix}
		0& h^{S}_2\\
		h^{S}_2 & 0
	\end{pmatrix}\\
	&+& (u_{xx} - u_{yy}) 
	\begin{pmatrix}
		h^{S}_2 & 0\\
		0 & -h^{S}_2
	\end{pmatrix}
	\bigg],
\end{eqnarray} 
and the modification for the nearest-neighbor hopping term along $\bm{a}_{1}'$ direction is
\begin{eqnarray}\label{eq:hdeltax}\nonumber
	E_{s}(\overleftrightarrow{\bm{u}},\bm{\delta}_{j} = \bm{a}_{1}') 
	& = & \sigma_{0} \otimes \bigg[ (u_{xx} + u_{yy})
	\begin{pmatrix}
		P^{(A_1)}_{11} & P^{(A_1)}_{12}\\ -P^{(A_1)}_{12} & P^{(A_1)}_{22}\\
	\end{pmatrix} + (u_{xx} - u_{yy})
	\begin{pmatrix}
		P^{(E)}_{11} & P^{(E)}_{12}\\
		-P^{(E)}_{12} & P^{(E)}_{22}\\
	\end{pmatrix}
	\\
	&+& 2u_{xy}
	\begin{pmatrix}
		0 & N^{(E)}_{12}\\
		N^{(E)}_{12} & 0\\
	\end{pmatrix}\bigg].
\end{eqnarray}
The hopping terms along other directions can be obtained by symmetry~\cite{Kaxiras2018, Benjamin2020}
\begin{eqnarray}\label{eq:strain_symmetry}
	E_{s}(\overleftrightarrow{\bm{u}},\hat{g}\bm{\delta}_{j}) = D(\hat{g}) E_{s}(\hat{g}^{-1}\overleftrightarrow{\bm{u}},\bm{\delta}_{j}) D^{\dagger}(\hat{g}).
\end{eqnarray}
Without loss of generality, we adopt the strained tight-binding parameters up to nearest-neighbor hopping terms for the monolayer TMD \cite{Kaxiras2018} estimated from first-principle calculations. Substituting the strain configuration given by Eq.~\ref{strain_tensor} into Eqs.~\ref{eq:InvariantForm}, \ref{eq:hdeltax} and \ref{eq:strain_symmetry}, we can obtain how the hopping terms are modified under strain. Notice that a significant difference from previous works \cite{Kaxiras2018, Benjamin2020} is that the strain tensor  from  Eq.~\ref{strain_tensor} are not uniform but  spatially dependent with moir\'e periodicity.

\begin{center}
	\begin{table}[ht]
		\caption{Strained tight-binding parameters up to nearest-neighbor hopping terms for monolayer TMD adapted from Refs.\cite{Kaxiras2018}.  All parameters are set in units of eV.} 
		\centering 
		\begin{ruledtabular}
			\begin{tabular}{c c c c c c c c c} 
				$E^{S}_{2}$ & $h^{S}_{2}$   & $P^{(A1)}_{11}$ & $P^{(A1)}_{12}$ & $P^{(A1)}_{22}$& $P^{(E)}_{11}$&$P^{(E)}_{12}$&$P_{22}^{(E)}$&$N_{12}^{(E)}$\\ [1ex]\hline 
				-1.012 &  -0.220  &-1.127& 0.325 & 1.617 &-0.966&-0.044&1.179&-0.776 \\ [0.5ex] 
			\end{tabular}
		\end{ruledtabular}
		\label{table:02} 
	\end{table}	
\end{center}

Finally, we take into account the moir\'e potential  introduced by the WSe$_2$ layer as $V(\bm{r})=\sum\limits_{\bm{q}}V_{\bm{q}}e^{i\bm{q}\cdot\bm{r}}$=2$V_0\sum\limits_{j=1,3,5}\cos(\bm{G}_j\cdot\bm{r}+\phi)$ with $\bm{G}_j=\frac{4\pi}{\sqrt{3}L_M}(\sin\frac{(j-1)\pi}{3},\cos\frac{(j-1)\pi}{3})$ which gives a spatial-dependent on-site energy 
$H_{V}=\sum\limits_{\bm{k}, \bm{R}_i} c^{\dagger}_{\bm{k},\bm{R}_{i}}V(\bm{R}_{i}) c_{\bm{k},\bm{R}_{i}}$. Note that as we mentioned, to be consistent with the main text, the moir\'e potential is only added on the valence bands. To be specific, we adopted $V_0=10$ meV, $\phi=0.3\pi$  in the following calculations.

Now the total moir\'e strained tight-binding  Hamiltonian reads
\begin{eqnarray}
	H_t = \sum_{\bm{k}, \bm{R}_i}\sum_{\bm{\delta_j}} c^{\dagger}_{\bm{k},\bm{R}_{i}}[ E_{0}(\bm{\delta}_{j}) + E_{s}(\overleftrightarrow{\bm{u}},\bm{\delta}_{j})]e^{i\bm{k}\cdot\bm{\delta_{j}}} c_{\bm{k},\bm{R}_{i}+\bm{\delta_j}} + H_{SOC}  + H_{V}.\label{Eq_tb}
\end{eqnarray}
By numerically diagonalizing $H_t$, we can investigate how the moir\'e bands are modified under the periodic strain fields $\bm{u}(\bm{r})$ induced by lattice relaxation.  The advantage of this tight-binding model description is that it directly maps out the effects of the atomic displacements induced by the lattice relaxation on the moir\'e bands, as we will see next.

\subsection{Moir\'e band structures and band topology from the strained tight-binding Hamiltonian}

In this section, we present the moir\'e band structures and a switch in band topology  for MoTe$_2$/WSe$_2$ heterobilayer layer under lattice relaxation using the effective tight-binding model Eq.~(\ref{Eq_tb}). As we discussed, the strain tensor can be decomposed as the scalar  part $u_0(\bm{r})=u_{xx}(\bm{r})+u_{yy}(\bm{r})$ and the vector part $(u_1(\bm{r}),u_2(\bm{r}))=(u_{xx}(\bm{r})-u_{yy}(\bm{r}),-2u_{xy}(\bm{r}))$.  As expected, we find the scalar part $u_0(\bm{r})$ is mainly to modify the onsite-energy and thus effectively shift the moir\'e potential, including the strength $V_0$ and the phase $\phi$, while the vector part $(u_1(\bm{r}),u_2(\bm{r}))$ is to generate the pseudomagentic fields that could drive a topological band inversion. As our main focus is the effects from strain $(u_1(\bm{r}),u_2(\bm{r}))$, we will fix the strength of the scalar part with the maximal $u_0$ within the moir\'e unit cell$\sim 0.5$\%, being comparable to the previous first principle calculation \cite{Fu_zhang2021,Li_Crommie2021}.

To demonstrate the lattice relaxation could switch the topology of top moir\'e bands, we  artificially tuned the strength of $u_{xx}(\bm{r})-u_{yy}(\bm{r})$ and $u_{xy}(\bm{r}))$  , i.e., $(u_1(\bm{r}),u_2(\bm{r}))\mapsto \lambda (u_1(\bm{r}),u_2(\bm{r}))$ and change $\lambda$ gradually. The results are summarized in Fig.~\ref{fig:FIGS5}.  As $u_{xx}(\bm{r})-u_{yy}(\bm{r})$ and $u_{xy}(\bm{r}))$ exhibit similar strain strength, without loss of generality, we will represent the strain strength with the maximal value of  $u_{xy}(\bm{r})$ in within the moir\'e unit cell, and  this value would be referred as the strain  strength $u$ in the following discussions. 

Fig.~\ref{fig:FIGS5}(b) shows the moir\'e band structures with the strain strength $u=0, 0.75\%, 1.5\%$ at $K$ valley, which is compatible with the band structures obtained from the continuum model of main text.  Here, we can distinguish the bands from the two different valleys ($\pm K$) as they exhibit opposite spin polarization. It can be seen that the energy change caused by the strain $(u_1(\bm{r}),u_2(\bm{r}))$ ($\sim$ meV) on the band structures is much smaller than the band width (tens of meV). However, being consistent with the main text, we find this small change is enough to drive a topological band inversion at moir\'e $\kappa$ valley (see Fig.~\ref{fig:FIGS5}(c),(d) and (e)). The gap at $\kappa$ and the Chern number of top moir\'e  band as a function of the strain strength $u$ is displayed in Fig.~\ref{fig:FIGS5}(f) and Fig.~\ref{fig:FIGS5}(g), respectively. It clearly shows that the top moir\'e band becomes topological after the gap is inverted by the strain. The topological gap, orders of one meV, is also consistent with the  experiments \cite{Fai_ex2021}. The results from the strained tight-binding model that takes account of the atomic displacement induced by the lattice relaxation directly are thus in agreement with our results of main text

%

\subsection{From tight-binding model to continuum model}

To further demonstrate this consistency, let us estimate the gap according to the pseudomagnetic field generated by the strain configuration  $(u_1(\bm{r}),u_2(\bm{r}))$. As we mentioned in Eq.~(\ref{guage}), to know the pseudomagnetic field, we need to obtain the parameter $\beta$ that characterizes how strong the strain couples with the electron's momentum. It can be obtained by comparing the energy dispersion of   the strained tight-binding model and the effective continuum model Hamiltonian. 

According to symmetry analysis, the continuum Hamiltonian near $K$ valley under strain takes the form
\begin{eqnarray}
	H_{eff} = -\frac{\hbar^2k^2}{2m^{*}} +f_0 (u_{xx} + u_{yy}) + f_1 a [(u_{xx} - u_{yy})k_x  - 2u_{xy}k_y].\label{continuum}
\end{eqnarray}
The values of the parameters can be determined by fitting the energy dispersion  of the tight-binding model near $K$ valley, which gives $m^{*} = 0.6357 m_{e}$,  $f_1 \approx 0.5826 $ eV. Note that the second term in Eq.~(\ref{continuum}) is not relevant to our discussion and will be compensated by the chemical potential term. Comparing with the definition of the gauge field $\bm{A} = \frac{\sqrt{3}\beta\hbar}{2ea}(u_{xx}-u_{yy}, -2u_{xy})$, we find $\beta=\frac{2m^{*}a^{2}}{\sqrt{3}\hbar^2}f_{1}=0.72$.

After getting the value of $\beta$, we can obtain the pseduomagnetic field $B(\bm{r})$ from each strain configuration  $(u_1(\bm{r}),u_2(\bm{r}))$ according to Eq.~(\ref{guage}) and Eq.~(\ref{B_field}).  According to the main text, the magnitude of the pseduomagnetic field $B_{ps}=B(\bm{r}=0)/3$ and the gap change ratio is $0.05$ meV/T. The gap at the moir\'e $\kappa$ valley is thus expected to be
\begin{equation}
	\epsilon_g(B_{ps})\approx |\epsilon_g(0)-0.05B_{ps}|.\label{Eq_gap}
\end{equation}  In this way, we estimated the gap at $\kappa$ as a function of strain strength of ($u_{xx}-u_{yy},u_{xy})$, as shown in Fig.~\ref{fig:FIGS5}(f) (red line). It clearly shows the gap directly calculated from the strained tight-binding model matches with the pseudomagnetic field picture of the main text.
	
\end{document}